\documentclass[a4paper,fleqn,usenatbib]{mnras}
\usepackage{newtxtext,newtxmath}

\usepackage[T1]{fontenc}
\usepackage{ae,aecompl}

\usepackage{graphicx}	
\usepackage{amsmath}	
\usepackage{amssymb}	
\usepackage{longtable}
\usepackage{ltxtable}
\usepackage{lscape}
\usepackage{rotating}
\usepackage{multicol}
\usepackage{enumitem}
\usepackage{color}
\usepackage{cite}
\usepackage{natbib} 
\usepackage{journal_names}
\usepackage{afterpage}
\usepackage{caption}
\usepackage[flushleft]{threeparttable}
\usepackage{subfig}
\usepackage{multirow}
\usepackage{comment}

\title[MYSO Jets]{A search for non-thermal radio emission from jets of massive young stellar objects}

\author[W. O. Obonyo et al.]{
W. O. Obonyo,$^{1}$\thanks{E-mail: pywoo@leeds.ac.uk}
S. L. Lumsden,$^{1}$
M. G. Hoare,$^{1}$
S. J. D. Purser,$^{3}$
S. E. Kurtz$^{2}$
\newauthor
and K.G. Johnston$^{1}$
\\
$^{1}$School of Physics and Astronomy,
The University of Leeds,
Woodhouse Lane, Leeds LS2 9JT, United Kingdom.
\\
$^{2}$Instituto de Radioastronom\'ia y Astrof\'isica, Universidad Nacional Aut\'onoma de M\'exico, Apartado Postal 3-72, Morelia 58089, M\'exico\\
$^{3}$Astronomy \& Astrophysics, Dublin Institute for Advanced Studies (DIAS), 10 Burlington Road, Dublin 4, Ireland.
}

\date{Accepted XXX. Received YYY; in original form ZZZ}

\pubyear{2015}

\begin{document}
\label{firstpage}
\pagerange{\pageref{firstpage}--\pageref{lastpage}}

\maketitle

\begin{abstract}
Massive young stellar objects (MYSOs) have recently been shown to drive jets whose particles can interact with either the magnetic fields of the jet or ambient medium to emit non-thermal radiation. 
We report a search for non-thermal radio emission from a sample of 15 MYSOs to establish the prevalence of the emission in the objects. 
We used their spectra across the L-, C- and Q-bands along with spectral index maps to characterise their emission. 
We find that about 50\% of the sources show evidence for non-thermal emission with 40\% showing clear non-thermal lobes, especially sources of higher bolometric luminosity. The common or IRAS names of the sources that manifest non-thermal lobes are; V645Cyg, IRAS 22134+5834, NGC 7538 IRS 9, IRAS 23262+640, AFGL 402d and AFGL 490. All the central cores of the sources are thermal with corresponding mass-loss rates that lie in the range $\mathrm{\sim 3 \times 10^{-7}}$ to $\mathrm{7 \times 10^{-6}\,M_\odot yr^{-1}}$. Given the presence of non-thermal lobes in some of the sources and the evidence of non-thermal emission from some spectral index maps, it seems that magnetic fields play a significant role in the jets of massive protostars. Also noted is that some of the sources show evidence of binarity and  variability. 

\end{abstract}

\section{Introduction}
Stellar jets and outflows are ever-present in both low and high mass young stellar objects \citep{2018A&ARv..26....3A}. They are detectable at radio wavelengths where the effect of dust extinction is minimal. Earlier observational studies of the jets from massive stars were
limited to a few nearby objects e.g. HH 80-81 \citep{1998ApJ...502..337M}, Cepheus A HW2 \citep{2006ApJ...638..878C}, IRAS 16547-4247 \citep{2003ApJ...594L.131B} and G345.5+01.46 \citep{2010ApJ...725..734G}. However, with the upgrade of both the Australia Telescope Compact Array (ATCA) and the Very Large Array(VLA), it is now possible to study more distant objects in larger numbers. Such studies were conducted by \citet{2012ApJ...753...51G}, \citet{2016MNRAS.460.1039P}, \citet{2016ApJS..227...25R} and \citet{2016A&A...585A..71M} who confirmed that jets are ubiquitous in massive protostars. The jets are indicators of disk-fed accretion \citep{2001ARA&A..39..403R}, tracers of accretion history \citep{2015ApJ...800...86K} and the vents through which part of the high radiation pressure from massive protostars are
released (\citealt{2014ASSP...36..379K},  \citealt{2007ApJ...660..479B}). They also remove excess angular momentum from the protostar to the ambient medium, allowing for further accretion \citep{2018A&ARv..26....3A}.

Models reveal a close connection between protostellar jets and accretion disks \citep{2007prpl.conf..277P}. This implies that the presence of a jet is an indicator of the existence of an accretion disk and vice versa. Indeed, a few massive young stellar objects that drive jets are also known to harbour accretion disks e.g. IRAS 13481-6124 (\citealt{2010Natur.466..339K}, \citealt{2016A&A...589L...4C}), AFGL 2591 \citep{2012A&A...543A..22W} 
and G023.01$-$00.41 \citep{2018arXiv180509842S}. These disks, if optically thick, might explain how the objects accrete mass in the presence of high radiation pressure and strong UV fields \citep{2009ApJ...702..567V}. 

Typically, a massive young stellar object (MYSO) is understood to consist of a central object, a disk and a jet. However, it is only the jet that is expected to emit non-thermal radiation (\citealt{2013EPJWC..6103003C}, \citealt{2010ApJ...723..449L}, \citealt{2007ApJ...668L.159V}).
Charged jet particles can be accelerated (\citealt{1978MNRAS.182..147B}, \citealt{1987PhR...154....1B}) to relativistic levels at shocks where they may interact with magnetic fields giving rise to non-thermal (synchrotron) emission. The shocks are therefore traced by the non-thermal radio knots and Herbig-Haro (HH) objects. Some of the theories put forward to explain how shocks are formed in YSOs include episodic accretion \citep{2015MNRAS.450..295D} and interaction between jet material with the ambient medium and/or internal working surfaces within the jet \citep{2007A&A...472..855R}. Star formation models that incorporate disks can explain how the shocks are formed in MYSOs. For instance, an unstable self-gravitating disk can fragment into clumps that are accreted episodically towards a protostar, causing accretion bursts \citep{2017MNRAS.464L..90M}. Similarly, a wobbling magnetised disk has a potential of producing knots at working surfaces \citep{1993ASSL..186..301R} if jet material spirals through its hourglass-like field lines (\citealt{1982MNRAS.199..883B}, \citealt{1986ApJ...301..571P}, \citealt{2005IAUS..227..163P}).
Recent observations of HH 80-81 (\citealt{2017ApJ...851...16R}, \citealt{2018MNRAS.474.3808V}) and the triple radio source in the Serpens star-forming region \citep{2016ApJ...818...27R} have confirmed that massive protostars emit both thermal and non-thermal radiation. They used spectral indexing to classify the emissions. The study by \citet{2018MNRAS.474.3808V} demonstrated the value of studying the spectral index for non-thermal emission at lower radio frequencies. 

Unlike thermal jets and outflows which have been well modelled (\citealt{1986ApJ...304..713R}), models of non-thermal jets can be improved with an increase in observational data. Numerical simulations can also be used to provide clues on how both magneto-hydrodynamic (MHD) and radiative-hydrodynamic (RHD) jets \citep{2005AdSpR..35..908D} are launched, accelerated and collimated. In-depth study of the objects can reveal whether they represent different evolutionary phases. Claims have been made that a MYSO can drive either a MHD jet e.g. HH 80-81 \citep{2010Sci...330.1209C}, a RHD jet e.g.  W75N(B) \citep{2015Sci...348..114C} or a RHD equatorial wind e.g. S140 IRS 1 \citep{2006ApJ...649..856H}. 

Motivated by the need to establish the prevalence of non-thermal emission in protostellar jets, we observed a sample of MYSOs at L-band (frequency range 1-2\,GHz and central frequency 1.5\,GHz) where synchrotron emission is dominant and easily detectable. We also used previous C- and Q-band observations from the JVLA's A configuration by Purser et al. 2019 (in prep). The C- and Q-band observations were of bandwidths 2\,GHz and 8\,GHz, centred at 6\,GHz and 44\,GHz, respectively. 

\section{Sample Selection}
A total of 15 objects (listed in Table \ref{table_samplet}) were selected from a sample of 63 MYSOs that were previously observed in C-band at 6\,GHz using the NRAO's \footnote{National Radio Astronomy Observatory} Jansky Very Large Array (JVLA) telescope (details in Purser et al. 2019 in prep). We selected the 15 brightest objects from that sample to provide the best possibility for detection in L-band at 1.5\,GHz.

\begin{table*}
\caption{Table showing object names, pointing centres, distance from the Sun and Bolometric luminosities as taken from the \citet{2011MNRAS.418.1689U} catalogue. Local rms noise levels of the maps and properties of their restoring beams (size in arcsec and position angle in degrees) at L-band are also shown.}

\begin{tabular}{l l c c c c c c c }
\hline
RMS Name & Common &   RA   &    Dec   &   d   & $ L_{Bol} $ & Field rms  & Beam size &   PA\\
           & Name  & (J2000) &  (J2000)   & (kpc) & $ (L_{\odot})$& ($\mathrm{\mu Jy/beam}$) & maj ($^{\prime\prime}$) $\times$ min ($''$) & $ (^o)$\\

\hline
G083.7071+03.2817  & -               & 20$^h$33$^m$36.51$^s$ & +45$^\circ$35$^{\prime}$44.0$^{\prime\prime}$  & 1.4 & 3900 & $25$ & $1.45 \times 1.13$ & 83.0   \\
G094.2615-00.4116  & IRAS 21307+5049 & 21$^h$32$^m$30.59$^s$ & +51$^\circ$02$^{\prime}$16.0$^{\prime\prime}$  & 5.2 & 9000 & $33$ &$1.47 \times 1.09$  & 82.3   \\
G094.4637-00.8043  & IRAS 21334+5039 & 21$^h$35$^m$09.14$^s$ & +50$^\circ$53$^{\prime}$08.9$^{\prime\prime}$  & 4.9 & 21000& $25$ &$1.47	\times 1.11$  & 82.6   \\
G094.6028-01.7966  & V645Cyg              & 21$^h$39$^m$58.25$^s$ & +50$^\circ$14$^{\prime}$20.9$^{\prime\prime}$  & 4.9 & 43000& $24$ &$1.49 \times 1.13$  & 83.1  \\
G103.8744+01.8558  & IRAS 22134+5834 & 22$^h$15$^m$08.97$^s$ & +58$^\circ$49$^{\prime}$07.3$^{\prime\prime}$  & 1.6 & 6800 & $23$ &$1.53	\times 1.08$  & 78.2  \\
G108.5955+00.4935C & IRAS 22506+5944 & 22$^h$52$^m$38.09$^s$ & +60$^\circ$01$^{\prime}$01.1$^{\prime\prime}$ & 4.3 & 3000 & $17$ &$1.30	\times 1.10$  &-26.7  \\
G110.0931-00.0641  & IRAS 23033+5951 & 23$^h$05$^m$25.16$^s$ & +60$^\circ$08$^{\prime}$15.4$^{\prime\prime}$  & 4.3 & 11850& $18$ &$1.31	\times 1.16$  &-19.8  \\
G111.2552-00.7702  & IRAS 23139+5939 & 23$^h$16$^m$10.40$^s$ & +59$^\circ$55$^{\prime}$28.2$^{\prime\prime}$  & 3.5 & 9870  & $18$ &$1.39	\times 1.20$  &-22.9 \\
G111.5671+00.7517  & NGC 7538 IRS 9  & 23$^h$14$^m$01.76$^s$ & +61$^\circ$27$^{\prime}$19.9$^{\prime\prime}$  & 2.7 & 44620 & $29$ &$1.32	\times 1.22$  &-17.9  \\
G114.0835+02.8568  & IRAS 23262+6401  & 23$^h$28$^m$27.76$^s$ & +64$^\circ$17$^{\prime}$38.5$^{\prime\prime}$  & 4.2 & 7130  & $20$ &$1.33	\times 1.06$  &-13.5 \\
G126.7144-00.8220  & S187 IR & 01$^h$23$^m$33.17$^s$ & +61$^\circ$48$^{\prime}$48.2$^{\prime\prime}$  & 0.7 & 2600  & $20$ &$1.47	\times 1.06$  &-25.1 \\
G136.3833+02.2666  & IRAS 02461+6147 & 02$^h$50$^m$08.57$^s$ & +61$^\circ$59$^{\prime}$52.1$^{\prime\prime}$  & 3.2 & 7800  & $40$ &$1.50 \times 1.15$  &-05.1 \\
G138.2957+01.5552  & AFGL 402d & 03$^h$01$^m$31.32$^s$ & +60$^\circ$29$^{\prime}$13.2$^{\prime\prime}$  & 2.9 & 17000 & $25$ &$1.45	\times 1.12$  &25.8  \\
G139.9091+00.1969A & AFGL 437s & 03$^h$07$^m$24.52$^s$ & +58$^\circ$30$^{\prime}$43.3$^{\prime\prime}$  & 3.2 & 11000 & $25$ &$1.29	\times 1.06$  &21.2  \\
G141.9996+01.8202  & AFGL 490 & 03$^h$27$^m$38.76$^s$ & +58$^\circ$47$^{\prime}$00.1$^{\prime\prime}$  & 0.8 & 5500  & $28$ &$1.30	\times 1.06$  &27.4  \\
\hline
\end{tabular}
\label{table_samplet}
\end{table*}

The 15 objects chosen are all from the RMS catalogue \footnote{http://rms.leeds.ac.uk/cgi-bin/public/RMS\textunderscore DATABASE.cgi} \citep{2013ApJS..208...11L}. Other selection criteria include; a distance $d\lesssim 7$ kpc from the Sun and bolometric luminosities $L_{bol} \gtrsim 2500L_{\odot}$. Given that the typical sizes of MYSO jets lie in the range $\sim 0.1 \,$pc to $2.4\,$pc \citep{1994ASPC...62..237M}, we should resolve some of the the lobes within the jets at the frequency of observation and telescope configuration (i.e resolution of $1.2^{\prime\prime}$).



\section{Observations and data reduction}
The 15 objects were observed on the \rm{$8^{th}$}, \rm{$19^{th}$} and \rm{$22^{nd}$} of August 2015 using the JVLA under project code 15A-218. The observation made use of twenty seven antennae of the A configuration giving a resolution $\theta_{HPBW}$ of  $\sim 1.2^{\prime \prime}$ at the central frequency of L-band. 
The continuum data were subdivided into 16 spectral windows (SPW), each of bandwidth 64\,MHz. The spectral windows were further subdivided into 64 channels, each of width 1\,MHz to ease flagging procedure and control the spread of RFI across the observed frequencies. The sources observed at L- were also observed at C- band in 2012 and some of them at Q- band in 2014/15 (Purser et al. 2019 in prep) using the JVLA's A configuration. 
Synthesised beams of C- and Q-bands are typically 0.33$^{\prime\prime}$ and 0.04$^{\prime\prime}$ respectively. Noise levels in C- and Q-band maps lie in the range $\sim6.5-10\,\mu$Jy/beam and $\sim35-60\,\mu$Jy/beam at full $uv-$range i.e 13-800$\mathrm{k\lambda}$ and 100-5300$\mathrm{k\lambda}$ respectively.

The L-band sources were observed with a phase calibrator cycle time of $\sim 3 - 7$ minutes to correct for instrumental and atmospheric effects on their phases and flux densities. Each of the science objects had a total integration time of $\sim 30 - 40$ minutes. Flux calibrators, $\mathrm{0137+331=3C48}$ and $\mathrm{1331+305=3C286}$ were also observed for flux scaling.

Calibration and imaging were done using  NRAO's CASA (Common Astronomy Software Applications; \citealt{2007ASPC..376..127M}). If an image had a bright object in its field, a phase-only self-calibration of between 1-3 iterations was carried out to improve it.
While CLEANing the data, their $uv$ visibilities were weighted using Briggs weighting with a robustness parameter of 0.5 \citep{1995AAS...18711202B} resulting in synthesised beams whose average sizes are $\sim 1.4^{\prime\prime} \times 1.1^{\prime\prime}$. The rms noise levels of the cleaned maps of full $uv-$ coverage, $\mathrm{\sim2-220\,k\lambda}$, lie in the range $18 - 40 \, \rm{\mu Jy/beam}$ with objects close to bright and extended sources having higher rms noise. A list showing the objects, their local rms noise and sizes of synthesised beams used in deconvolution is shown in Table \ref{table_samplet}.

\section{Near Infrared (NIR) Emission}
Massive protostars are embedded in natal clouds containing dust particles that can be distributed into various geometries by jets and outflows from the protostars, creating cavities and shells within the clouds. Light from the embedded sources can then scatter off the dust in the cavities at NIR wavelengths, tracing their structure. Molecular hydrogen emission at 2.12\,$\mu$m from shocks produced by jets are also seen. As a result, infrared images were relied upon to provide helpful hints for the presence of protostellar outflows and jets.

Most of the infrared images used in the study were generated by the Wide Field Camera (WFCAM; \citealt{2007A&A...467..777C}) of the United Kingdom Infra-red Telescope (UKIRT; \citealt{2007MNRAS.379.1599L}) as part of the UKIDSS Galactic Plane Survey \citep{2008MNRAS.391..136L}. UKIRT is a 3.8-m telescope whose typical angular resolution is $\sim$0.8$^{\prime\prime}$ \citep{2008MNRAS.391..136L}. WFCAM, whose pixel scale is 0.4$^{\prime\prime}$, observes at five IR wave bands; Z, Y, J, H and K of wavelength range 0.83$-$2.37\,$\mu$m. Only three of the bands;\ J, H and K of central wavelengths 1.2483\,$\mu$m, 1.6313\,$\mu$m and 2.2010\,$\mu$m respectively \citep{2006MNRAS.367..454H} were used in the study. In addition, 2MASS (Two Micron All Sky Survey) images of some of the objects were inspected, especially if their declinations are above the region of the sky covered by UKIDSS i.e $\delta >$60$^o$, even-though the images are of a lower resolution ($\sim 2^{\prime \prime}.5$) and sensitivity i.e a pixel size of $2^{\prime\prime}$ \citep{2006AJ....131.1163S}. Also used were findings from H$\mathrm{_2}$ emission observations of the MYSOs e.g. \citet{2015MNRAS.450.4364N}, \citet{2017ApJ...844...38W}, \citet{2010MNRAS.404..661V} and \citet{1998AJ....115.1118D}.

\section{Results and discussion}
\subsection{L-band results}
L-band images of all the sources are shown in Figure \ref{L_band_images}. Positions, fluxes, sizes and position angles of all their components are presented in Table \ref{table_results}. The components can be classified as cores or lobes. A core is detectable in the near/mid infrared, brighter at higher radio frequencies, e.g. Q-band,  and is likely to be located between lobes, if the components show a linear orientation. Lobes, on the other hand, are seen a few arc-seconds away from the cores, aligned in the direction of the outflow cavities. Thermal lobes are detectable at higher frequencies while non-thermal lobes are more prominent at lower radio frequencies.

The L-band images display a variety of morphologies ranging from emission that comes from cores with radio lobes and partially resolved cores to point-like counterparts. G094.4637, G094.6028, G111.5671, G114.0835, G138.2957 and G141.9996 have radio lobes, suggesting that they harbour cores that drive the jets. G110.0931, on the other hand, is resolved, displaying an extended and jet-like structure but does not have radio lobes at L-band. However, it encloses multiple components at higher resolution, in C-band. Some of the components may be lobes. Other sources whose L-band emission encloses multiple components at the higher resolution, in C-band, are G094.2615, G094.4637 and G139.9091. Four sources; G083.7071-A, G108.5955C, G111.2552 and G139.9091 are partially resolved while three; G103.8744-A, G114.0835-A and G126.7144 are point-like. 

The cores of all the protostars were detected at L-band except for G103.8744, G136.3833 and G138.2957. None of the detections in the field of G103.8744 appear to be its core; they are all located a few arcseconds away from the position of the MYSO. G136.3833's field has a bright source that is approximately $6^{\prime\prime}$ to the west. This source does not manifest an explicit association with the MYSO. G138.2957, on the other hand, has two non-thermal lobes that are approximately 2$^{\prime\prime}$ apart. Each of the lobes are located $\sim 1^{\prime\prime}$ away from the position of the core (see Figure \ref{L_band_images}). All the L- and C-band emission have a positional match that agree within $\sim 0.3^{\prime\prime}$. However, NIR positions, taken from 2MASS, show slight offsets from the radio emission, perhaps because the emission is from a section of an outflow cavity that is further away from the source and is less obscured. In some cases, no radio emission coincides with the NIR position e.g. G103.8744 where the peak of the radio sources are positioned a few arcseconds away from the peak of the IR emission.

\begin{figure*}
    \centering
    
    \includegraphics[width = 0.43\textwidth]{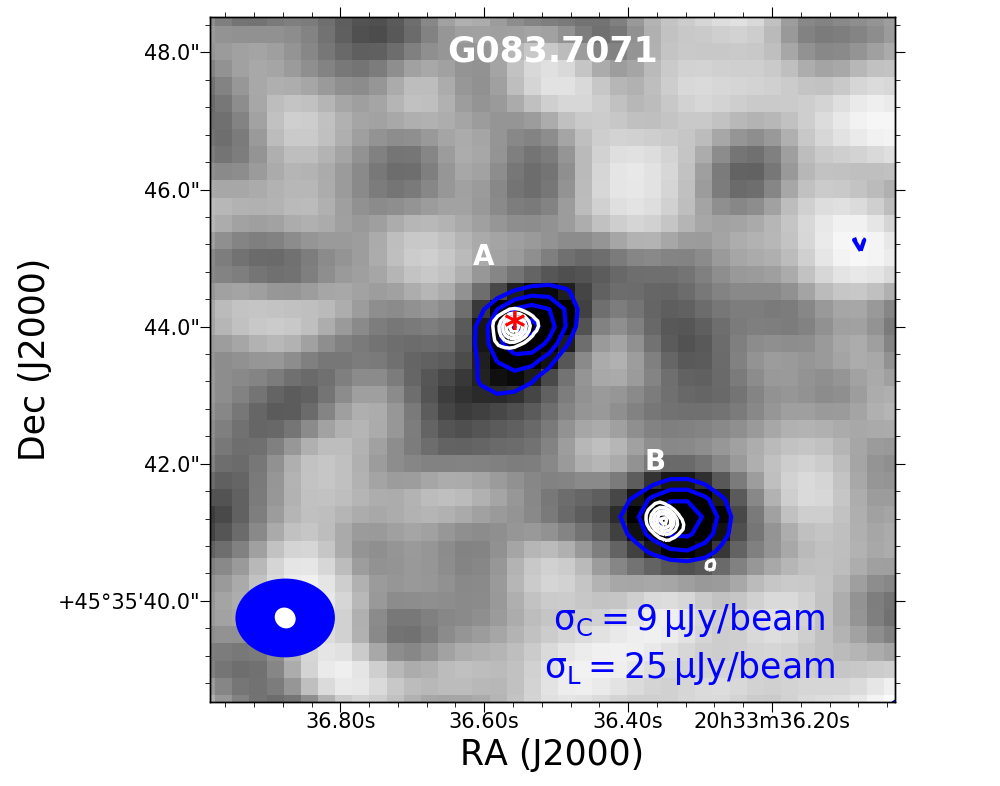}
    \includegraphics[width = 0.43\textwidth]{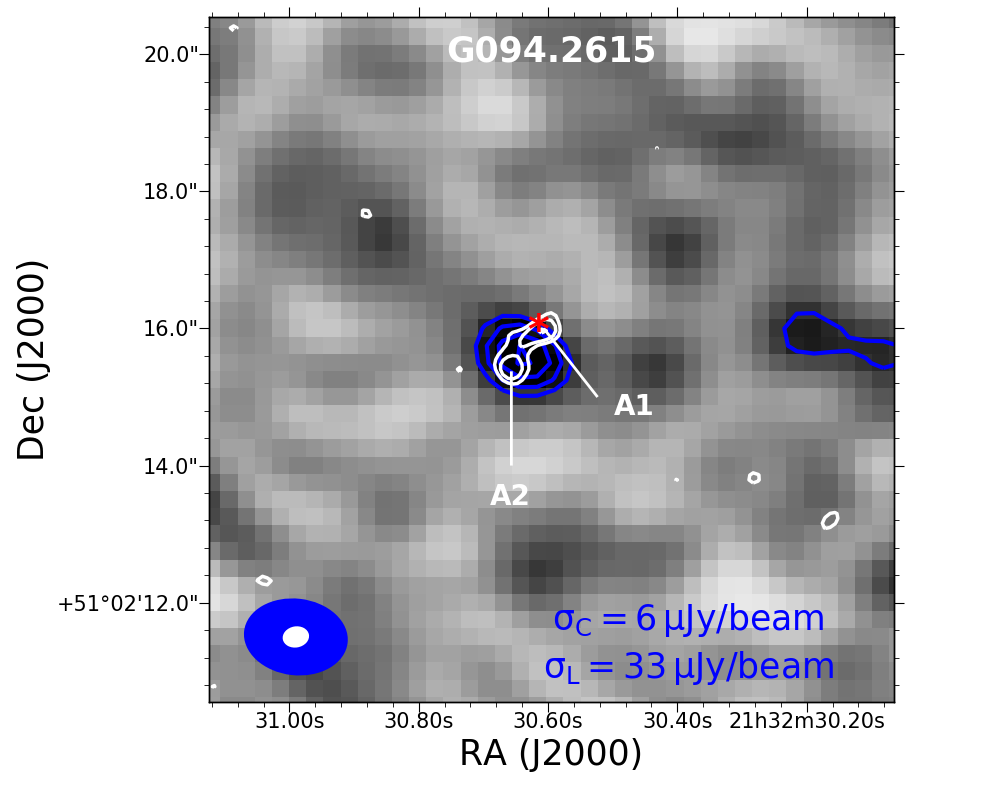}
    \includegraphics[width = 0.43\textwidth]{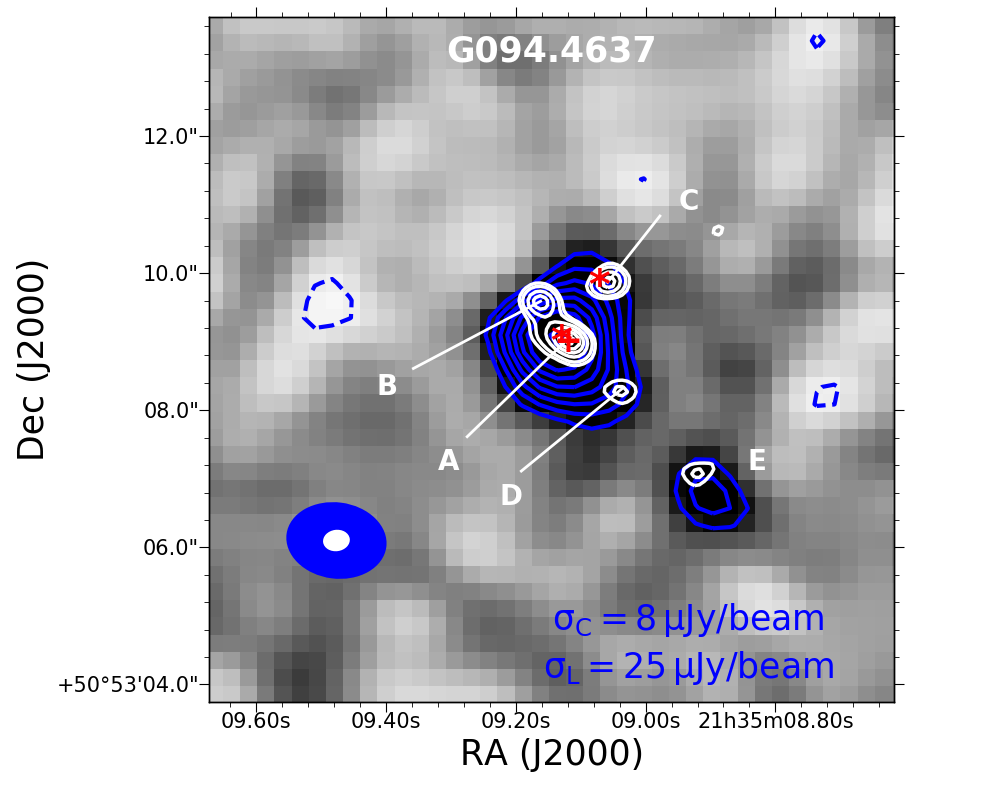}
    \includegraphics[width = 0.43\textwidth]{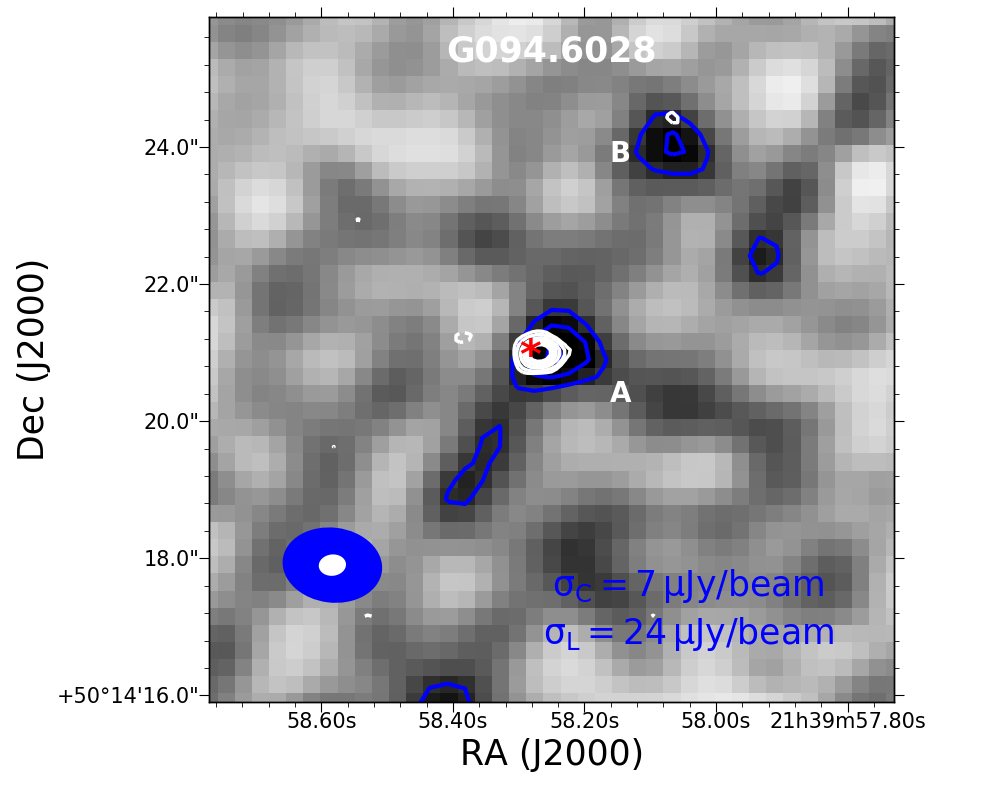}
    \includegraphics[width = 0.43\textwidth]{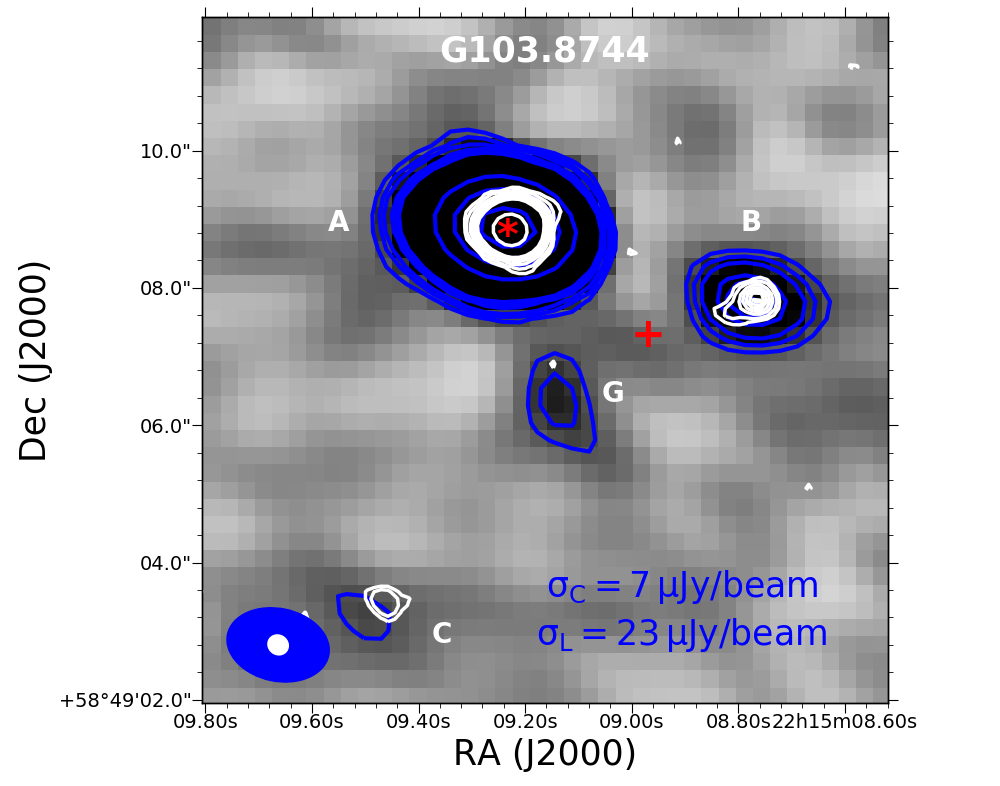}
    \includegraphics[width = 0.43\textwidth]{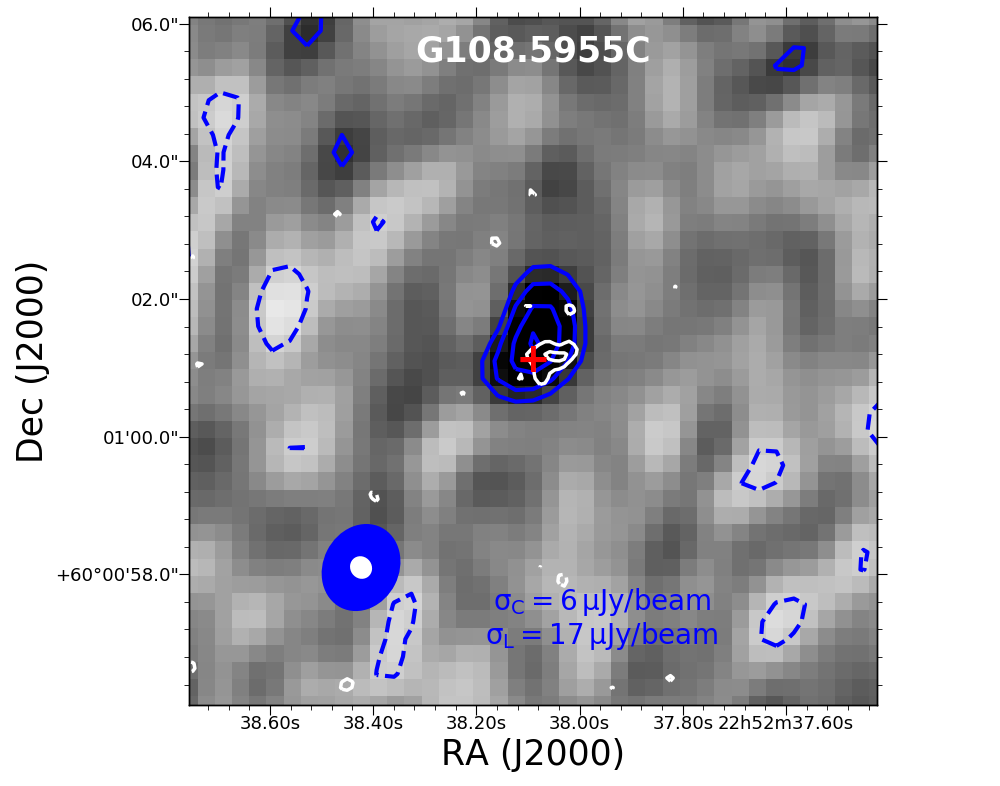}
    \caption{L-band maps of the sources together with their L-  and C-band contours of levels $-3\sigma$(dashed), $3\sigma$, $4\sigma$, $5\sigma$,... and $-3\sigma$(dashed), $3\sigma$, $5\sigma$, $9\sigma$, $13\sigma$, $17\sigma$, $21\sigma$,... shown in blue and white colours respectively. All the cores were detected at L- band except for G138.2957 and G136.3833, both of which were detected at C-band. The red asterisks show the Q-band locations of the MYSOs' cores. The locations agree with the positions of the IR emission from the MYSOs within $1^{\prime\prime}$. IR locations of MYSOs whose Q-band positions are unavailable are indicated with a plus sign. Synthesised beams for both L- and C- (Purser et al. 2019 in prep) bands are shown on the lower left corner of each frame.}
    \label{L_band_images}
\end{figure*}

\begin{figure*}\ContinuedFloat
    \includegraphics[width = 0.43\textwidth]{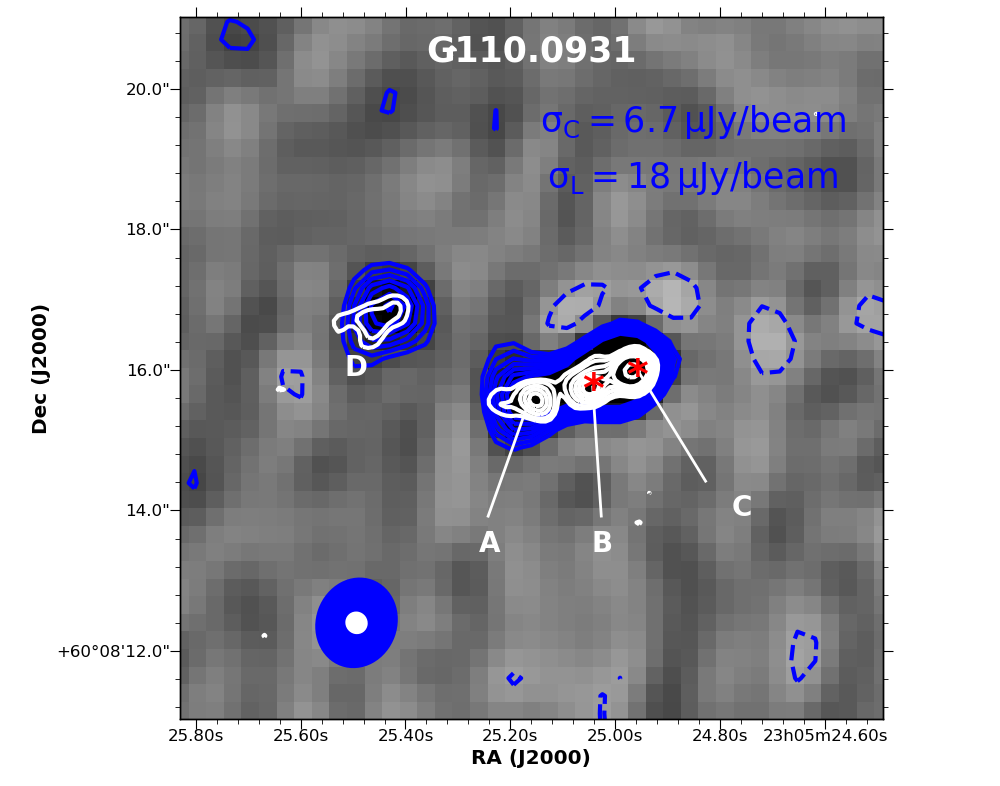}
    \includegraphics[width = 0.43\textwidth]{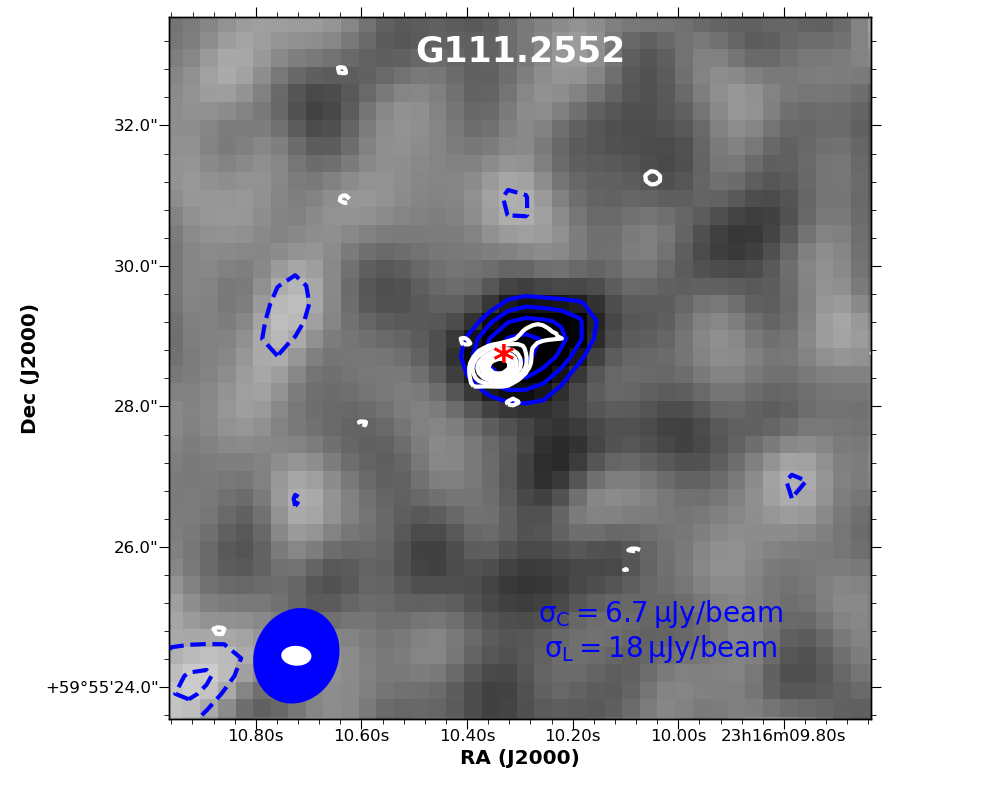}
    \includegraphics[width = 0.43\textwidth]{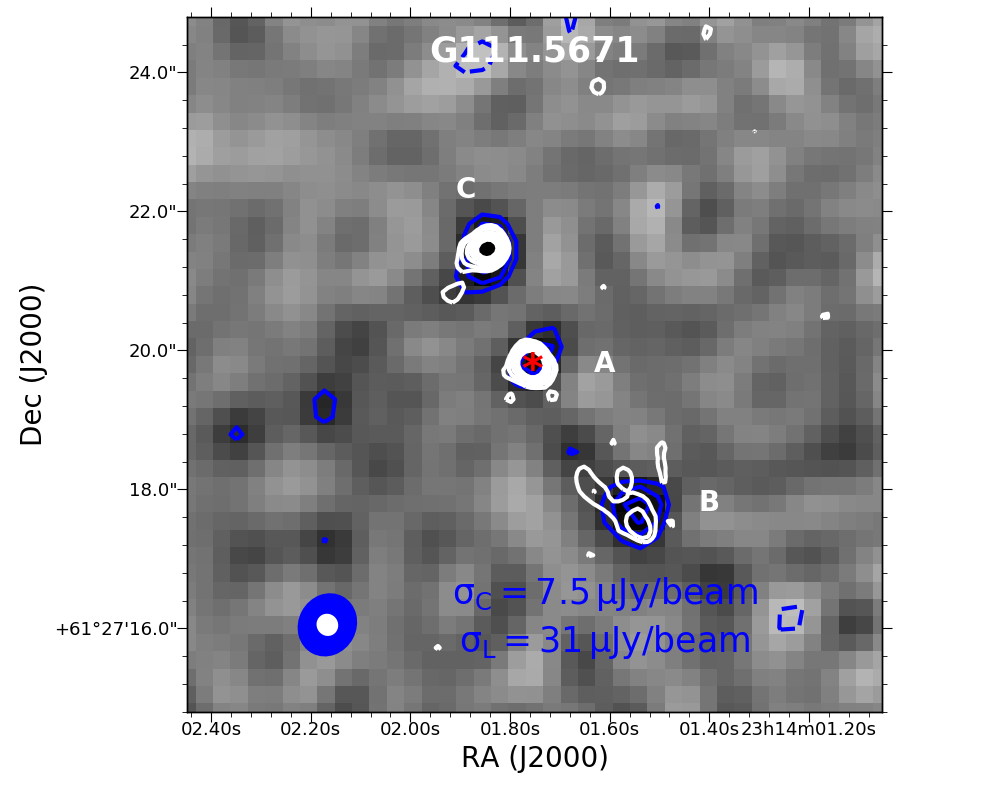}
    \includegraphics[width = 0.43\textwidth]{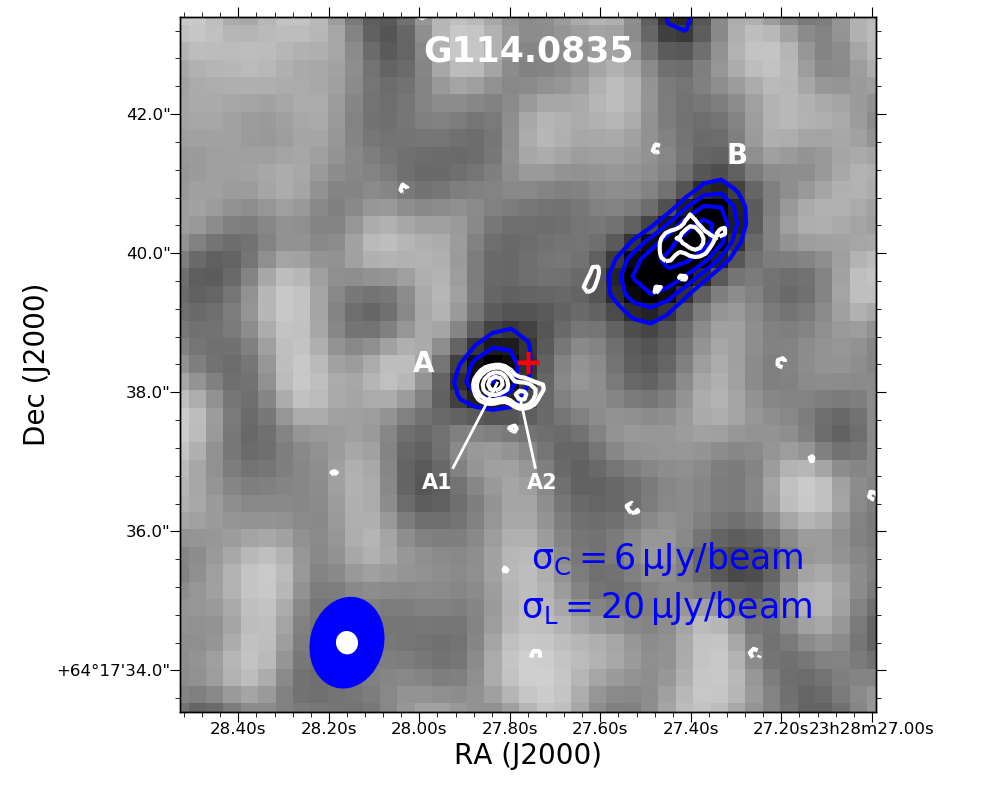}
    \includegraphics[width = 0.43\textwidth]{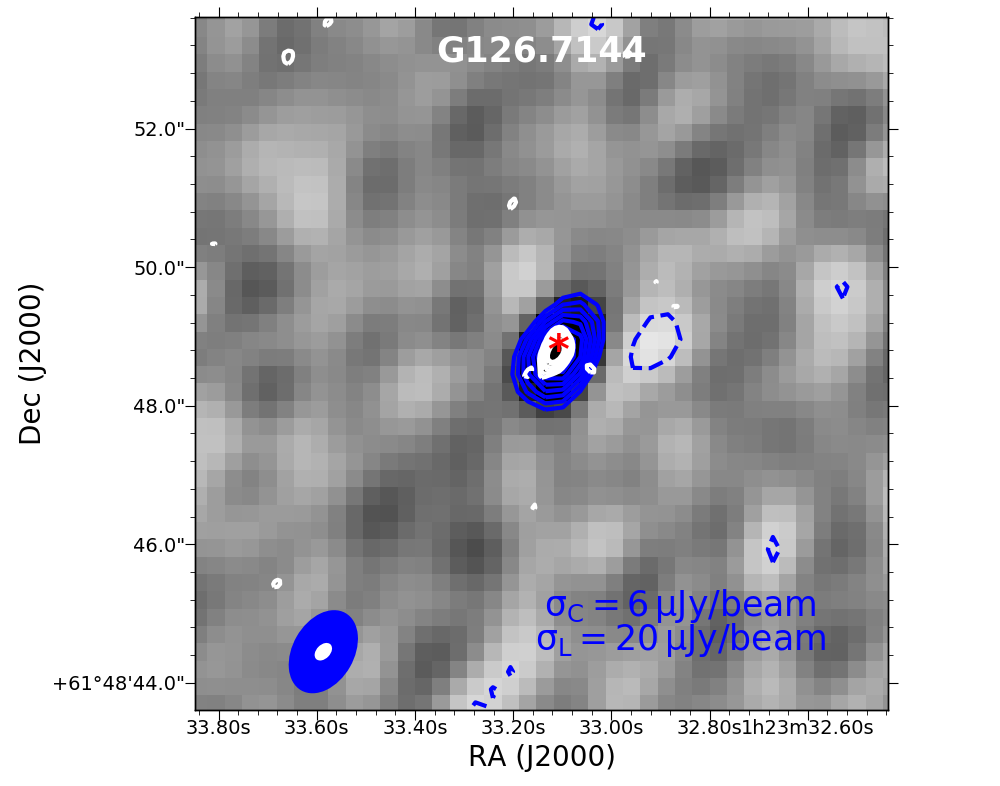}
    \includegraphics[width = 0.43\textwidth]{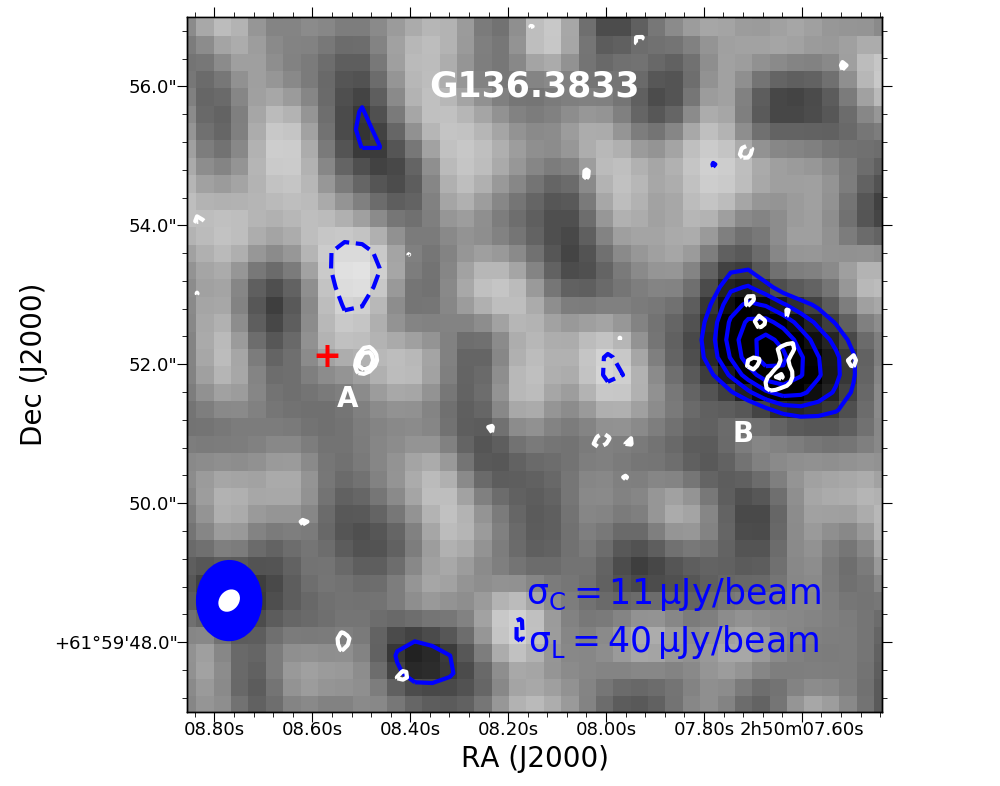}
    \caption{Figure \ref{L_band_images} continued.}
\end{figure*}

\begin{figure*}\ContinuedFloat
    \includegraphics[width = 0.43\textwidth]{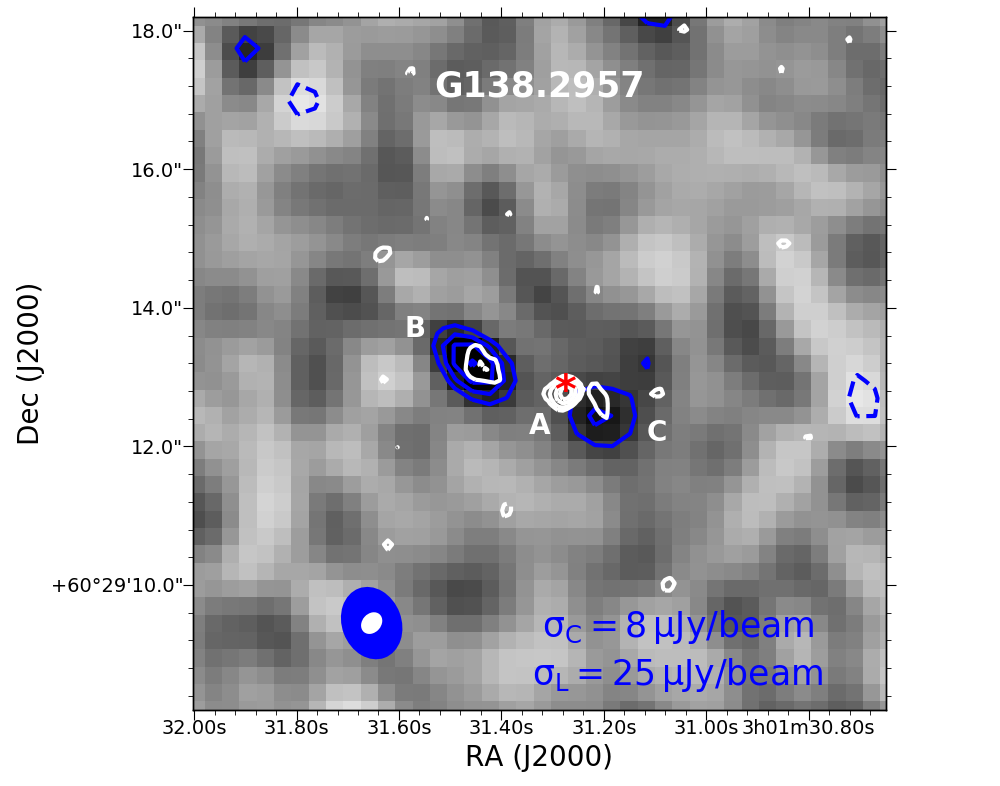}
    \includegraphics[width = 0.43\textwidth]{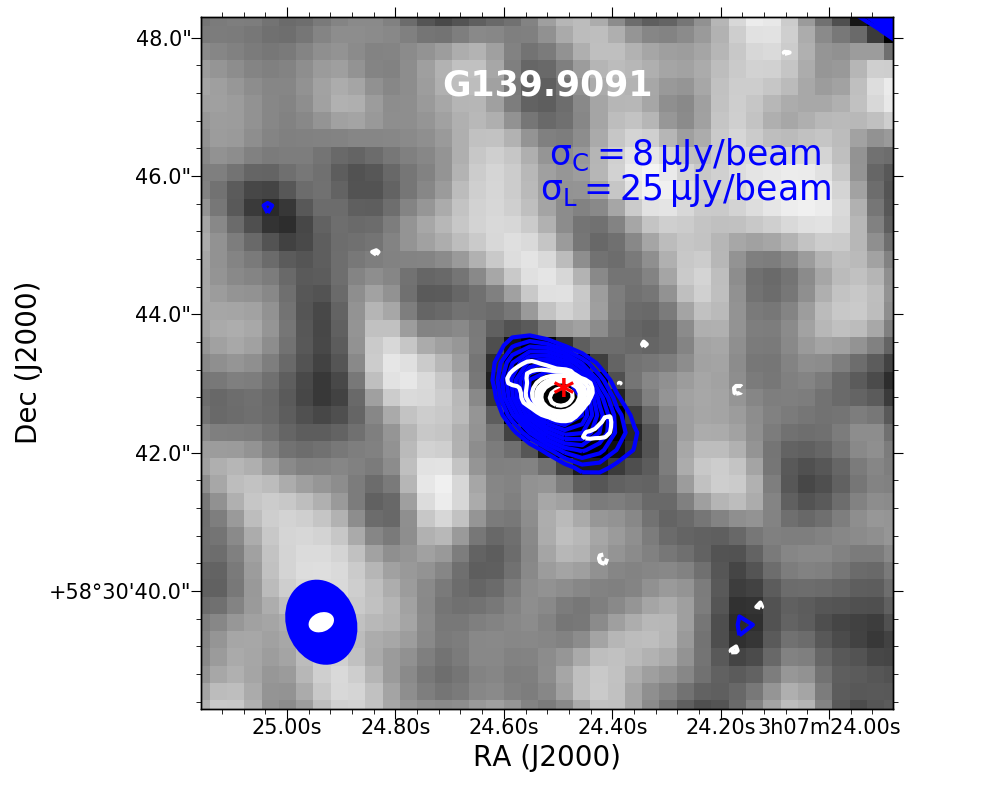}
    \includegraphics[width = 0.43\textwidth]{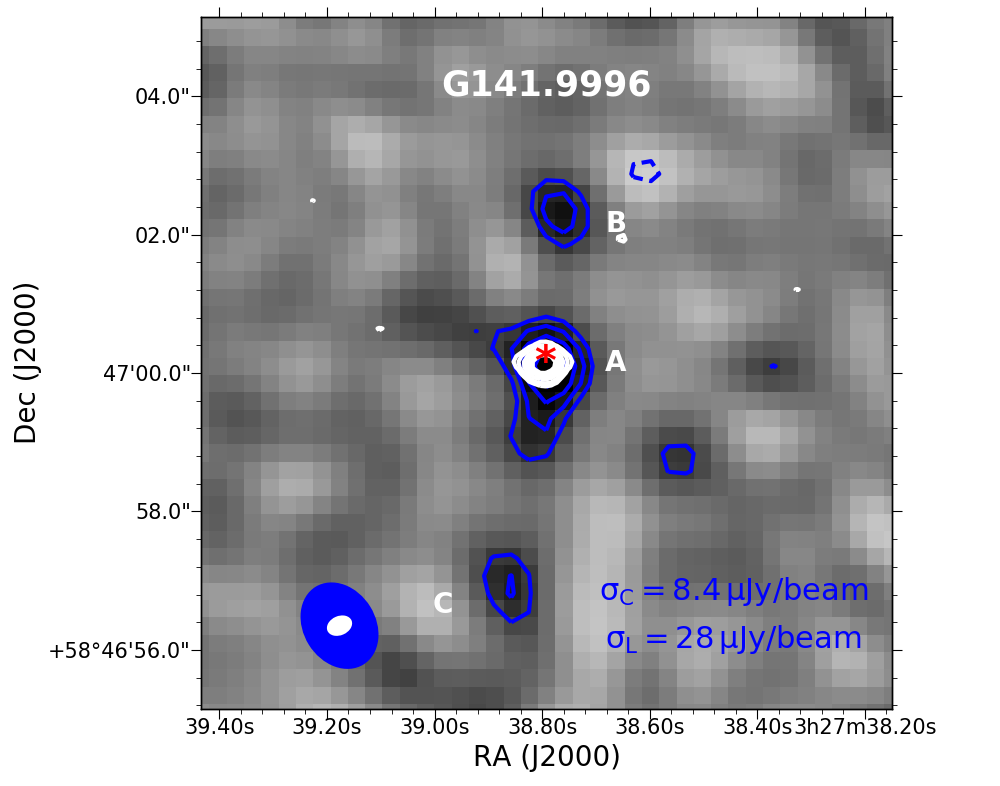}
    \caption{Figure \ref{L_band_images} continued.}
\end{figure*}

The flux densities and peak fluxes of the sources that lie within a radius of 5-10$^{\prime\prime}$ from the MYSOs' position were estimated by enclosing a box around a source. Uncertainties on the fluxes were calculated by averaging fluxes from five different locations of a source's field which were devoid of emission, estimated using a box of the same size. An additional error, approximated at 10\% of a source's flux density was incorporated, in quadrature, to include the uncertainty associated with calibration of absolute flux. 
Sizes of the sources $\theta_S$ were calculated as $\theta_S = \sqrt{\theta_{FWHM} ^2 - \theta_{B} ^2}$, where $\theta_S$ is the deconvolved size of the source, $\theta_{FWHM}$ is the full width at half maximum of a source's flux distribution and $\theta_{B}$ is the full width at half maximum of the beam. 
These quantities, together with their position angles, calculated by inspection, are presented in Table \ref{table_results}. A source's radio emission was considered real if its peak flux is at least three times higher than the root mean square noise in the field (i.e at least $3\sigma$ detection). Sources were considered resolved if their structures are extended with respect to the beam (i.e larger than the beam) otherwise they are compact. 



\begin{table*}

\caption{A table showing the position, flux density, peak flux, size and position angle of emission from the components of the MYSOs at L-band. Sources whose sizes are not shown are point-like at the resolution of 1.2$^{\prime\prime}$. Any other sources that are located near a MYSO and are likely to be associated with it are also listed. The source naming scheme is adopted from Purser et al. 2019(in prep) for consistency.}

\begin{tabular}{ l l c c c c c }
\hline
    Object & RA & DEC & Integrated & Peak flux & Deconvolved Source Size & Position Angle \\
    
     Name & (J2000) & (J2000)& Flux (mJy) & (mJy) & $\mathrm{\theta_{maj}('') \times \theta_{min}('')}$ & ($^o$)  \\
\hline 
G083.7071-A  & 20$^h$33$^m$36.56$^s$ & +45$^\circ$35$^{\prime}$43.9$^{\prime\prime}$ & $0.17\pm0.02$ & $0.13\pm0.02$ & $2.4\pm0.2\times 1.6\pm0.3$ & $130\pm10$  \\
G083.7071-B  & 20$^h$33$^m$36.34$^s$ & +45$^\circ$35$^{\prime}$41.1$^{\prime\prime}$ & $0.09\pm0.02$ & $0.09\pm0.03$ & $-$ & $-$  \\
G094.2615(A1+A2) & 21$^h$32$^m$30.61$^s$ & +51$^\circ$02$^{\prime}$15.8$^{\prime\prime}$ & $0.11\pm0.02$ & $0.13\pm0.03$ &  $2.0\pm0.3\times 1.1\pm0.2$& $48\pm7$   \\
G094.4637($\mathrm{A+B+C+D}$)  & 21$^h$35$^m$09.13$^s$ & +50$^\circ$53$^{\prime}$09.1$^{\prime\prime}$ & $0.38\pm0.04$ & $0.27\pm0.03$ &	$1.8\pm0.4\times 1.6\pm0.4$ & $-$  \\
G094.4637-E    & 21$^h$35$^m$08.91$^s$ & +50$^\circ$53$^{\prime}$07.0$^{\prime\prime}$ & $0.03\pm0.01$	& $0.04\pm0.02$ & $ - $      &   $ - $   \\

G094.6028-A  & 21$^h$39$^m$58.27$^s$ & +50$^\circ$14$^{\prime}$20.9$^{\prime\prime}$ & $0.08\pm0.02$ & $0.12\pm0.02$ & $-$ & $-$ \\
G094.6028-B  & 21$^h$39$^m$58.07$^s$ & +50$^\circ$14$^{\prime}$24.0$^{\prime\prime}$ & $0.07\pm0.02$ & $0.10\pm0.02$ & $-$ & $-$ \\

G103.8744-A  & 22$^h$15$^m$09.25$^s$ & +58$^\circ$49$^{\prime}$08.8$^{\prime\prime}$ & $2.21\pm0.10$ & $1.90\pm0.05$ & $3.3\pm0.3\times 2.3\pm0.2$ & $70\pm15$  \\
G103.8744-B  & 22$^h$15$^m$08.79$^s$ & +58$^\circ$49$^{\prime}$07.8$^{\prime\prime}$ & $0.17\pm0.02$ & $0.23\pm0.02$ & $2.3\pm0.2\times 1.3\pm0.2$ & $63\pm16$  \\
G103.8744-G  & 22$^h$15$^m$09.16$^s$ & +58$^\circ$49$^{\prime}$06.6$^{\prime\prime}$ & $0.09\pm0.03$ & $0.15\pm0.02$ & $-$ & $-$  \\
G108.5955C       & 22$^h$52$^m$38.06$^s$ & +60$^\circ$01$^{\prime}$01.1$^{\prime\prime}$ & $0.20\pm0.03$ & $0.16\pm0.02$ & $1.9\pm0.4\times1.1\pm0.3$	&$160\pm20$    \\ 
G110.0931        & 23$^h$05$^m$25.04$^s$ & +60$^\circ$08$^{\prime}$15.8$^{\prime\prime}$ & $0.59\pm0.08$ & $0.38\pm0.04$ &	$2.6\pm0.3\times1.2\pm0.2$	&$108\pm8$  \\
G110.0931-D        & 23$^h$05$^m$25.43$^s$ & +60$^\circ$08$^{\prime}$16.9$^{\prime\prime}$ & $0.19\pm0.03$ & $0.17\pm0.03$ &	$-$	&$-$  \\
G111.2552        & 23$^h$16$^m$10.34$^s$ & +59$^\circ$55$^{\prime}$28.6$^{\prime\prime}$ & $0.16\pm0.03$ & $0.12\pm0.02$ & $1.7\pm0.5\times1.1\pm0.4$ &$121\pm11 $   \\
G111.5671-A  & 23$^h$14$^m$01.76$^s$ & +61$^\circ$27$^{\prime}$19.8$^{\prime\prime}$ & $0.12\pm0.03$	& $ 0.17\pm0.02$  &	$ - $   & $ - $      \\
G111.5671-B  & 23$^h$14$^m$01.54$^s$& +61$^\circ$27$^{\prime}$17.5$^{\prime\prime}$ & $0.20\pm0.03$& $0.17\pm0.03$ &$-$ & $-$ \\
G111.5671-C  & 23$^h$14$^m$01.85$^s$ & +61$^\circ$27$^{\prime}$21.4$^{\prime\prime}$ & $0.18\pm0.03$& $0.18\pm0.03$ &$ - $      &   $ - $     \\

G114.0835-A  & 23$^h$28$^m$27.82$^s$ & +64$^\circ$17$^{\prime}$38.1$^{\prime\prime}$ & $0.10\pm0.03$ &$0.10\pm0.01$ &$-$ & $-$  \\

G114.0835-B  &23$^h$28$^m$27.40$^s$ &+64$^\circ$17$^{\prime}$40.2$^{\prime\prime}$ & $0.21\pm0.04$ &$0.13\pm0.02$ &$2.5\pm0.4\times1.2\pm0.3$ &	$135\pm8$ \\

G126.7144 & 01$^h$23$^m$33.11$^s$ & +61$^\circ$48$^{\prime}$48.8$^{\prime\prime}$  & $0.47\pm0.05$ &$0.56\pm0.03$ &	$2.0\pm0.2\times1.2\pm0.3$ & $140\pm15$   \\
G136.3833-A & 02$^h$50$^m$08.49$^s$ & +61$^\circ$59$^{\prime}$52.1$^{\prime\prime}$  & $<0.01$ 	& $<0.12$      &	$ - $                           & $ - $        \\
G136.3833-B & 02$^h$50$^m$07.68$^s$ & +61$^\circ$59$^{\prime}$52.2$^{\prime\prime}$  & $0.65\pm0.05$ 	& $0.29\pm0.01$      &	$1.6\pm0.4\times0.8\pm0.3$                           & $60\pm12$        \\

G138.2957-A & 03$^h$01$^m$31.28$^s$ & +060$^\circ$29$^{\prime}$12.8$^{\prime\prime}$  & $<0.06$ & $<0.06$ & $- $  &	$-$ \\
G138.2957-B & 03$^h$01$^m$31.45$^s$ & +060$^\circ$29$^{\prime}$13.2$^{\prime\prime}$  & $0.13\pm0.04$ & $0.16\pm0.02$ & $1.5\pm0.2\times0.8\pm0.2 $  &	$48\pm8$ \\
G138.2957-C & 03$^h$01$^m$31.20$^s$ & +060$^\circ$29$^{\prime}$12.4$^{\prime\prime}$  & $0.10\pm0.03$ & $0.12\pm0.02$ & $-$  &	$-$ \\
G139.9091& 03$^h$07$^m$24.49$^s$ & +58$^\circ$30$^{\prime}$42.8$^{\prime\prime}$  & $0.41\pm0.04$ &$0.34\pm0.02$ & $2.5\pm0.3\times1.5\pm0.2 $ & $ 50\pm8 $    \\
G141.9996-A & 03$^h$27$^m$38.76$^s$ & +58$^\circ$47$^{\prime}$00.2$^{\prime\prime}$  & $0.31\pm0.04$ &$0.18\pm0.02$ & $2.3\pm0.5\times1.6\pm0.3$ & $180\pm21$  \\

G141.9996-B & 03$^h$27$^m$38.77$^s$ & +58$^\circ$47$^{\prime}$02.3$^{\prime\prime}$  & $0.06\pm0.03$ &$0.12\pm0.02$ & $-$ & $-$  \\
G141.9996-C & 03$^h$27$^m$38.86$^s$ & +58$^\circ$46$^{\prime}$57.0$^{\prime\prime}$  & $0.07\pm0.03$ &$0.11\pm0.02$ & $-$ & $-$  \\

\hline
\end{tabular}
\label{table_results}
\end{table*}

\begin{table*}

\begin{threeparttable}
\caption{Flux densities of the objects at different frequencies together with their spectral indices. Both L-and C-band fluxes shown are from images of similar range of uv distance (in $\mathrm{\lambda}$). The Q-band fluxes were taken from Purser et al. 2019 (in prep). For fluxes taken from literature, the reference is shown in brackets. Separable components of the MYSOs are also included in the table. The class tells whether an L-band emission traces a core or a lobe. Known HII regions, assumed to be sources that have stopped driving jets, are also indicated in the table.} 
\label{spectral_indicesT}

\begin{tabular}{ l c c c c c c c c }
\hline
\multirow{2}{*}{Object} & \multicolumn{4}{c}{Integrated Flux (mJy)} & \multicolumn{3}{c}{Spectral Index}  & Class \\
    \cline{2-5}
    \cline{6-8}

 & 1.5\,GHz & 5.8\,GHz & 44.0\,GHz & Other (Ref) &   $\alpha_{LC}$ &$\alpha_{\footnotesize{CQ}}$  & $\alpha_{LQ} $ &\\
    \hline
G083.7071-A & $0.17\pm0.02$ & $0.22\pm0.02$ & $0.68\pm0.06$ &$-$&$0.19\pm0.11$&$0.56\pm0.06$&$0.44\pm0.10$ & Core \\
G083.7071-B & $0.09\pm0.02$ & $0.20\pm0.02$ & $-$ & $-$& $0.60\pm0.13$&$-$ & $-$ & Core\\

G094.2615(A1+A2)& $0.11\pm0.03$ &$0.10\pm0.02$ & $0.24\pm0.08$ & &$-0.07\pm0.25$&$0.44\pm0.20$&$0.23\pm0.18$  & Core/Lobe\\

G094.4637(A+B+C+D)&$0.38\pm0.04$& $0.52\pm0.05$  &  $1.89\pm0.08$ &  &$0.23\pm0.12$&$0.64\pm0.05$&$0.53\pm0.09$& Core(s) + lobe(s)\\
G094.4637-E &$0.03\pm0.01$& $0.04\pm0.02$  & $-$    &$-$&$0.21\pm0.45$&$-$ &$-$& Lobe\\
 
G094.6028-A&$0.08\pm0.02$  &$0.32\pm0.03$ &$0.62\pm0.07$ & (a)&$1.04\pm0.20$&$0.32\pm0.09$&$0.44\pm0.15$ & Core\\
G094.6028-B&$0.07\pm0.02$  &$<0.04$ &$-$ &   &$<-0.42$&$ -$&$-$ & Lobe\\    

G103.8744-A& $ 2.21\pm0.10$ & $3.52\pm0.04$  & $1.56\pm0.34$ & (f) &$0.39\pm0.05$&$-0.25\pm0.06$&$-$ & HII-region\\   
G103.8744-B& $ 0.17\pm0.02$ & $0.25\pm0.06$  & $<0.11$ & $ - $ &$0.29\pm0.20$&$ - $&$ - $ & Unknown\\
G103.8744-C& $ < 0.06$ & $0.09\pm0.02$  & $-$ & $ - $ &$ >0.30 $&$-$&$-$ & Unknown\\
G103.8744-G& $ 0.09\pm0.03$ & $<0.05$  & $-$ & $ - $ &$<-0.44$&$-$&$-$ & Lobe\\

G108.5955C & $0.20\pm0.03$ &$0.26\pm0.04$ & $-$  & $-$& $0.20\pm0.16$&$-$& $-$ & Unknown\\

G110.0931(A+B+C) & $ 0.59\pm0.08 $ & $0.95\pm0.13$ & $1.03\pm0.20$ & (b)  &$0.36\pm0.14$&$0.05\pm0.07$& $0.17\pm0.07$& Core + lobe(s)\\
G110.0931-A &  $ - $  & $0.22\pm0.02$  & $-$ &  (b)& $-$&  $-0.14\pm1.49$& $-$  & Lobe\\  
G110.0931-B &  $-$  &  $0.31\pm0.02$ & $0.62\pm0.15$ & (b)&  $-$&  $0.33\pm0.14$ &$-$  & Core\\
G110.0931-C &  $-$  & $0.49\pm 0.01$ &$0.38\pm0.11$& (b) & $-$ &$-0.11\pm0.07$&$-$ & Unknown\\
G110.0931-D & $ 0.19\pm0.03 $ & $0.17\pm0.02$ & $-$ &$-$& $-0.08\pm0.15$&$-$& $-$& HII-region\\

G111.2552 &$ 0.16\pm0.03$&$0.36\pm0.04$ &$0.35\pm0.07$ & (c)& $0.61\pm0.16$& $0.73\pm0.08$&$0.67\pm0.04$ & Core\\

G111.5671-A & $0.12\pm0.03$ & $0.42\pm0.08$  &$2.46\pm0.10$ & (d)& $0.94\pm0.23$&$0.79\pm0.14$ &$0.84\pm0.09$& Core\\
G111.5671-B & $0.20\pm0.03$ & $ < 0.09 $  &$ - $ &$ - $& $<-0.60 $&$ $ &$ $& Lobe\\
G111.5671-C & $0.18 \pm 0.03$ & $0.40\pm0.09$  &$ - $ &$ - $& $0.60\pm0.21 $&$-$ &$ - $ & Lobe\\ 

G114.0835-A& $0.10\pm0.03$ &$0.17\pm0.03$ & $-$ & $-$ & $0.40\pm0.20$& $-$& $-$  & Core\\
G114.0835-B& $0.21\pm0.04$ &$0.12\pm0.02$ & $-$ & $-$ & $-0.42\pm0.19$& $-$& $-$  & Lobe \\

G126.7144& $0.47\pm0.05$ &$1.28\pm0.05$& $6.86\pm0.24$& $-$ & $0.75\pm0.08$&$0.83\pm0.02$&$0.82\pm0.02$  & Core \\

G136.3833-A& $<0.01$ &$0.04\pm0.01$ &$0.47\pm0.09$& $-$ & $>1.04$&$1.22\pm0.16$& $ >1.20$  & Core \\
G136.3833-B& $0.65\pm0.05$ &$0.26\pm0.02$ &$<0.14$& $-$ & $-0.68\pm0.08$&$<-0.31$& $ <-0.49$  & Unknown \\
 
G138.2957(A+B+C)& $0.23\pm0.05$ &$0.27\pm0.05$ & $0.55\pm0.09$& (e)& $0.12\pm0.15$& $0.35\pm0.09$ &$0.28\pm0.06$& Core + lobes\\
G138.2957-B& $0.13\pm0.04$ &  $-$ & $-$ & $-$& $-$&$-$ & $-$ & Lobe\\
G138.2957-C& $0.10\pm0.03$ &$-$ & $-$ &$-$& $-$& $-$ &$-$ & Lobe\\

G139.9091&$0.41\pm0.04$ &  $0.74\pm0.07$ & $1.39\pm0.14$ & $-$  & $0.44\pm0.10$&$0.31\pm0.07$&$0.36\pm0.04$  & Core\\

G141.9996-A& $0.31\pm0.04$ & $0.37\pm0.04$ &$1.04\pm0.14$ & $-$ &$0.13\pm0.13$&$0.51\pm0.09$& $0.37\pm0.12$ & Core \\

G141.9996-B& $0.06\pm0.03$ & <0.03 &$-$ & $-$ &$<-0.51$&$-$& $-$ & Lobe \\

G141.9996-C& $0.07\pm0.03$ & <0.03 &$-$ & $-$ &$<-0.63$&$-$& $-$ & Lobe \\

\hline
\end{tabular}
        \begin{tablenotes}
            \item[a] \footnotesize{\citet{1997ApJ...482..433D}}
            \item[b] \footnotesize{\citet{2012ApJ...755..100R}}
            \item[c] \footnotesize{\citet{2006AJ....132.1918T}}
            \item[d] \footnotesize{\citet{2005ApJ...621..839S}}
            \item[e] \footnotesize{\citet{1999RMxAA..35...97C}}
            \item[f] \footnotesize{\citet{2016A&A...587A..69W}}
        \end{tablenotes}
\end{threeparttable}

\end{table*}

\subsection{Spectral indices and maps of the sources}\label{spectralindexsec}
Spectral indexing provides a means of classifying radio emitters as either thermal ($\alpha \geq -0.1$) or non-thermal ($\alpha < -0.1$) e.g \citet{2017ApJ...851...16R}, \citet{2018MNRAS.474.3808V}. However, it is difficult to distinguish between optically thin thermal emission and non-thermal emission if the value of the spectral index is closer to -0.1 and the uncertainties are large. Non-thermal emission is thus claimed with certainty only when $\alpha <<$ -0.1 and the uncertainty on the index does not allow for overlap with -0.1.

The integrated fluxes of the sources at L-, C- and Q-bands, together with any others available in literature, were used to estimate their indices despite the fact that data collection was not contemporaneous and MYSOs may be variable e.g. S255 NIRS 3 \citep{2018A&A...612A.103C}. Moreover, the observations are of different resolutions and are clearly sensitive to different spatial scales. To ensure the accuracy of the method, both L- and C- band data were re-imaged using a common range of $uv-$wavelength i.e 15-200\,k$\lambda$, so as to use emissions from comparable spatial scales in estimating the spectral indices of the sources.
Nonetheless, full range of $uv-$wavelength was used in Q-band despite the fact that it traces a different spatial scale since the overlap between its uv-coverage and 15-200\,k$\lambda$ is insufficient.  Further caution was exercised by estimating L- and C-band fluxes from a similar enclosing box regardless of the morphology of a source to provide for uniformity. 

A high resolution observation at Q-band can result in missing flux \citep{2005A&A...437..947V}, but at the same time dust emission contribute about 43$\pm$11\,\% \citep{2017PhDT.......147P} of the flux at Q-band. 
For example, the flux of G111.5671 at 44\,GHz (Purser et al. 2019 in prep), observed using the VLA's A configuration is lower than C-configuration flux \citep{2005A&A...437..947V} by 18\%. Thus, the quantities of free-free emission at Q-band, essential for estimating the spectral indices, are unknown and some of the indices may have large uncertainties. However, the indices are used since the SEDs of most of the cores, shown in Figure \ref{spectral_index_plots}, do not show a strong evidence of unexpected behaviour.

The indices were estimated to be equivalent to the slopes of the lines of best fit of a plot of logarithm of flux density and frequency of an object, generated by minimising chi-squares (shown in Figure \ref{spectral_index_plots}). 
In cases where a source's flux density could not be estimated (i.e too faint to detect), an upper limit, evaluated as three times its field rms (i.e $3\times$ rms) was set as its peak flux. An upper limit on the integrated flux of such a source was estimated from a polygon whose size is similar to that of the source at a frequency where it is detected. 
Some sources are extended at L-band but are resolved into multiple cores and lobes e.g G094.4637 or a core and a lobe e.g G094.2615, in the higher resolution C- and Q-bands. For the
extended sources, the flux at L-band and the sum of the fluxes of
the components in the higher resolution observations were used to
estimate the index. Indices of the resolved components, on the other hand, were estimated from the fluxes at the frequencies where they are resolved. If a source was only detected at one frequency, a limit on the index was calculated. Table \ref{spectral_indicesT} shows a list of the objects, their spectral indices and the fluxes used in estimating the indices. Details of the components incorporated in the calculation of the spectral indices of L-band sources that harbour multiple components are indicated in Table \ref{spectral_indicesT}.

Spectral index maps of the MYSOs were generated from L- and C- band images using CASA task {\tt{immath}} in {\tt{spix}} mode to display the distribution of the indices across a source. The maps used are of $uv-$wavelength range 15-200\,k$\lambda$. Only pixels whose values are more than 3$\sigma$ in both L- and C-bands were used in creating the maps. The re-imaged C-band maps show varied morphologies, some of which are comparable to L-band emission e.g. G110.0931 while others are not e.g. G111.2552 and G108.5955 (see Figure \ref{L_band_spix_images}).
Uncertainty in the spectral index of a given pixel on the map was approximated by assuming that the error on the flux of each cell is equivalent to the rms of the field. This error was propagated while calculating the spectral index of corresponding L- and C-band pixels. The maps and the corresponding errors are shown in Figure \ref{L_band_spix_images}. Whereas the beams of L- and C-band maps that were used in generating the spectral index maps are largely similar, a few maps were generated from C-band images of smaller beams. The smaller C-band beams can result in slightly lower fluxes at C-band and more negative spectral indices on the pixels of non-thermal sources.



\subsubsection{Spectral indices of the cores}
The L-Q band spectral indices $\alpha_{LQ}$ show that all the detected cores are thermal radio emitters ($\alpha > -0.1$). To decipher the nature of the objects further, their L-, C- and Q-band fluxes, among others from the literature, were used to compare their spectral indices as derived from L- to C- bands $\alpha_{LC}$ with those from C- to Q- bands $\alpha_{CQ}$. Table \ref{spectral_indicesT} shows three possibilities: $\alpha_{LC}$ that is lower than $\alpha_{CQ}$ e.g. G094.2615, G094.4637 and G141.9996; $\alpha_{LC}$ that is higher than $\alpha_{CQ}$ e.g. G110.0931 and G094.6028-A and $\alpha_{LC}$ which is comparable to $\alpha_{CQ}$ e.g. G139.9091 and G126.7144. 

Sources whose $\alpha_{LC} < \alpha_{CQ}$ have overall spectral indices that are flattened by L-band fluxes. This may be due to steepening of $\alpha_{CQ}$ by dust emission at Q- or contribution of non-thermal emission at L-band. \citet{2017PhDT.......147P} estimated that dust emission is, on average, $\sim$ 43$\pm$11\,\% of the total emission at 44\,GHz, capable of steepening C-Q band slope, however the values of $\alpha_{CQ}$ are $\sim+$0.6, consistent with thermal jets. 

Spectra of sources whose $\alpha_{LC}$ are higher than $\alpha_{CQ}$ exhibit an HII region-like feature due to the apparent turnovers of their SEDs. In G110.0931, this can clearly be attributed to loss of flux at Q band where one of its components (A) was not detected. 
Finally, sources whose $\alpha_{LC}$ are comparable with $\alpha_{CQ}$ seem to have thermal cores that are separated from the lobes (e.g. in G111.5671-A) or cores with no clear lobes (e.g. in G111.2552 and G126.7144). In conclusion, L-Q band spectral indices ($\alpha_{LQ}$) of the MYSOs are largely positive ($\alpha > 0$) and the radiation from what one can consider a central component of a MYSO is also primarily thermal, making it a substantial radio emitter at higher frequencies.  The average $\alpha_{LC}$, $\alpha_{CQ}$ and $\alpha_{LQ}$ of the sources are 0.42$\pm$0.34, 0.43$\pm$0.33 and 0.42$\pm$0.27 confirming that the cores are thermal.

\subsubsection{Spectral indices of lobes}
Four sources, G094.6028, G103.8744, G111.5671 and G141.9996, have non-thermal lobes of spectral indices $\alpha \leq-$0.42. The lobes were detected at L- but not C-band.
G138.2957 has non-thermal lobes of spectral indices -0.46$\pm$0.31 and -0.38$\pm$0.33 to the east and west of its thermal core respectively (see Figure \ref{L_band_spix_images}). Two sources; G114.0835 and G136.3833 have nearby non-thermal sources that are aligned with the orientation of NIR emission in their fields. G094.4637 has a thermal lobe whose L- and C band emissions are displaced in the direction of the outflow. It also has a component, D that  is non-thermal. Besides the lobes, two sources; G094.2615 and G108.5955 show tentative indications of non-thermal emission to the south-east and east of their cores, respectively.

These findings suggest that six of the sources, equivalent to 40\,\%, have non-thermal lobes which indicate the presence of magnetic fields. It is, however, not clear if the magnetic fields are generated in the surrounding interstellar medium or by the protostars themselves. The non-thermal emission, present on an object's spectral index map, is mainly seen further away from the central source, confirming that the central sources are thermal while the jet lobes are not. Detailed comments on each object are given in the appendix.

\begin{figure*}
    \centering
    
    \includegraphics[width = 0.325\textwidth]{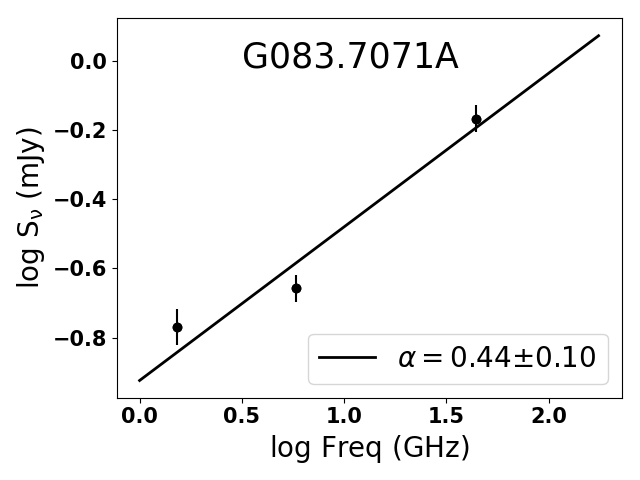}
    \includegraphics[width = 0.325\textwidth]{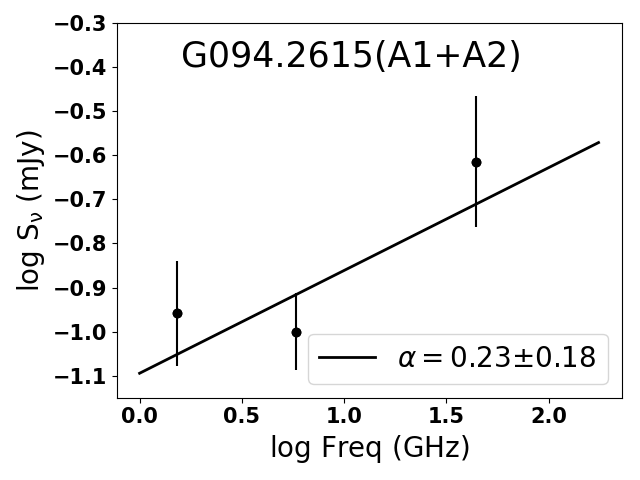}
    \includegraphics[width = 0.325\textwidth]{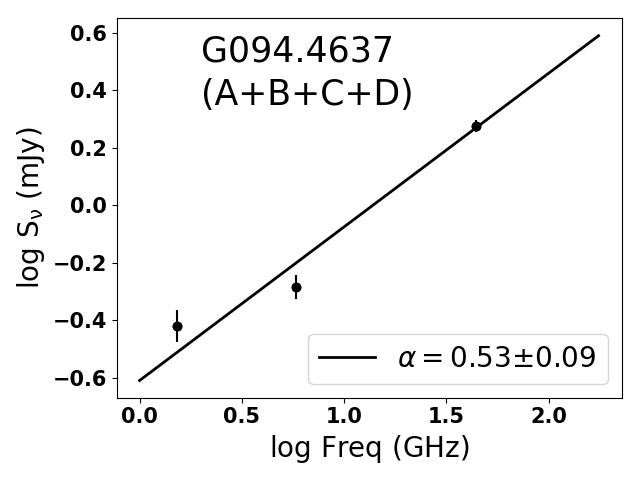}
    \includegraphics[width = 0.325\textwidth]{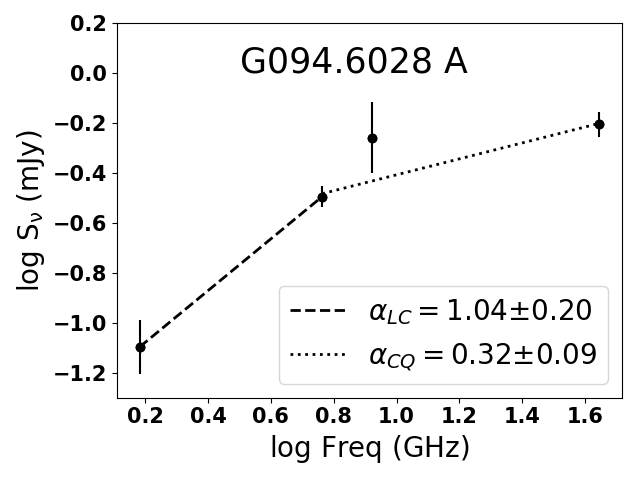}
    \includegraphics[width = 0.325\textwidth]{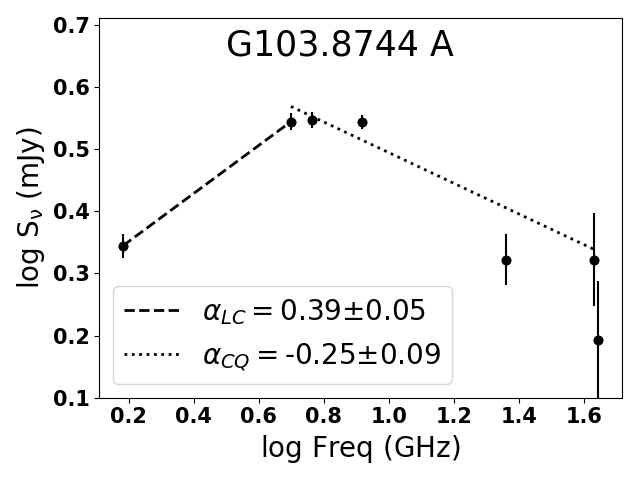}
    \includegraphics[width = 0.325\textwidth]{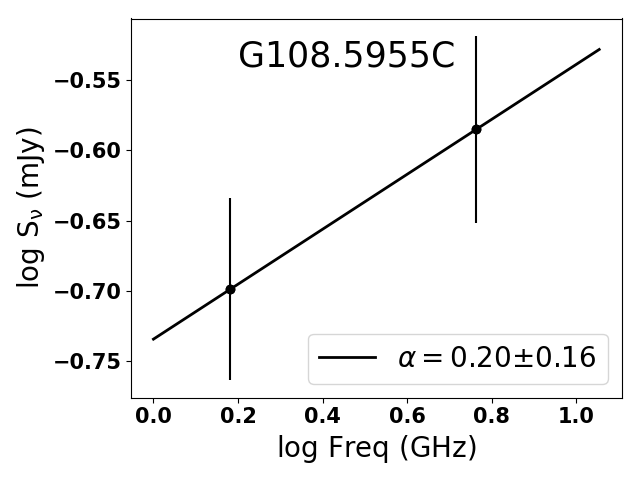}
    \includegraphics[width = 0.325\textwidth]{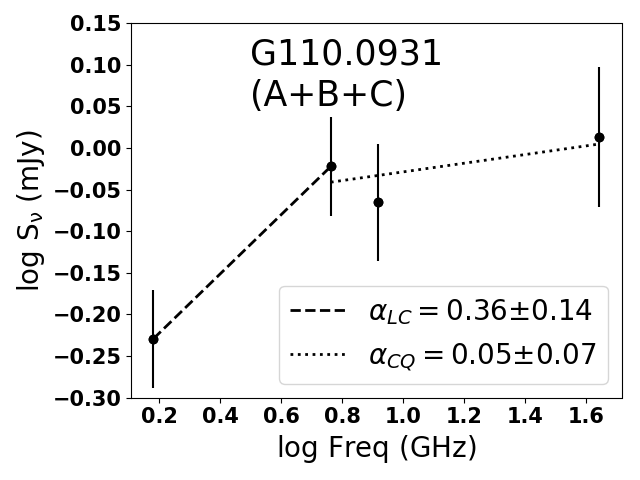}
    \includegraphics[width = 0.325\textwidth]{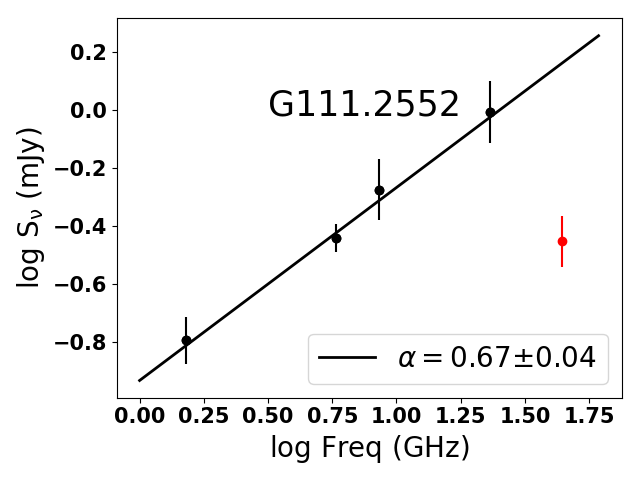}
    \includegraphics[width = 0.325\textwidth]{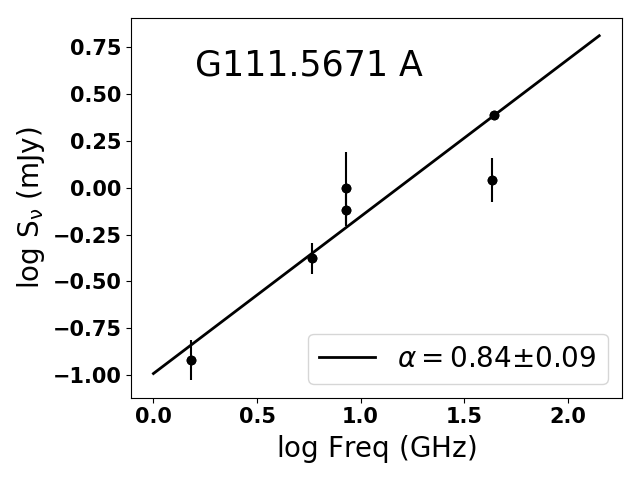}
    \includegraphics[width = 0.325\textwidth]{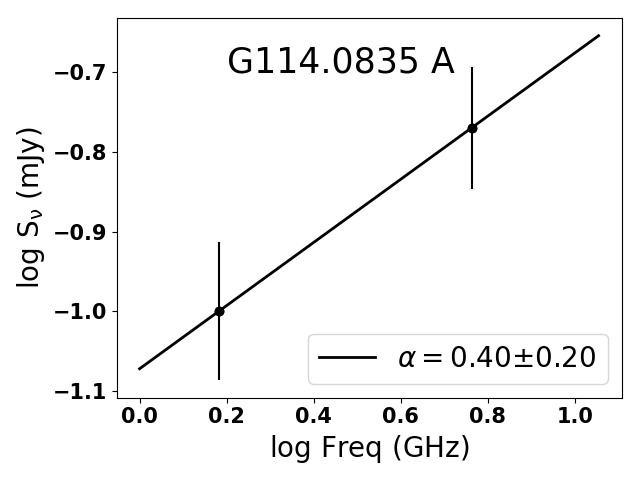}
    \includegraphics[width = 0.325\textwidth]{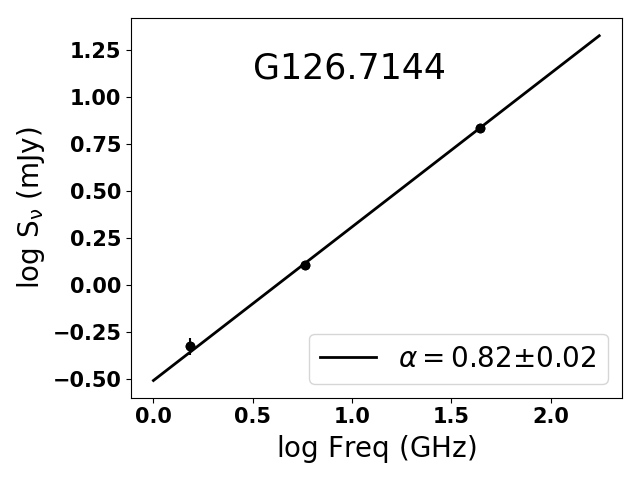}
    \includegraphics[width = 0.325\textwidth]{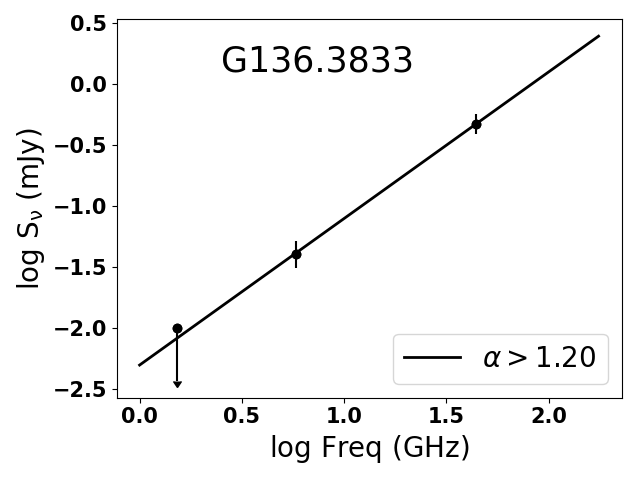}
    \includegraphics[width = 0.325\textwidth]{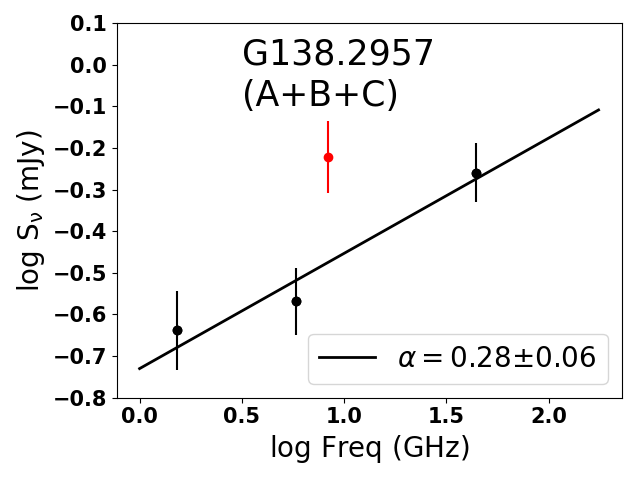}
    \includegraphics[width = 0.325\textwidth]{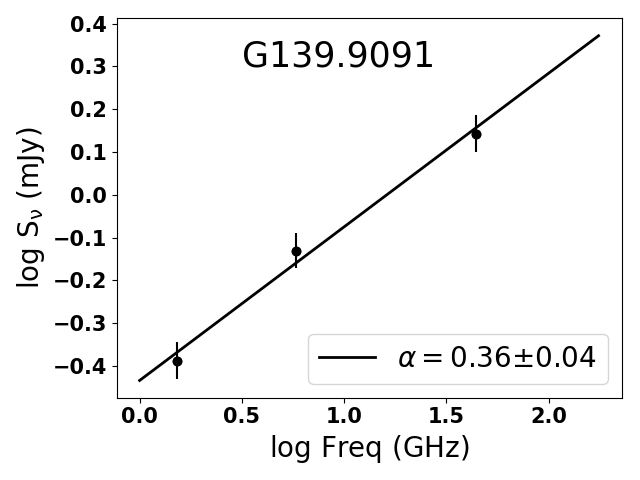}
    \includegraphics[width = 0.325\textwidth]{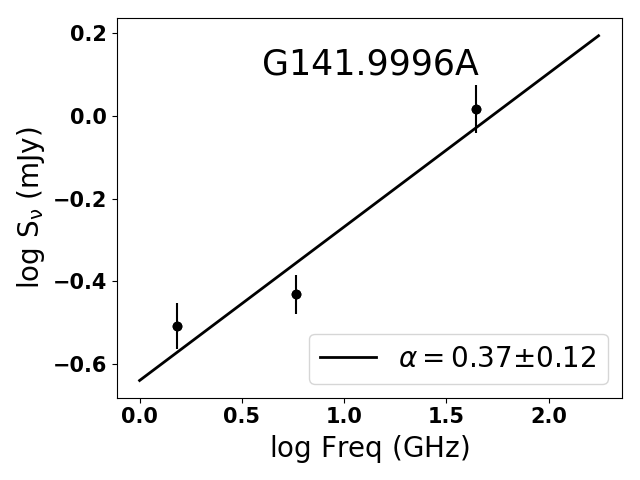}
    
    \caption{A plot showing how flux density of the sources at L-band, which harbour cores, varies with frequency. The radio emission of the MYSO, G103.8744 was not detected and the spectrum of a HII-region that is closest to its location was used. In some sources, additional data, available in the literature and listed in Table \ref{spectral_indicesT}, were used in estimating the indices. The red data points represent values that were considered to be too low or high due to variability and were not used in the fit. 
    }
    \label{spectral_index_plots}
\end{figure*}

\begin{figure*}
    \centering
    
    \includegraphics[width = 0.4\textwidth]{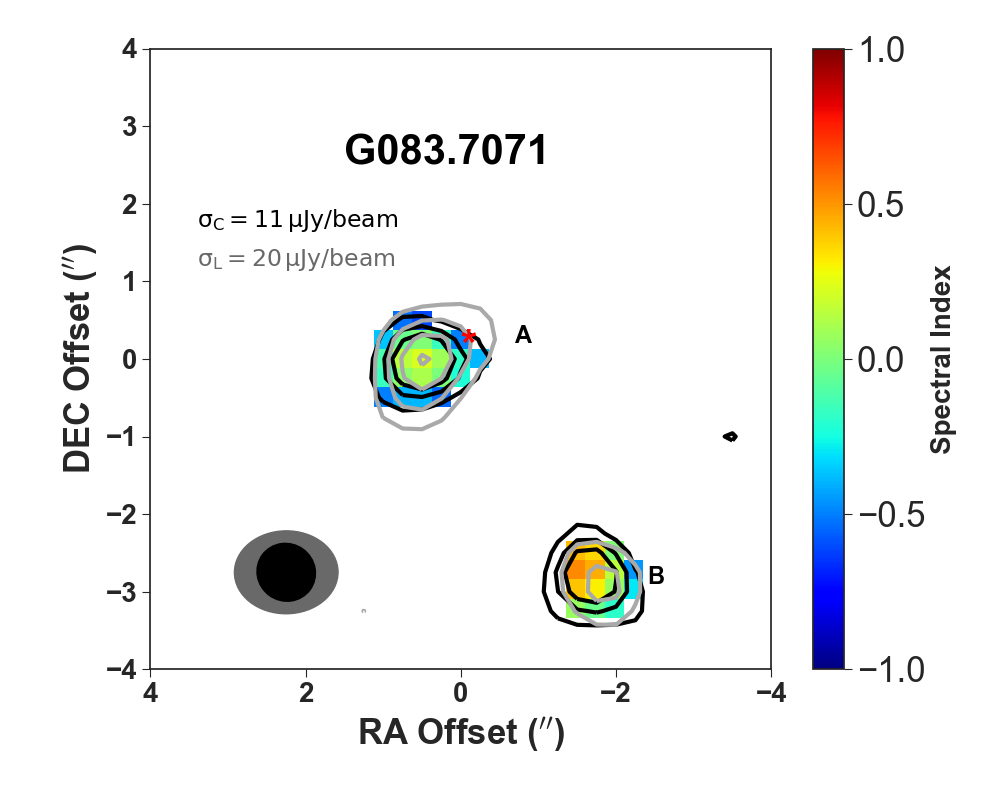}
    \includegraphics[width = 0.4\textwidth]{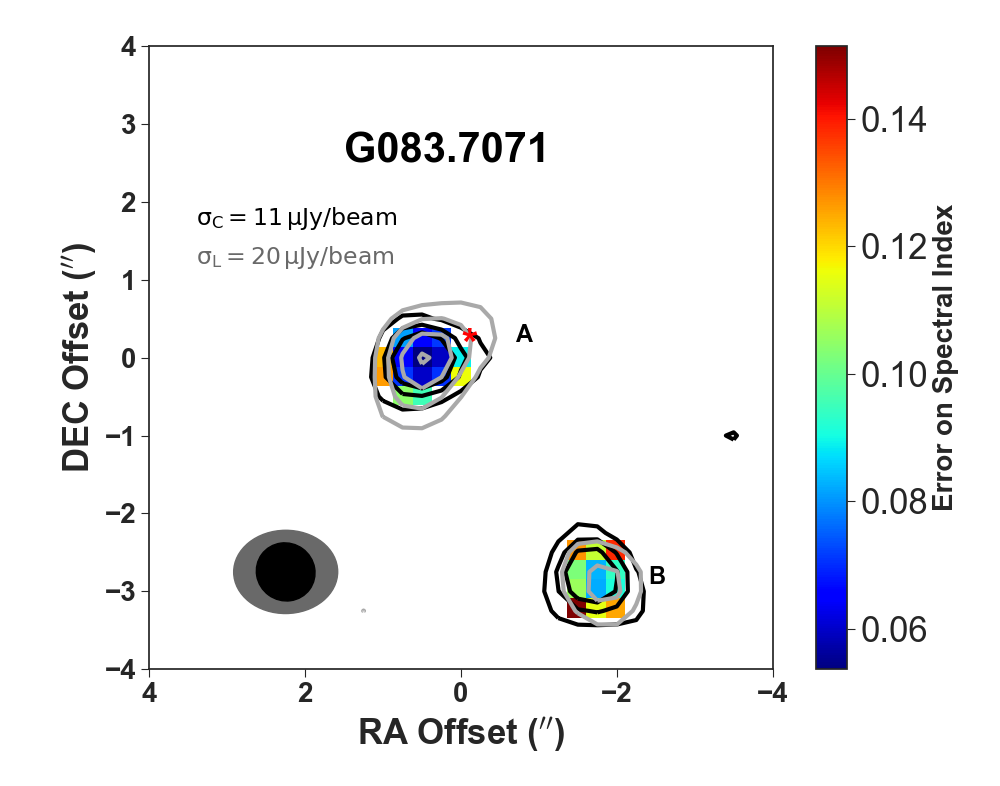}
    \includegraphics[width = 0.4\textwidth]{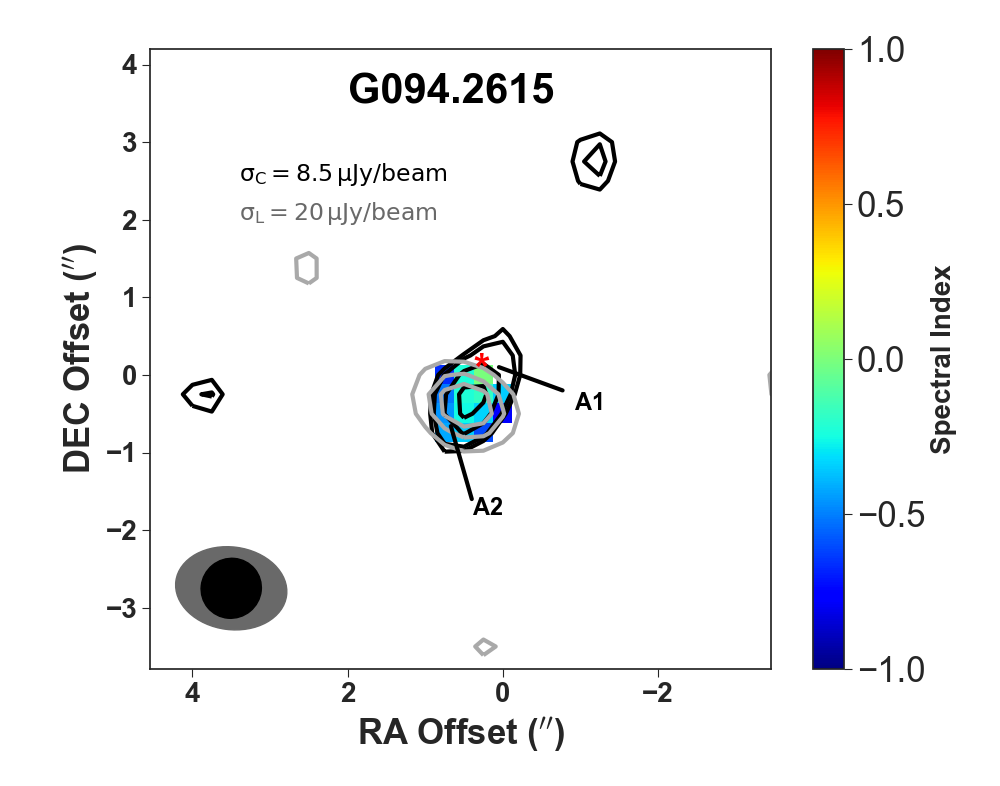}
    \includegraphics[width = 0.4\textwidth]{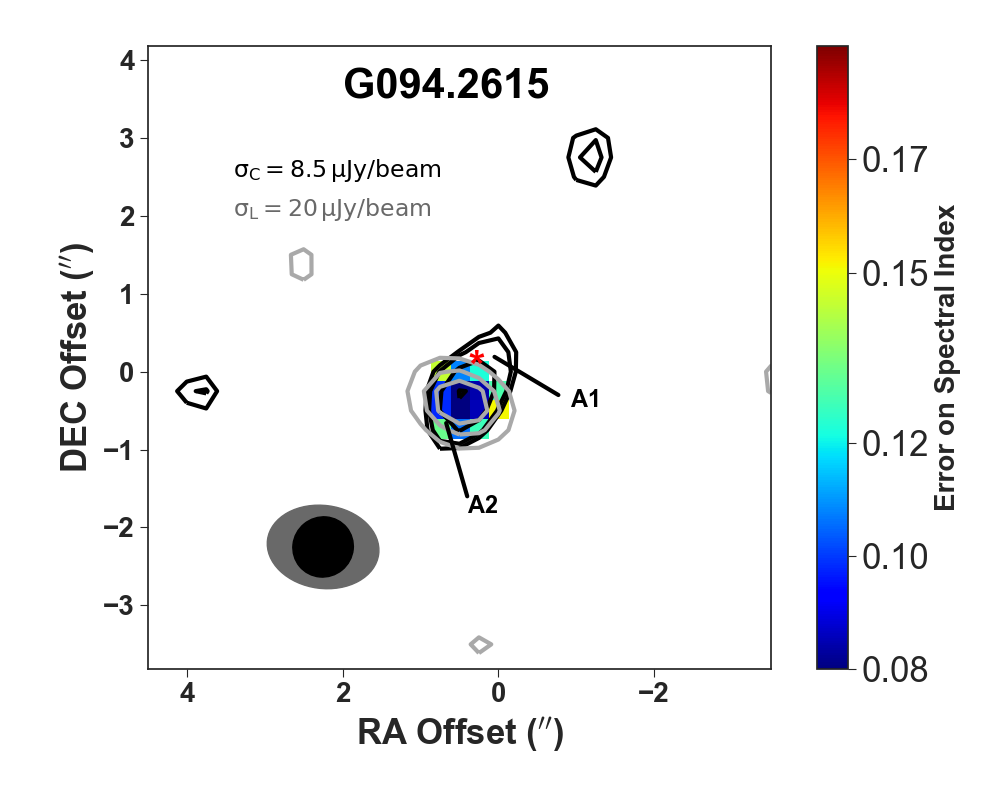}
    \includegraphics[width = 0.4\textwidth]{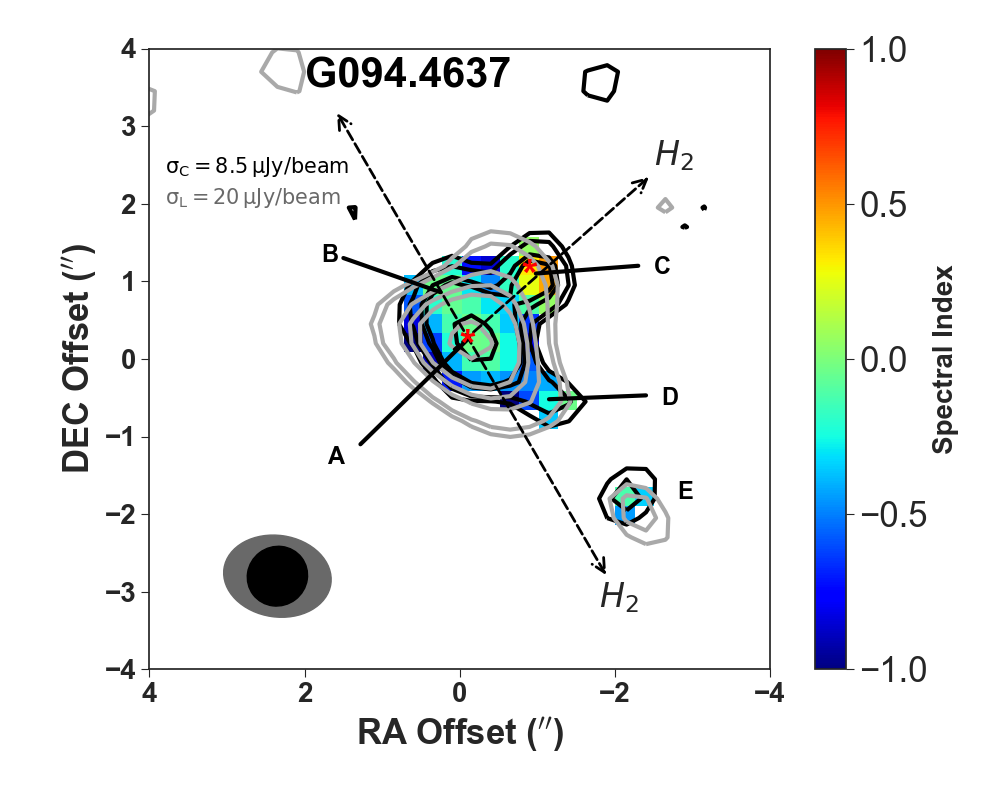}
    \includegraphics[width = 0.4\textwidth]{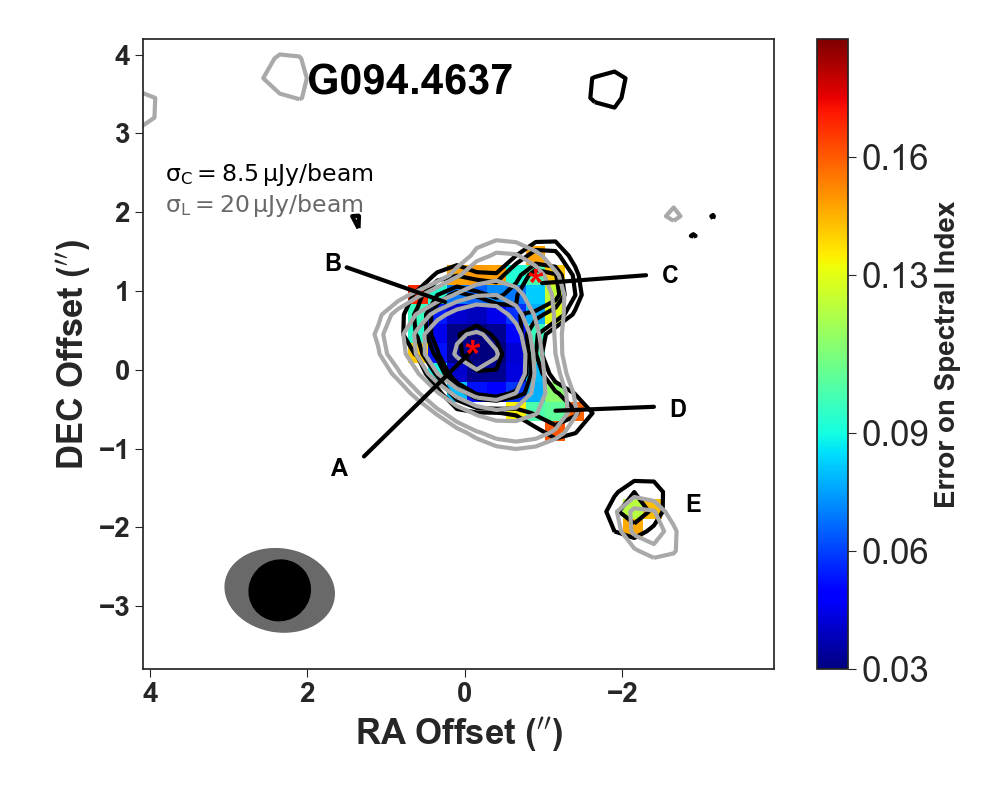}
    \caption{Spectral index maps of the sources shown on the left panels and their corresponding error maps on the right. L- and C-band contours of levels $\mathrm{3,5,7\sigma,...}$ and $\mathrm{3,7,11,15\sigma,...}$ are overlaid in grey and black colours respectively. The maps were generated from L- and C-band maps of similar uv-wave coverage (i.e $\mathrm{15-200\,k\lambda}$). Colour-scales show L-C band spectral indices $\alpha_{LC}$ of the sources and corresponding errors. Locations of sources that lie within the L-band emission are indicated by their names. The dashed lines are the approximate directions of known emissions that are associated with outflows. Asterisk and plus symbols are the Q-band and IR locations of the MYSO cores. The synthesised beams are shown in the lower left corner.}
    \label{L_band_spix_images}
\end{figure*}

\begin{figure*}\ContinuedFloat
    \includegraphics[width = 0.4\textwidth]{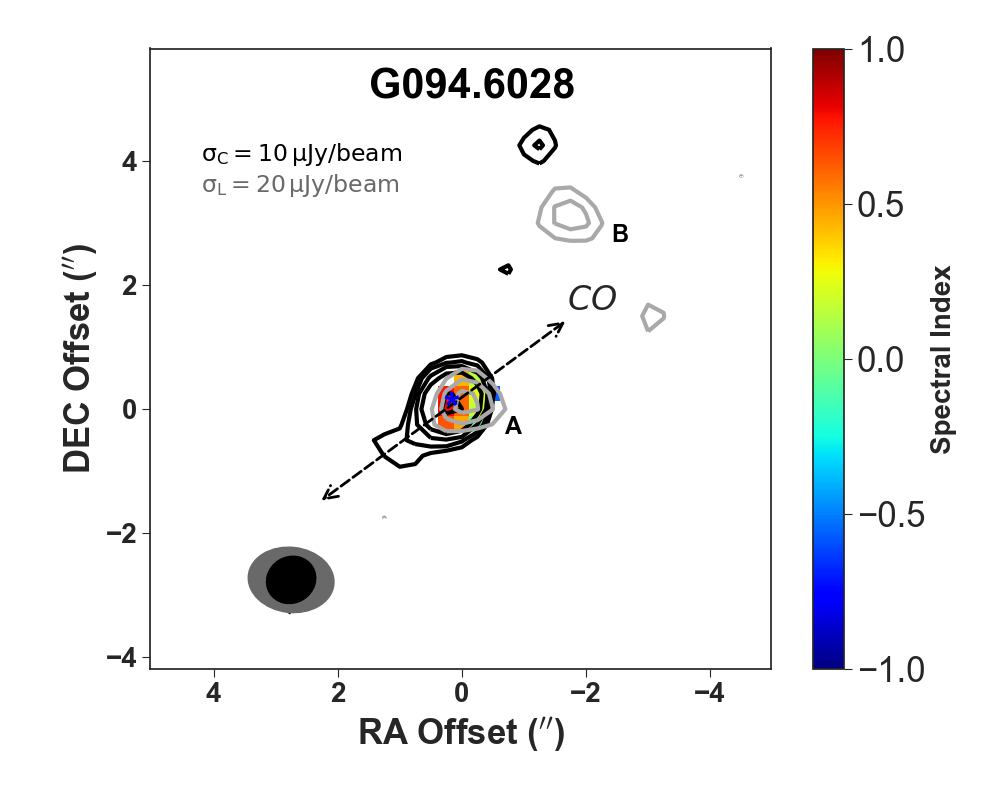}
    \includegraphics[width = 0.4\textwidth]{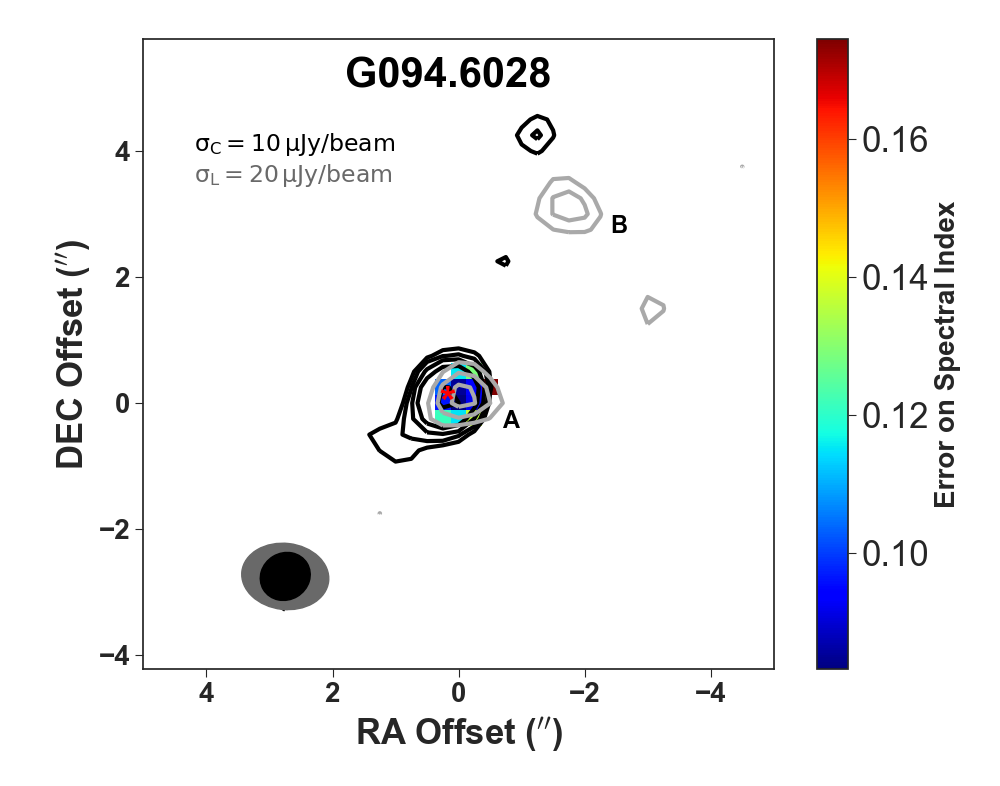}
    \includegraphics[width = 0.4\textwidth]{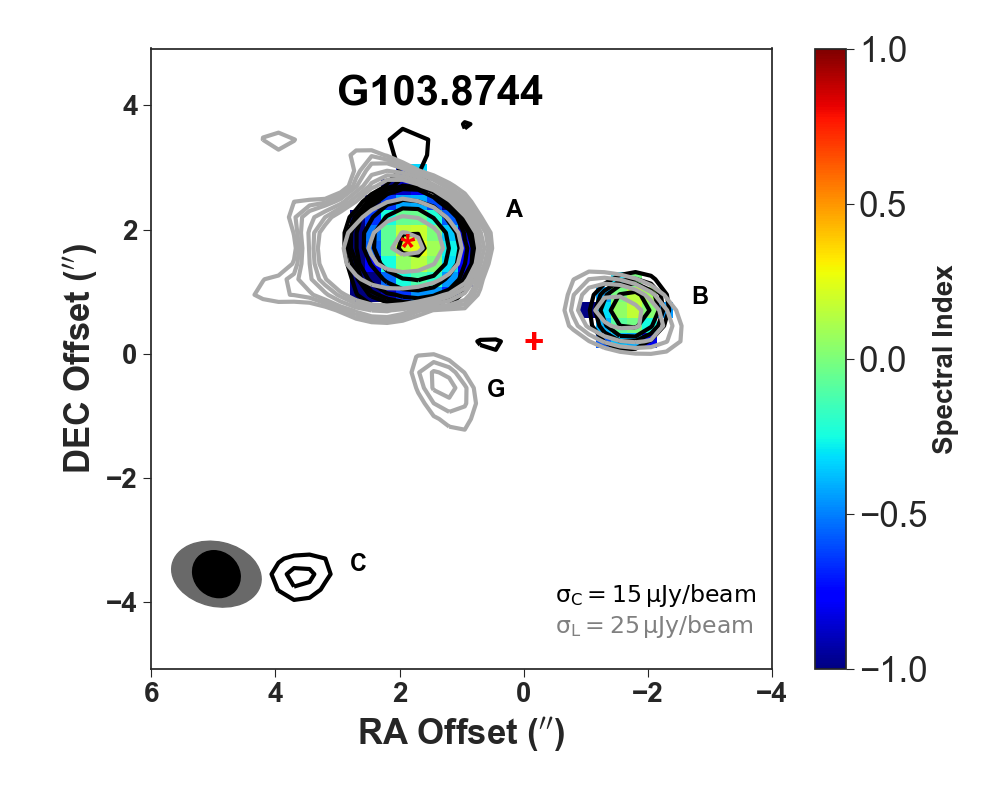}
    \includegraphics[width = 0.4\textwidth]{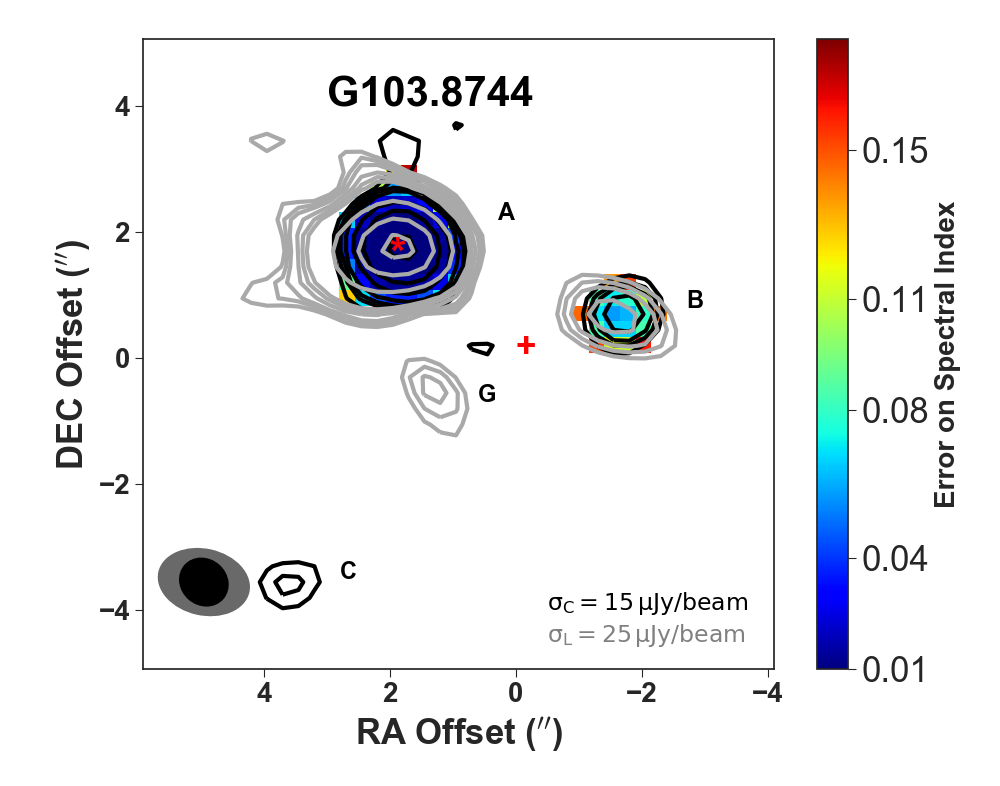}
    \includegraphics[width = 0.42\textwidth]{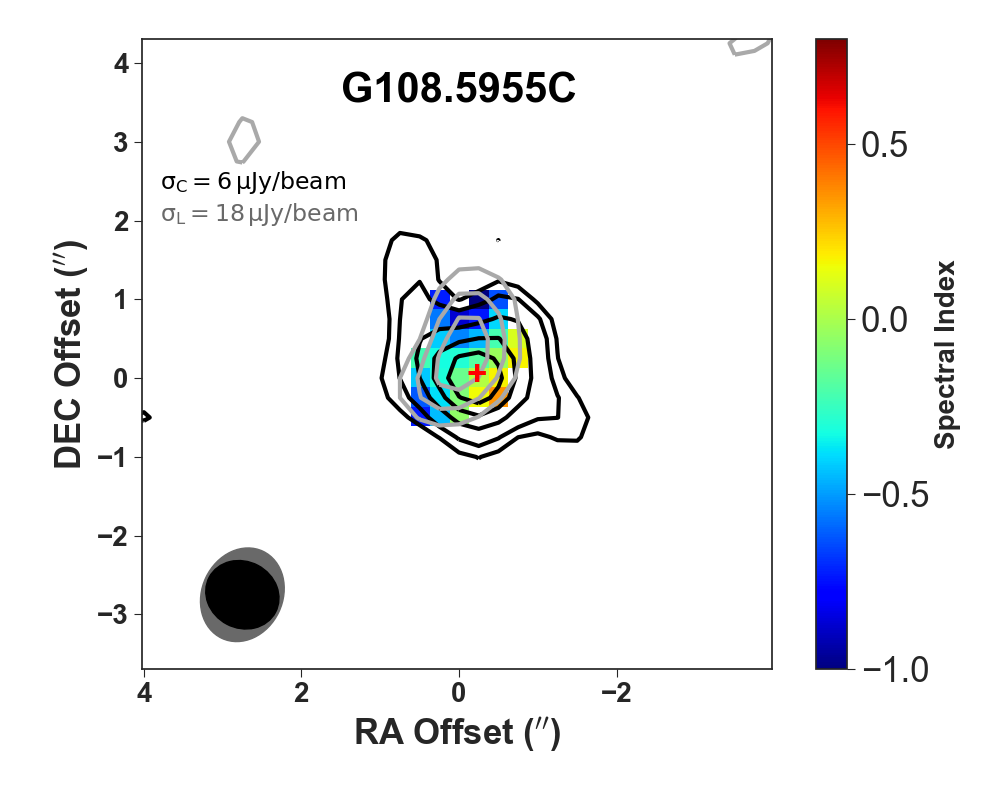}
    \includegraphics[width = 0.42\textwidth]{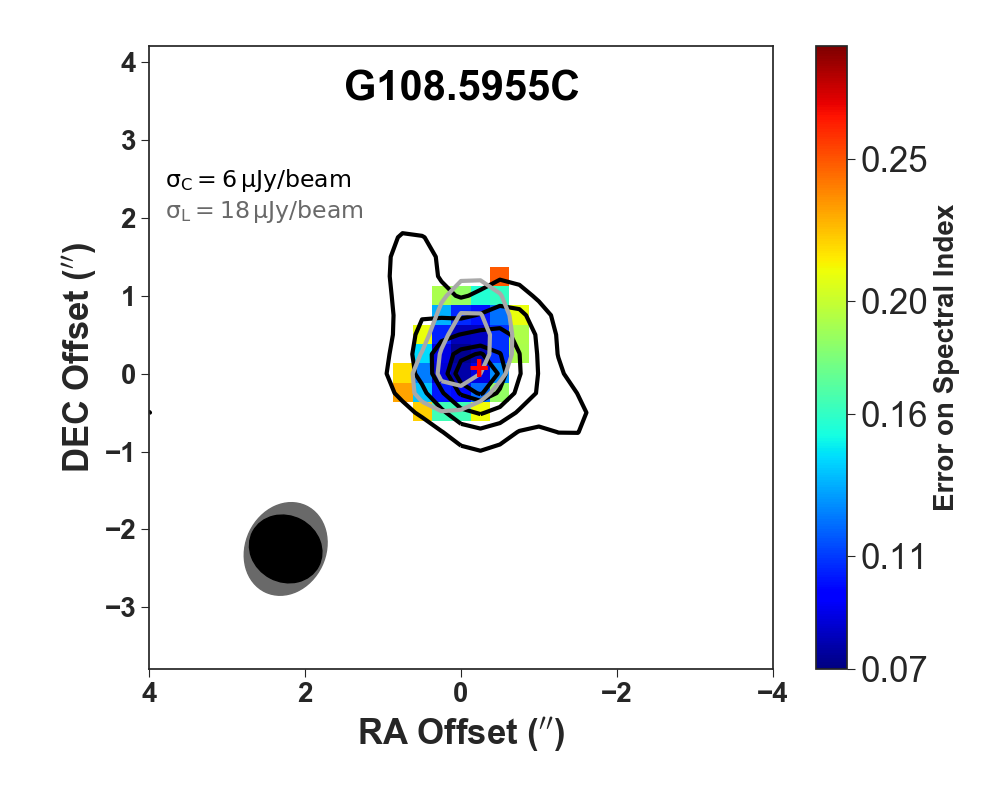}
    \caption{Figure \ref{L_band_spix_images} continued.}
\end{figure*} 

\begin{figure*}\ContinuedFloat
    \includegraphics[width = 0.42\textwidth]{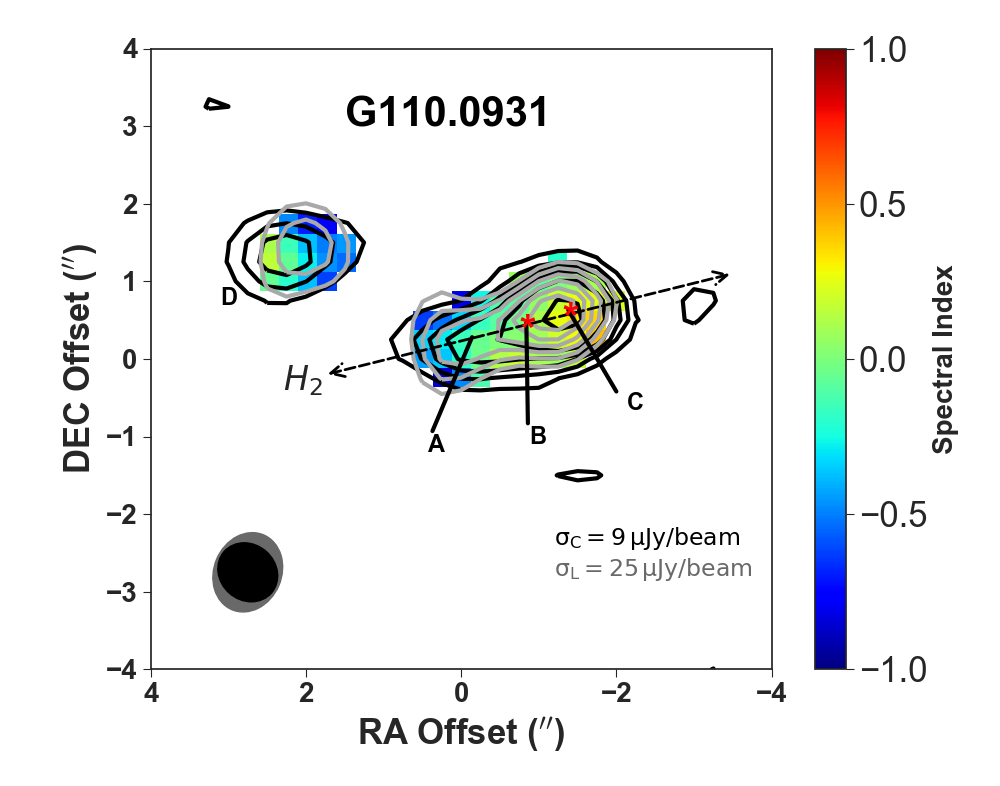}
    \includegraphics[width = 0.42\textwidth]{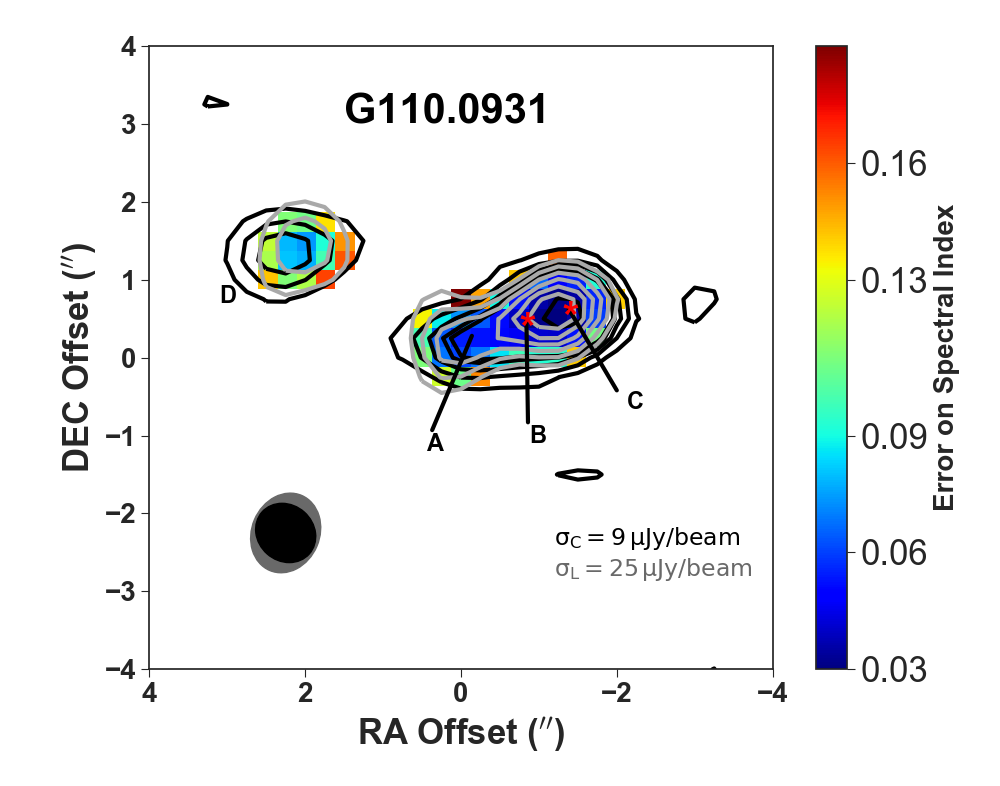}
    \includegraphics[width = 0.42\textwidth]{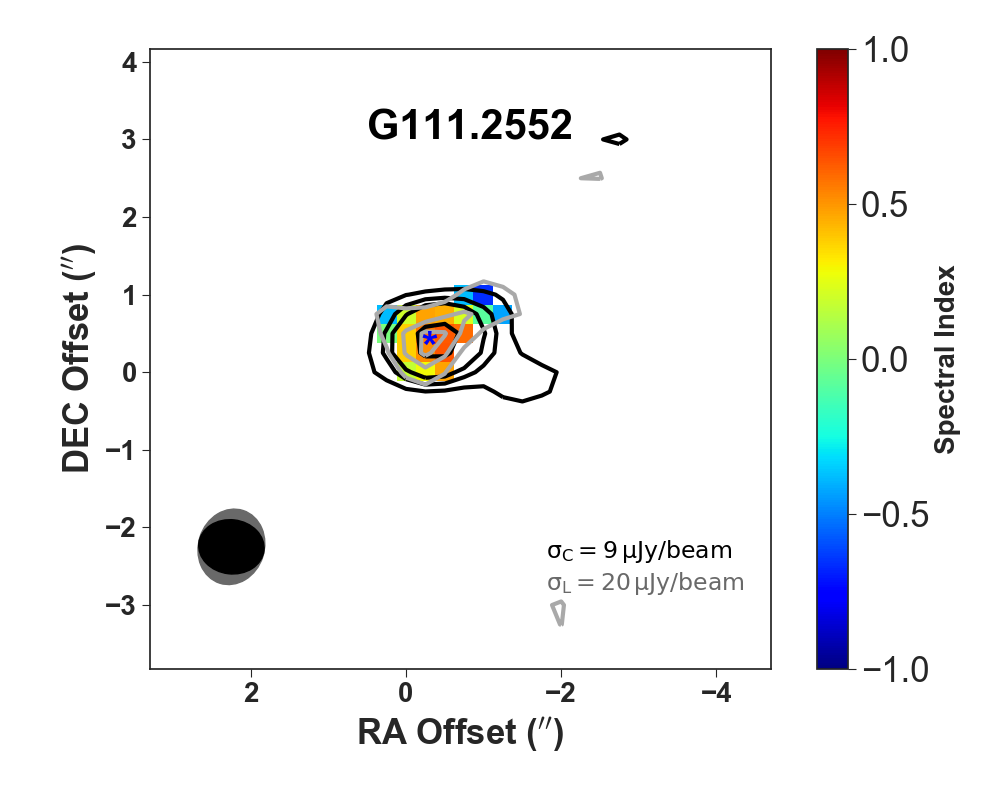}
    \includegraphics[width = 0.42\textwidth]{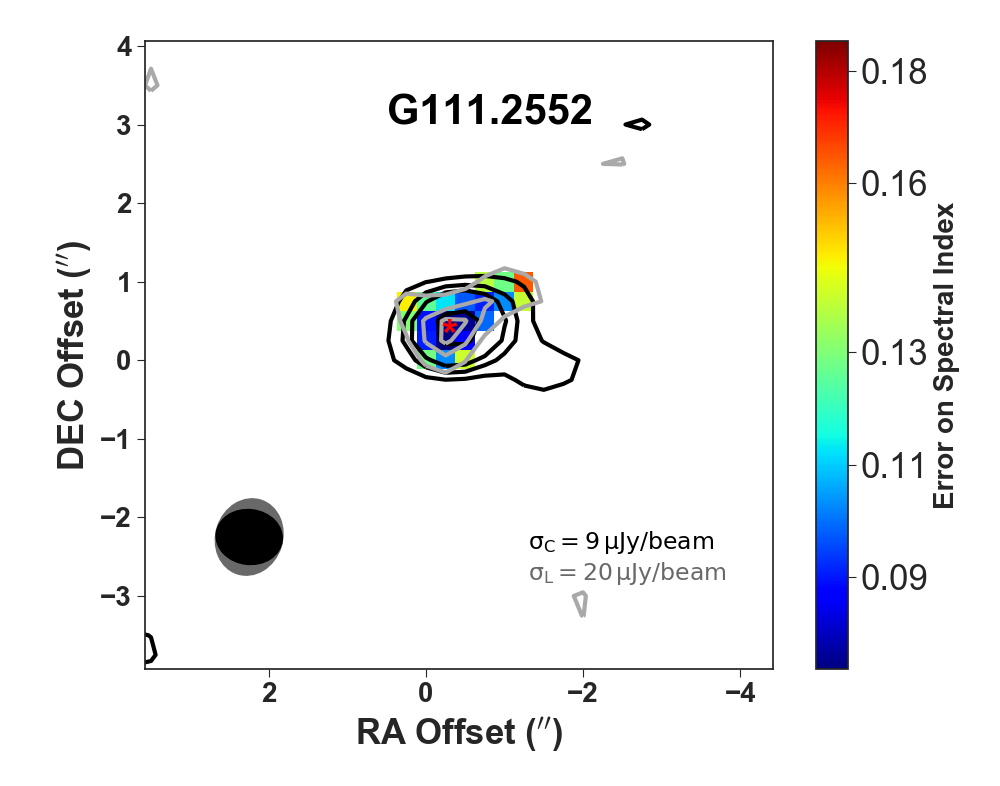}
    \includegraphics[width = 0.42\textwidth]{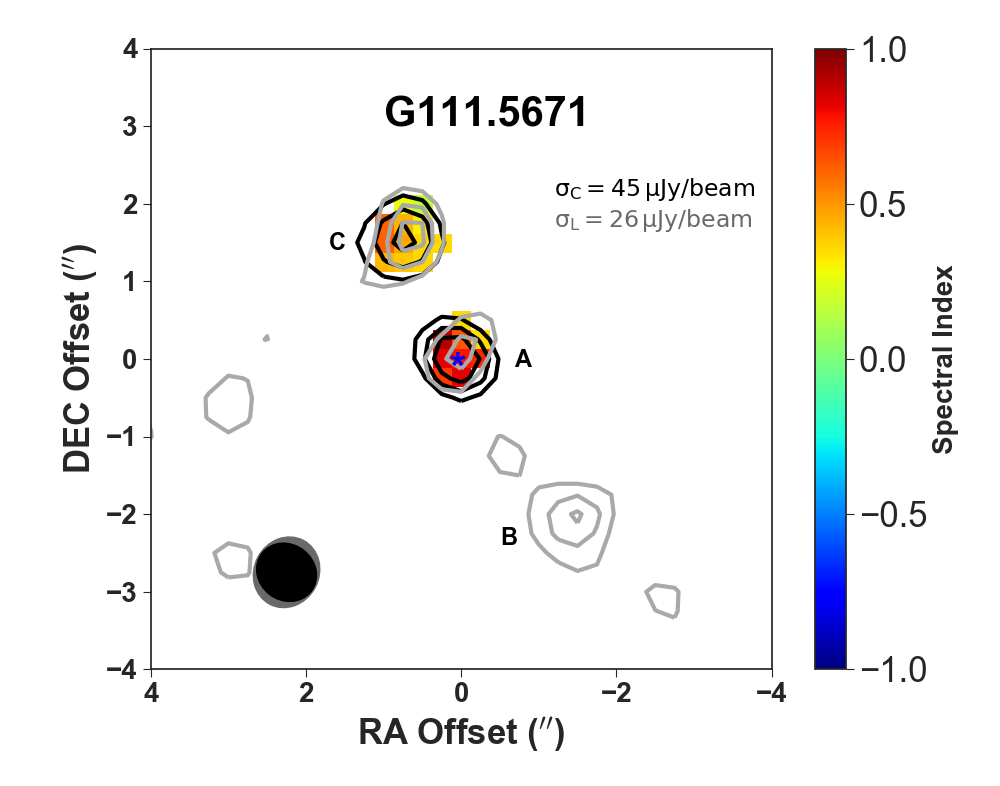}
    \includegraphics[width = 0.42\textwidth]{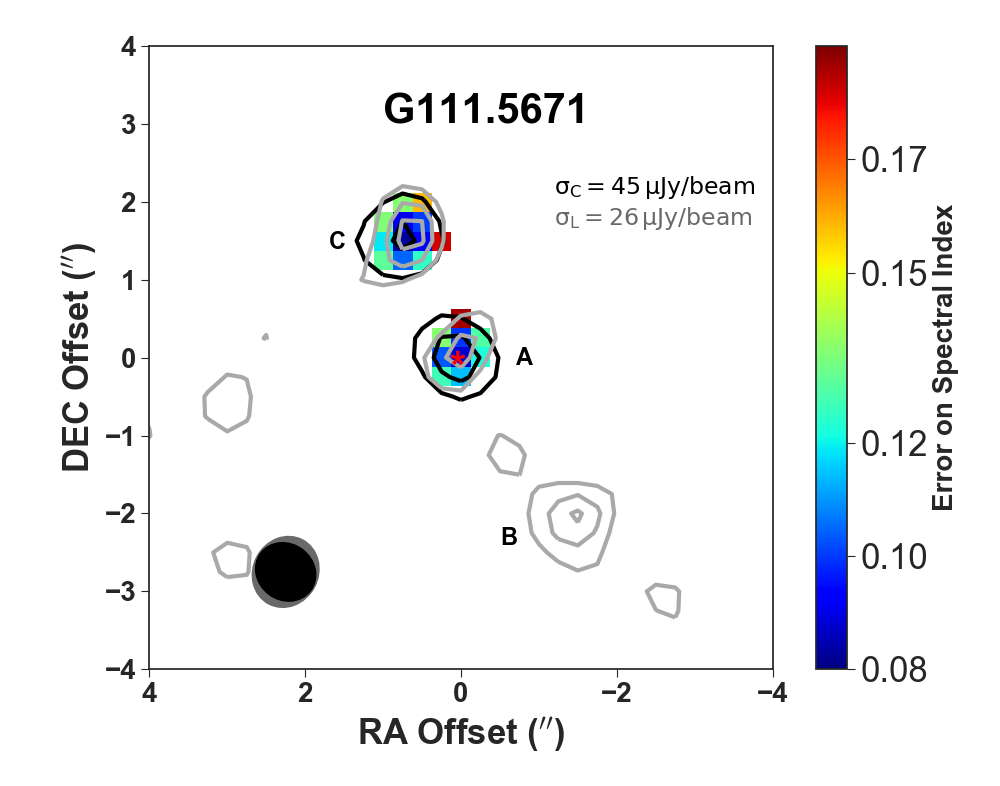}

    \caption{Figure \ref{L_band_spix_images} continued.}
\end{figure*} 

\begin{figure*}\ContinuedFloat
    \includegraphics[width = 0.42\textwidth]{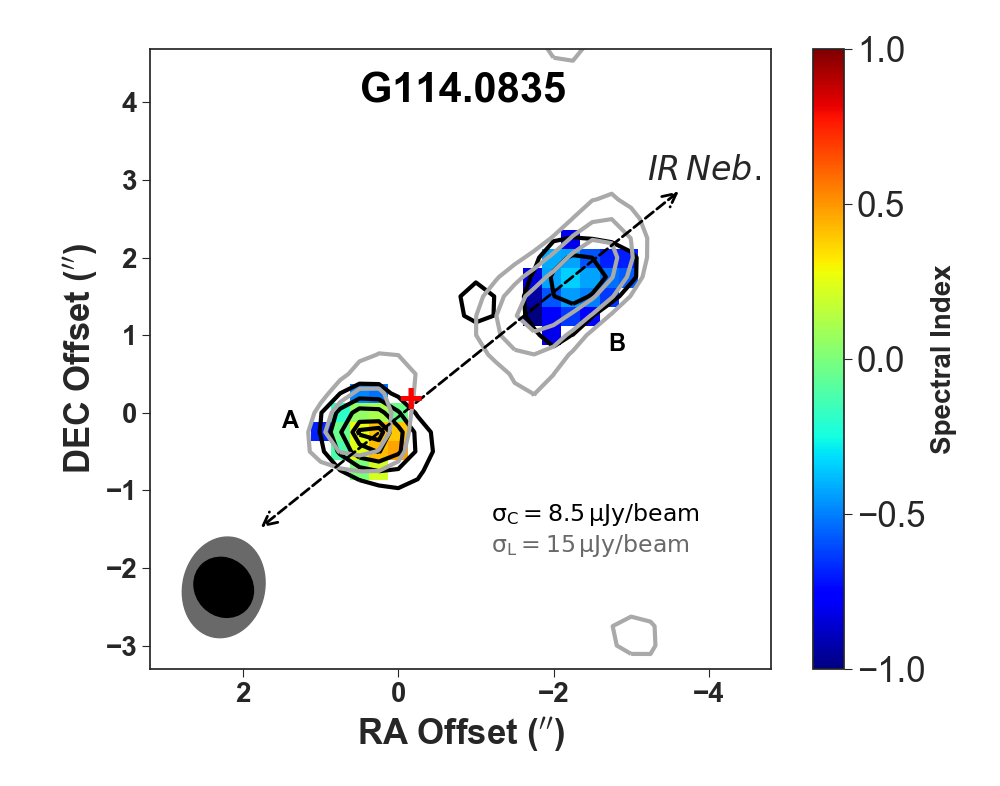}
    \includegraphics[width = 0.42\textwidth]{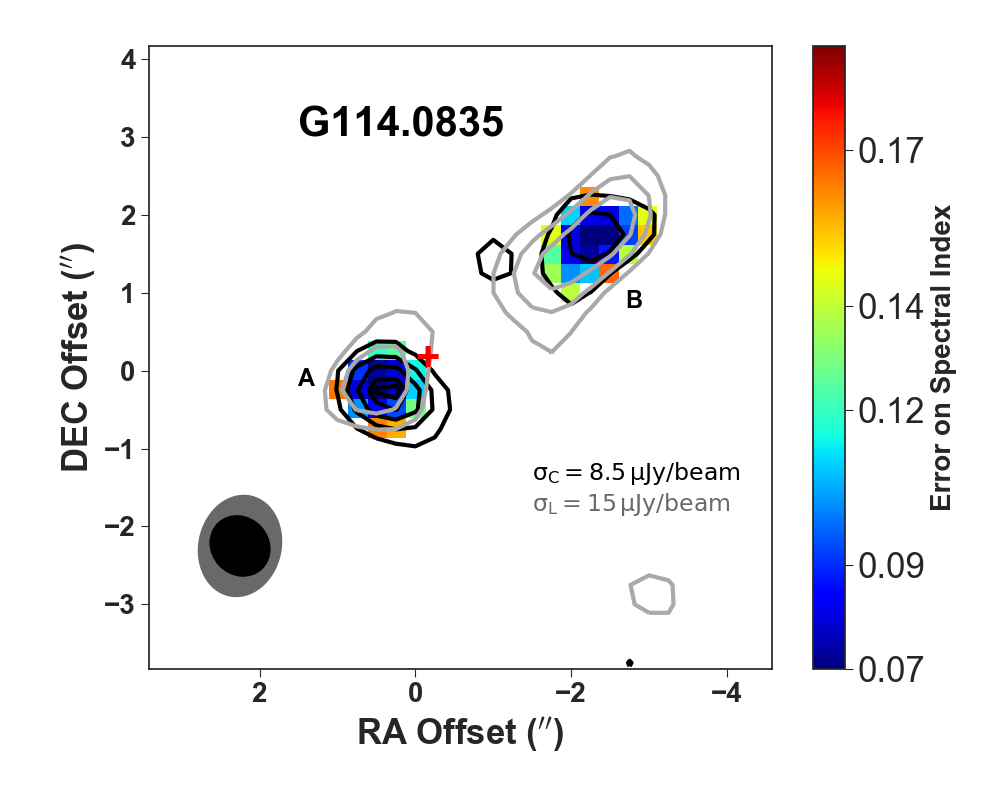}
    \includegraphics[width = 0.42\textwidth]{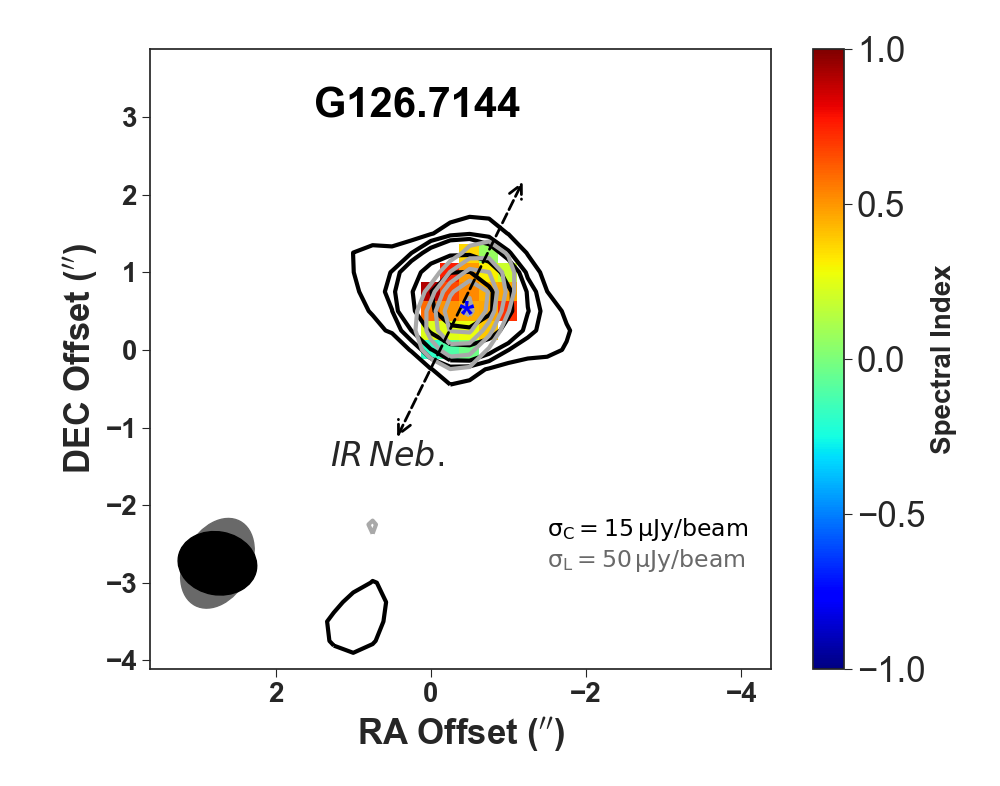}
    \includegraphics[width = 0.42\textwidth]{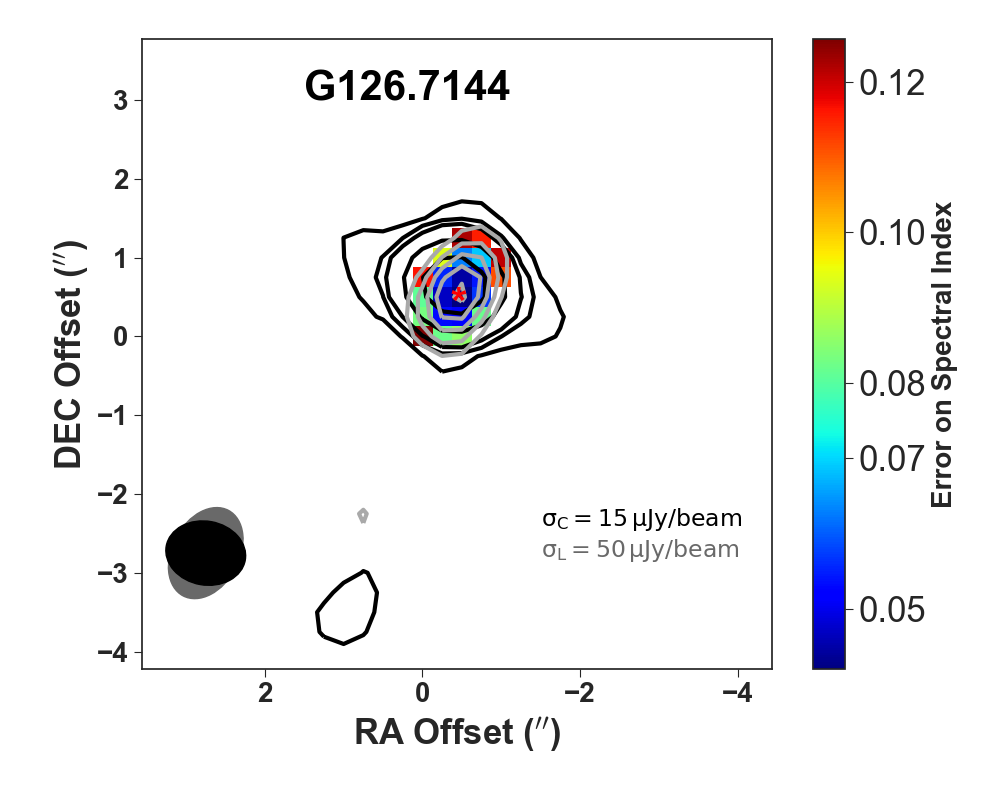}
    \includegraphics[width = 0.42\textwidth]{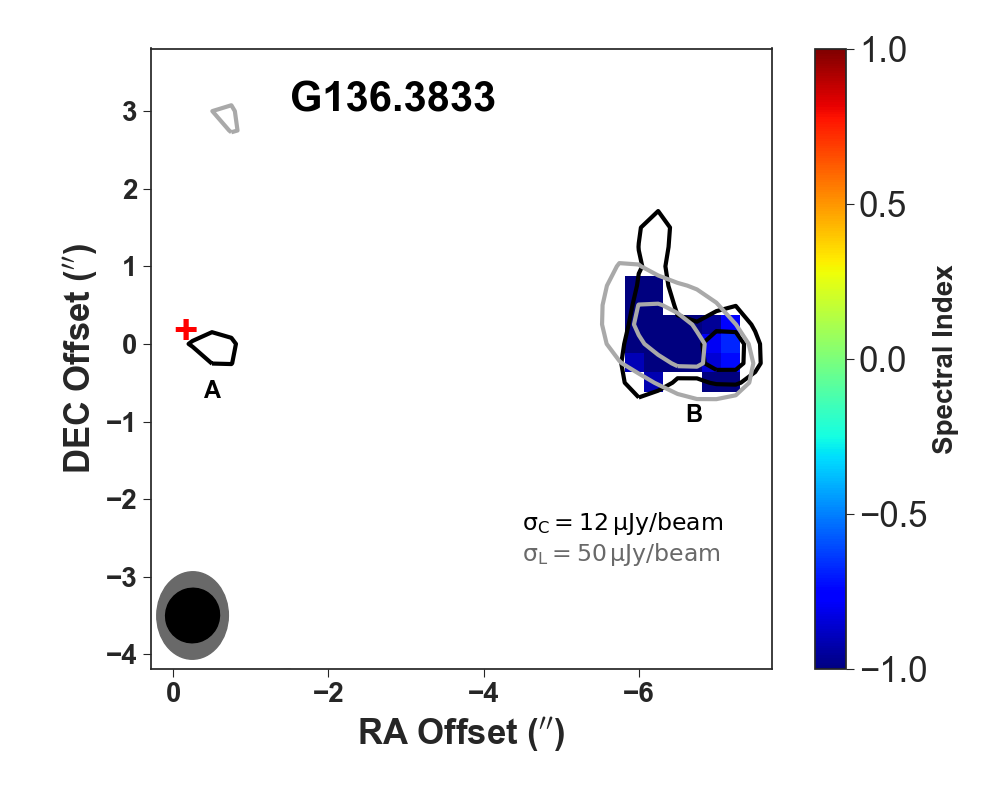}
    \includegraphics[width = 0.42\textwidth]{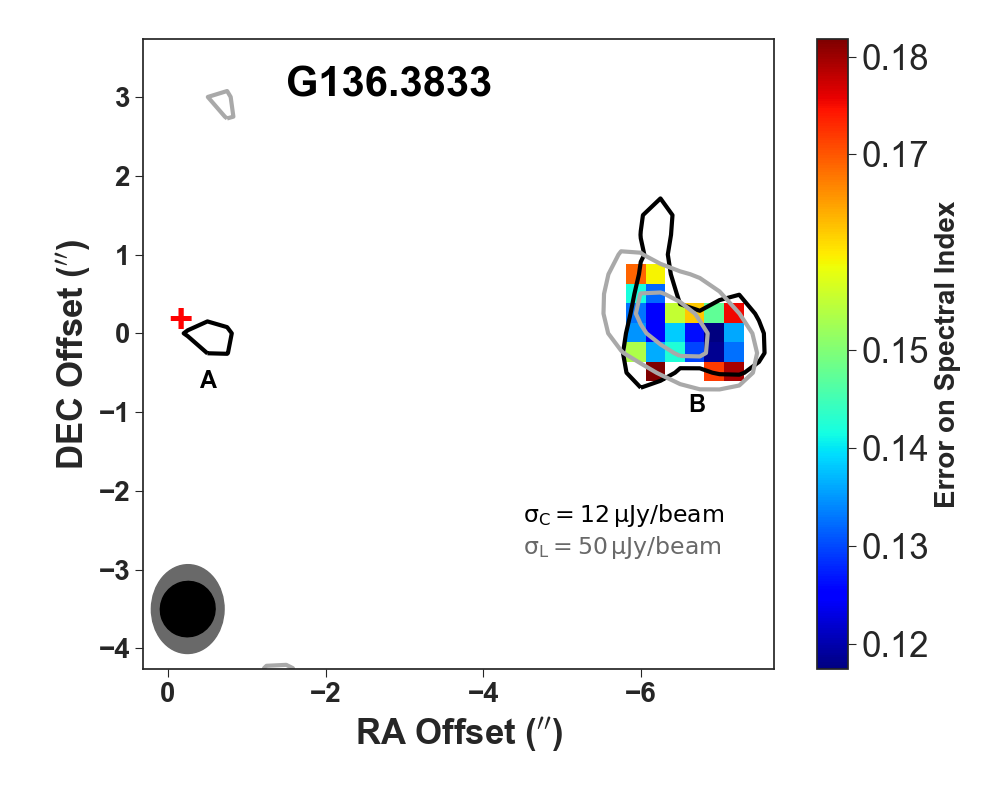}
    \caption{Figure \ref{L_band_spix_images} continued.}
\end{figure*}     

\begin{figure*}\ContinuedFloat
    \includegraphics[width = 0.42\textwidth]{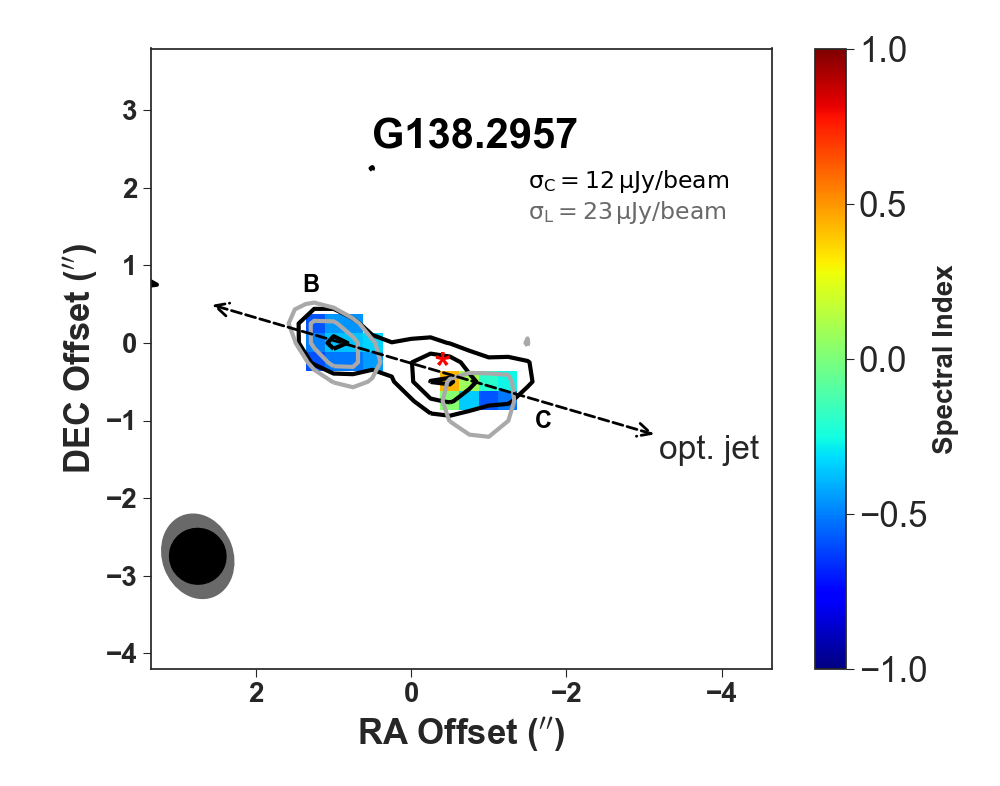}
    \includegraphics[width = 0.42\textwidth]{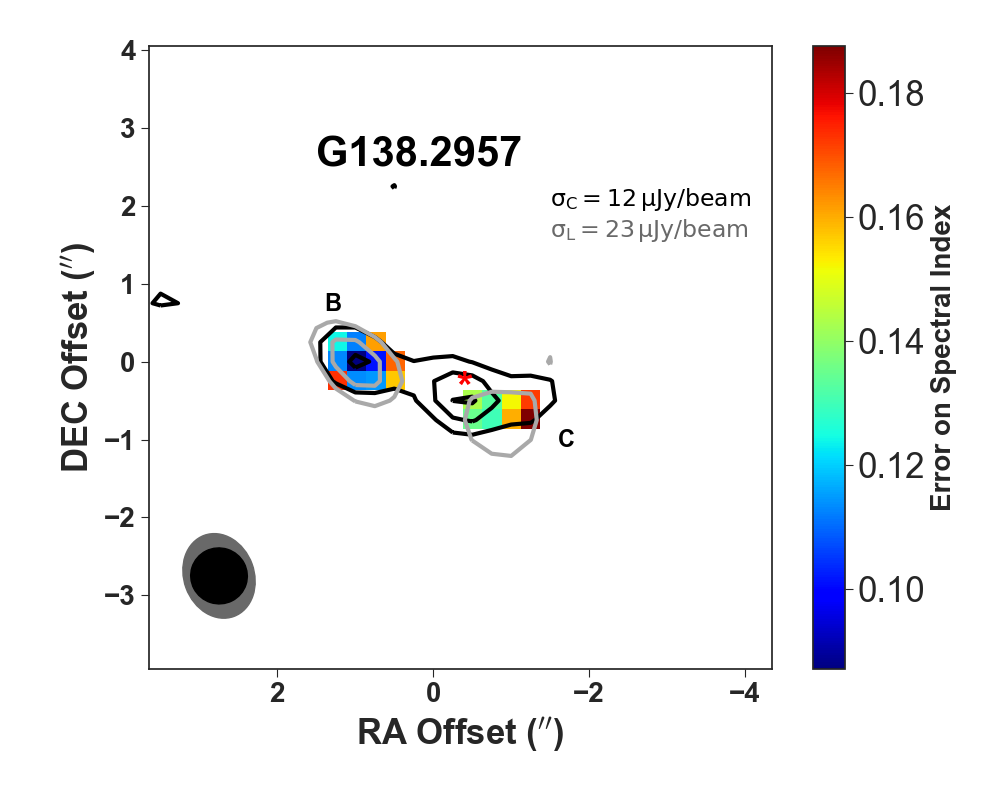}
    \includegraphics[width = 0.42\textwidth]{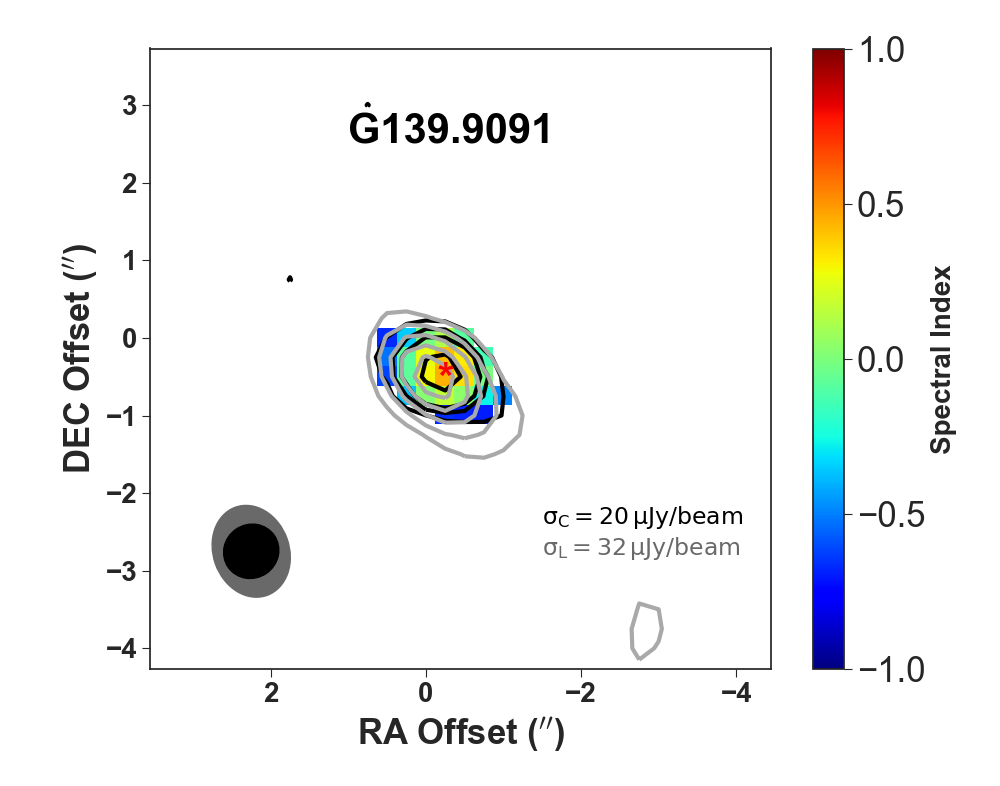}
    \includegraphics[width = 0.42\textwidth]{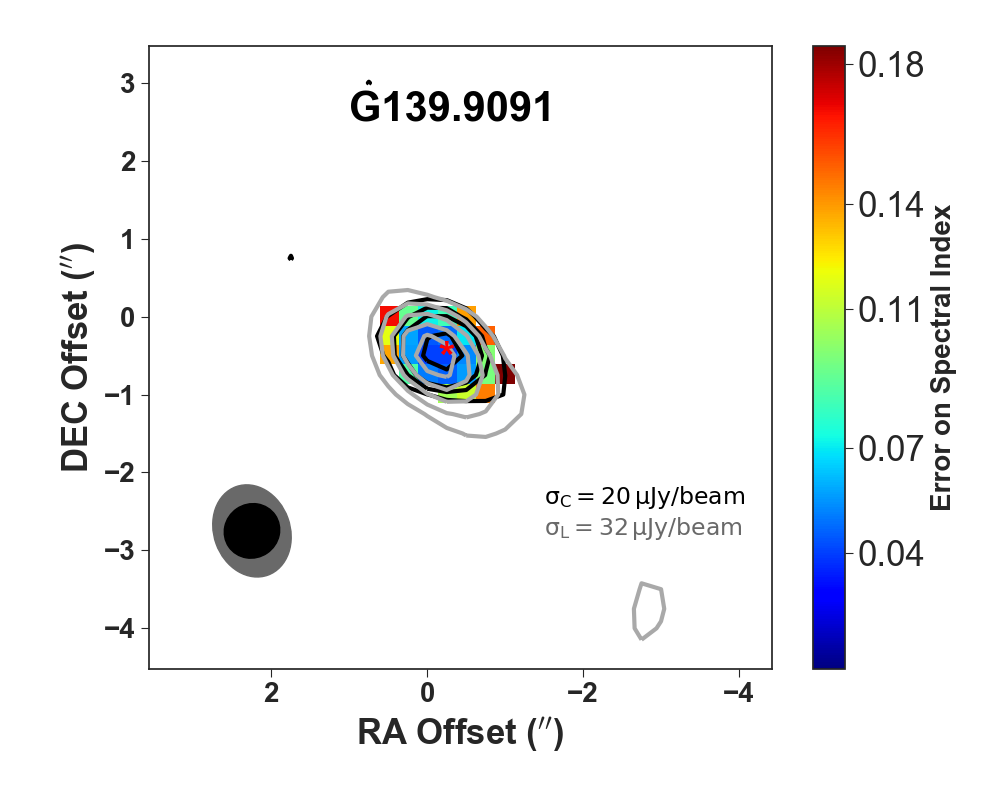}
    \includegraphics[width = 0.42\textwidth]{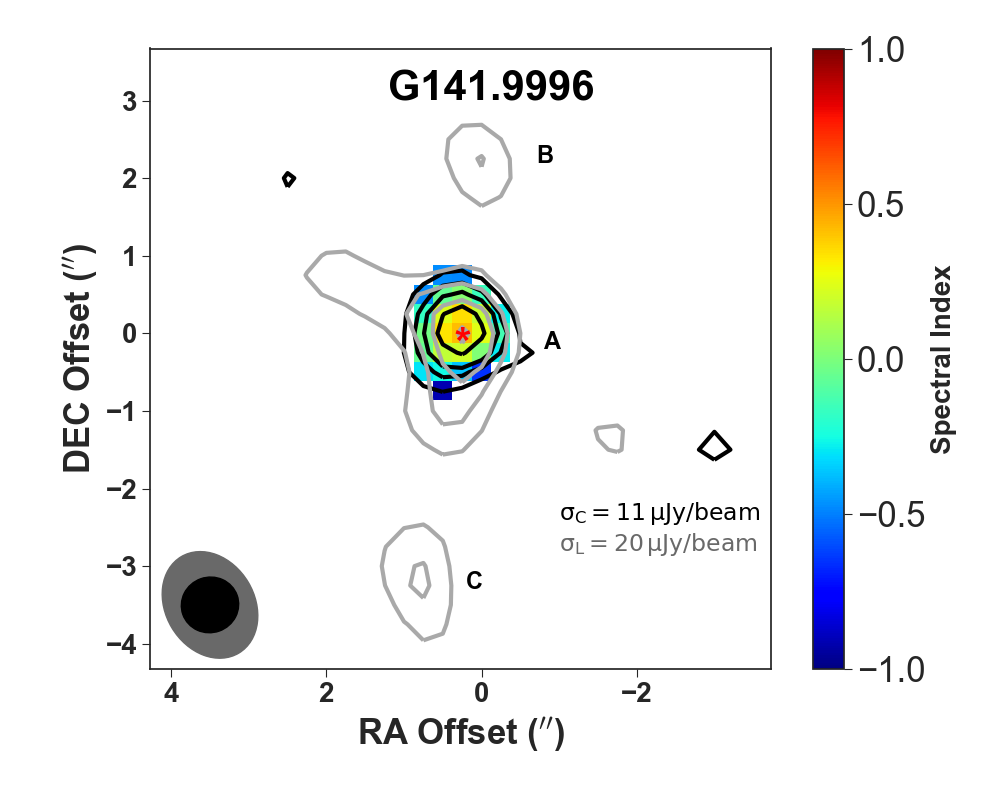}
    \includegraphics[width = 0.42\textwidth]{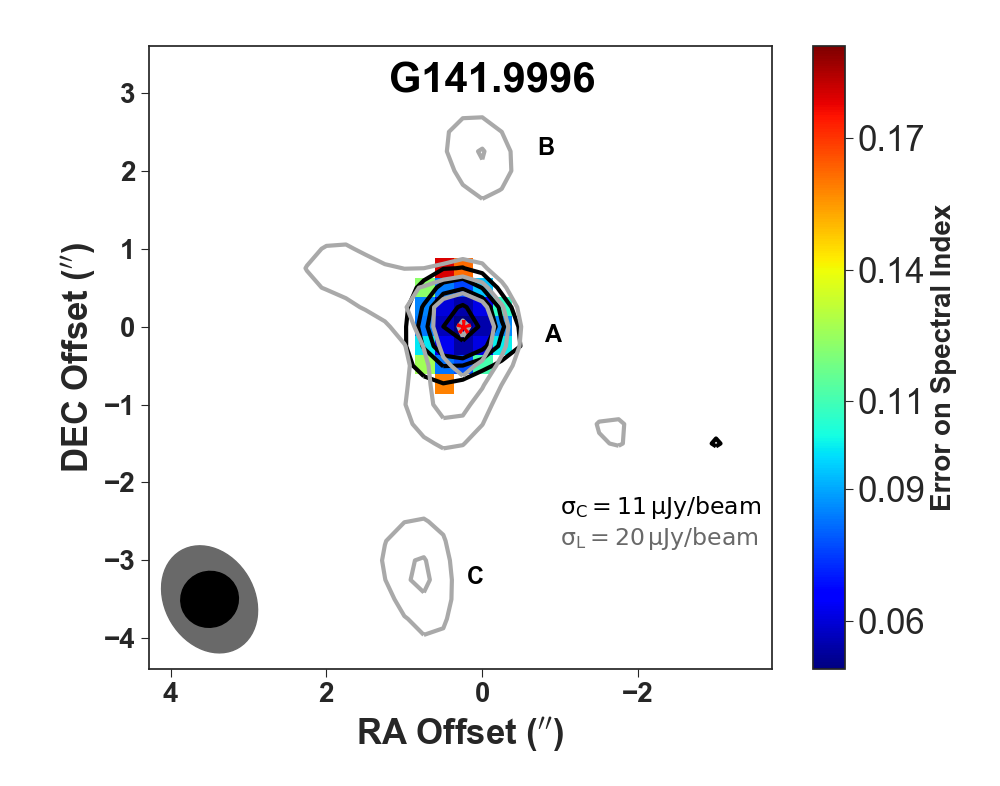}
    \caption{Figure \ref{L_band_spix_images} continued.}
\end{figure*}

\subsection{Nature of the cores and their radio luminosity at L-band}
The nature of the cores and their L-band emission were examined further by exploring the distribution of the cores on a bolometric-colour plot and the relationship between bolometric- and radio- luminosities at L-band.
\subsubsection{Relationship between Infrared colour and Bolometric luminosity}
\begin{figure}
    \centering
    
    \includegraphics[width = 0.48\textwidth]{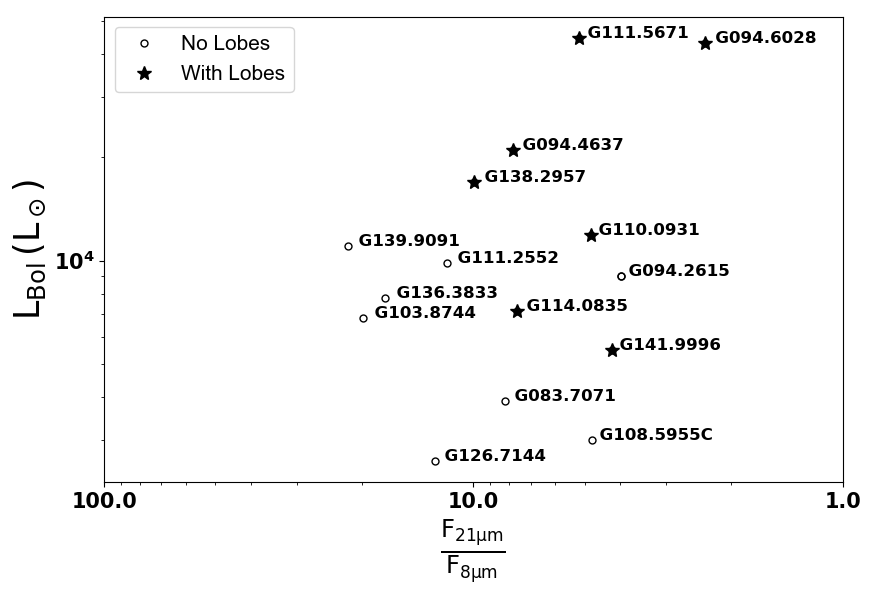}

    \caption{An illustration showing how mid-IR colour of MYSOs varies with bolometric luminosity. 
    }
    \label{evol_seq_obs}
\end{figure}

The infrared colour of young stellar objects is likely to be related to their evolution as they become bluer with age. A plot of mid-infrared colour versus bolometric luminosity of the sample, shown in Figure \ref{evol_seq_obs}, suggests that MYSOs of higher bolometric luminosities are more likely to have lobes compared to their lower luminosity counterparts, especially at lower values of $\mathrm{\frac{F_{21\mu m}}{F_{8\mu m}}}$. Whether this property of the jets is related to their evolution or driving power is not clear. However, the radio luminosity of a jet is correlated with its bolometric luminosity and force \citep{2015aska.confE.121A}, implying that sources with higher bolometric luminosity are likely to have more powerful lobes that are easily detectable in the radio compared to sources of lower luminosity. 

\subsubsection{Relationship between L-band radio luminosity and Bolometric Luminosity}
\begin{figure*}
    \centering
    \includegraphics[width = 0.8\textwidth]{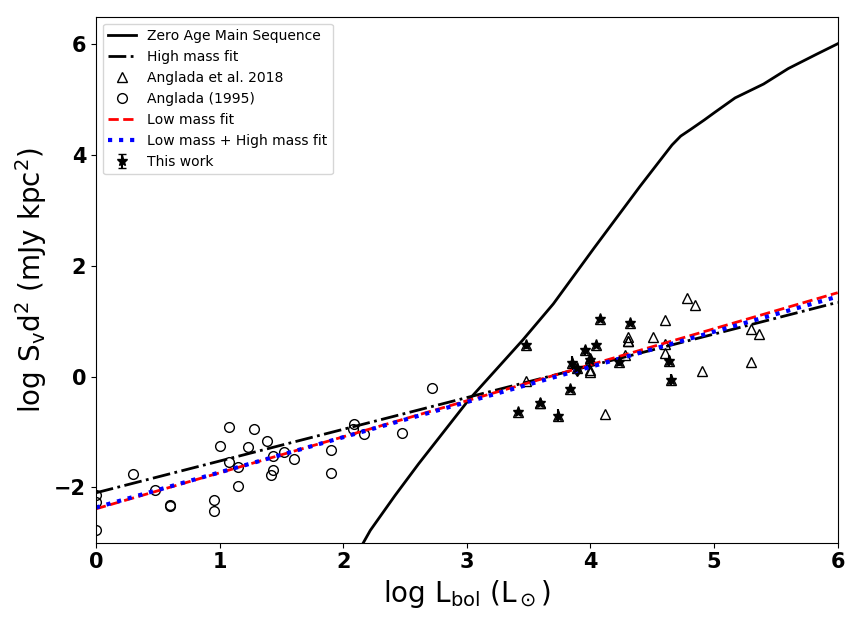}
    \caption{Relationship between L-band radio luminosity and bolometric luminosity for thermal radio sources. The asterisks represents the MYSOs under study while the empty circles and triangles represent the cores of low and high mass jets from \citet{1995RMxAC...1...67A} and \citet{2018A&ARv..26....3A} respectively.}
    \label{Lrad_vs_lbol_plot.png}
\end{figure*}

The radio luminosities of thermal radio jets show a correlation with bolometric luminosities. Different studies find the slope of the correlation to lie in the range $\mathrm{0.6 - 0.9}$, in a log-log plot. \citet{1992A&A...261..274C} and \citet{1995RMxAC...1...67A} estimated it as 0.8 and 0.6 respectively. Recently, \citet{2007ApJ...667..329S} found the slope to be $\sim$0.9 and 0.7 at 5\,GHz and 8.3\,GHz respectively, in low mass stars. At L-band, we find a slope of $\mathrm{0.64\pm0.03}$. This slope is for all the objects on the plot shown in Figure \ref{Lrad_vs_lbol_plot.png}, consisting of cores of L-band sources and thermal radio jets taken from the literature. Fluxes of the sources from the literature were scaled to L-band frequency using a spectral index of 0.6. High mass sources in the plot show a slope of $\mathrm{\sim 0.57\pm0.15}$, consistent with the lower end of the previous estimates i.e $\sim$ 0.6-0.9. In spite of the evidence of non-thermal emission at L-band, the cores are largely thermal, manifesting the properties of free-free emitters and showing no significant effect on the slope. The lower slope of high mass sources may be due to a smaller range of luminosity and uncertainties in their distances and fluxes. Clearly, the slopes are comparable to \citet{2018A&ARv..26....3A} who derived an empirical equation relating the two quantities at 8.3\,GHz, for objects of bolometric luminosities $\sim 1L_\odot \leq L_{bol} \leq 10^6\, L_\odot$ (see equation \ref{radio_lum_eq}): 
\begin{equation}
\label{radio_lum_eq}
    \Big(\frac{S_\nu d^2}{mJy\,kpc^2}\Big) = 10^{-1.90\pm0.07}\Big(\frac{L_{bol}}{L_\odot}\Big)^{0.59\pm0.03}
\end{equation}
The slopes, as seen in Figure \ref{Lrad_vs_lbol_plot.png}, demonstrates the similarity between low and high mass protstellar jets, owing to the comparable radio-bolometric luminosity correlations, implying that they have a common ionization, and perhaps driving mechanism. 

\subsection{Derived properties of the radio jets}
Quantities such as the jets' mass loss rates, injection radius and opening angles can be estimated from the fluxes and angular sizes of their cores. Fluxes and angular sizes of the cores at L-band, together with their spectral indices, were used to calculate some of these properties in accordance with the \citet{1986ApJ...304..713R} model.

\subsubsection{Mass loss rate}\label{section_mass_loss_rate}
\citet{1986ApJ...304..713R} modelled free-free emission from an ionised jet. The model can be used to calculate the mass loss rate through the jet if its ionization fraction, electron temperature, terminal velocity and an inclination angle to the line of sight are known. The model estimates the mass loss rate through a conical jet of pure hydrogen, electron temperature $T_e  = 10^4$\,K and inclination angle $i = 90^o$, in units of $\rm{10^{-6}M_\odot yr^{-1}}$, to be:  
\begin{equation}
    \dot M_{-6} = 0.938v_t x_o^{-1}\Big(\frac{\mu}{m_p}\Big)S^{0.75}_{mJy} \upsilon_o^{-0.75\alpha} d_{kpc}^{1.5} \upsilon_m^{0.75\alpha-0.45} \theta_o^{0.75} F^{-0.75}
\end{equation}
\noindent where $\alpha$, $\mu$, $m_p$, $v_t$, $x_o$, $S_{mJy}$, $\upsilon_o$, $\upsilon_m$, $d_{kpc}$ and  $\theta_o$ are the derived spectral index, mean particle mass per hydrogen atom, proton mass, terminal velocity of the jet in $10^3\, \mathrm{kms^{-1}}$, ionization fraction, observed flux density in mJy, frequency of observation (here L-band) in units of 10 GHz, turn-over frequency in units of 10 GHz, object's distance in kpc and the opening angle (in radians given as $\theta_o = 2 \tan^{-1}(\frac{\theta_{min}}{\theta_{maj}})$, \citealt{2000prpl.conf..815E}) respectively. $F$ is a quantity that depends on the opacity and spectral index of a jet given as \citep{1986ApJ...304..713R}:

\begin{equation}
    F\equiv F(q_\tau, \alpha) = \frac{2.1^2}{q_\tau(\alpha - 2)(\alpha + 0.1)}
\end{equation}
\noindent where $q_\tau$ is a quantity that describes how opacity varies along the jet. $F$ was calculated using the parameters of a standard spherical jet model if the spectral index of a source lies in the range $0.4 \leq \alpha \leq 0.8$ and standard collimated model if $ 0 < \alpha < 0.4$. $q_\tau$ was taken to be -3 if $0.4 \leq \alpha \leq 0.8$ and -2 for $ 0 < \alpha < 0.4$. The ionization fraction of MYSO jets is determined by collisions within the jets (\citealt{2002RMxAC..13....8B}, \citealt{1994ApJ...436..125H}, \citealt{1999A&A...342..717B}) and perhaps ionising radiation from the central source. Typical values of jet ionization fraction $x_o$ are $\sim0.02$ to $0.4$ (\citealt{2002RMxAC..13....8B}, \citealt{1999A&A...342..717B}), a quantity that varies along a jet in accordance with recombination models. The maximum value of ionization fraction in the literature ($x_o = 0.4$) was used in the calculation of mass loss rate in this work as all the YSOs considered here are massive and are expected to have more powerful jets that are better collisional ionisers.
Terminal velocity of jet materials (or simply jet velocity) was estimated from its range i.e. $\sim270$\,km/s in low-mass stars to $\sim1000$\,km/s in high-mass stars \citep{2015aska.confE.121A}. The average velocity of optical jets is $\sim 750\,$km/s \citep{1994ASPC...62..237M}, a value comparable to \citet{2011ApJ...732L..27J} and \citet{1998ApJ...502..337M} estimates of $\sim 500$\,km/s in Cepheus A HW2 and HH 80-81 respectively. Whereas the average velocity in optical jets is $\sim750$\,km/s, a value of $v_t = 500$\,km/s, estimated via proper motion of a radio jet \citep{1998ApJ...502..337M} was adopted as a reasonable approximation of the sample's jets' terminal velocity, in calculating the mass loss rate.  A turn-over frequency of 50\,GHz was assumed as the value is generally expected to be higher than 40\,GHz \citep{2018A&ARv..26....3A}.
Finally, a semi-opening angle of 31$^\circ \pm$3, the average semi-opening angle for nine sources whose angles were calculated directly, was adopted for sources whose angles could not be estimated from $\theta_o = 2 \tan^{-1}(\frac{\theta_{min}}{\theta_{maj}})$. The range of the calculated semi-opening angles, $25-35^o\circ$, is comparable to the estimate by \citet{2016A&A...585A..71M} i.e 20$^\circ \pm5^\circ$.

\begin{figure*}
    \centering
    
    \includegraphics[width = 0.85\textwidth]{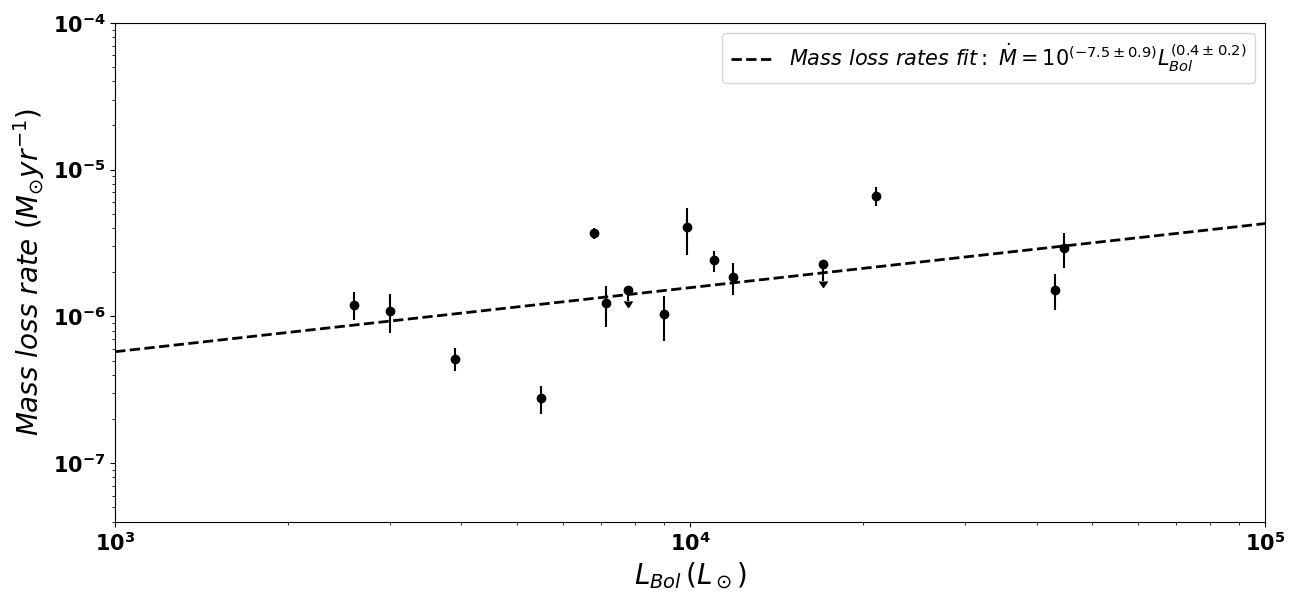}

    \caption{A plot showing how mass loss rates of the sources vary with their bolometric luminosity.}
    \label{mass_loss_and_accretion_rates}
\end{figure*}

The sample's mass loss rates $\dot M$ lie in the range $\sim 2.8\pm0.6 \times 10^{-7}$ to $6.6\pm0.9 \times 10^{-6}\,\mathrm{M_\odot yr^{-1}}$ (see Figure \ref{mass_loss_and_accretion_rates}), typical of rates from MYSO (\citealt{1993ASSL..186..231P}, \citealt{1994ASPC...62..237M}, \citealt{2016MNRAS.460.1039P}) and $\sim 10^2 - 10^3$ higher than low mass counterparts. Mass loss rates of the MYSOs were found to be related to their bolometric luminosity by the equation $    \dot M \simeq 10^{-7.5\pm 0.9} L_{bol}^{0.4\pm0.2}$. Although low mass YSOs are expected to have lower mass loss rates ($\sim 10^{-10} - 10^{-8}\,M_\odot yr^{-1}$; \citealt{1995RMxAC...3...93H}, \citealt{2000prpl.conf..759K}, \citealt{1994ASPC...62..237M}), some of them have rates that are comparable with high mass ones, in effect implying that distinguishing such jets from high mass ones \citep{1997ApJ...476..771C} is not straightforward. 
  
\subsubsection{Accretion rate}
Stellar jet outflows are considered to be closely related to accretion inflows \citep{2007prpl.conf..277P}, therefore it is reasonable to estimate accretion rates of YSOs from their mass loss rates. \citet{1995ApJ...452..736H} empirically established a correlation between accretion and mass loss rates of low mass stars where they demonstrated that  $0.00008 \leq \frac{\dot M}{\dot M_{acc}} \leq 0.4$ with an average value of $0.01$. \citet{2005MNRAS.356..167M} also modelled MHD jets and found that the stellar angular momentum problem in YSOs can only be solved if jet outflow rates are $\sim 10\%$ of accretion rates although magnetic field strength can influence the ratio. This value lies within \citet{1995ApJ...452..736H}'s range, albeit a factor of 10 above the average.  

The average value of $\frac{\dot M}{\dot M_{acc}} \sim 0.01$ in \citet{1995ApJ...452..736H}'s work was therefore used in estimating accretion rates resulting in $\dot M_{acc} \sim 10^{-4.6\pm0.3} - 10^{-3.2\pm0.2}\,M_\odot yr^{-1}$ which is consistent with high rates theoretically expected from the sources i.e $\mathrm{\sim 10^{-4} - 10^{-3}\,M_\odot yr^{-1}}$ (\citealt{2002Natur.416...59M}). The accretion rate of one of the sources, G094.6028, $\mathrm{1.5\pm0.4 \times 10^{-4}\ M_\odot yr^{-1}}$ was found to be comparable to $\mathrm{\sim 1.2\pm0.06 \times 10^{-4}\ M_\odot yr^{-1}}$ \citep{2017MNRAS.472.3624P}, derived using hydrogen Bracket gamma spectral lines.

\section{Conclusions}
We studied a sample of fifteen MSYOs, at least forty percent of which show evidence of non-thermal emission. Six of the sources; G094.6028, G103.8744, G111.5671, G114.0835, G138.2957 and G141.9996, manifest non-thermal radio lobes, similar to those seen in HH objects. The study, therefore, suggests that massive stars form via accretion disks in a manner similar to low mass stars but with high accretion rates. 

All the MYSOs have thermal central sources which are considered the drivers of the jets. The average spectral index of the thermal sources is $\sim 0.42\pm0.27$, clearly similar to the theoretical 0.6, expected from ionized winds.
They drive the jets at a rate higher than low mass counterparts with mass loss rates $\mathrm{\sim 3 \times 10^{-7}}$ to $\mathrm{7 \times 10^{-6}\,M_\odot yr^{-1}}$. 

Clearly, lobes of some MYSO jets confirm the presence of synchrotron emission, implying the presence of magnetic fields and charged particles at relativistic velocities. This finding suggests that magnetic fields play a significant role in the jets, perhaps in driving them from the core (\citealt{1982MNRAS.199..883B}, \citealt{1994ApJ...429..797S}). However, a thermal jet can also interact with ambient magnetic field producing non-thermal emission. 
A study of 46 young MYSOs by \citet{2016MNRAS.460.1039P} also found that about $50\%$ of the sample may be associated with non-thermal lobes, again emphasising the significance of magnetic fields in jets of massive protostars. 

It was also found that sources of higher bolometric luminosity are more likely to have lobes compared to lower luminosity counterparts and that some of them may be variable.

\section*{Acknowledgements}
WOO gratefully acknowledges the studentship funded by the UK's Science and Technology Facilities Council (STFC) through the DARA project, grant number ST/M007693/1. We also thank the referees for their very helpful suggestions.

\bibliographystyle{apj}
\bibliography{references}

\appendix

\section{Discussion of individual objects}

\subsection{G083.7071}
\begin{figure}
    \centering
    \includegraphics[width = 0.5\textwidth]{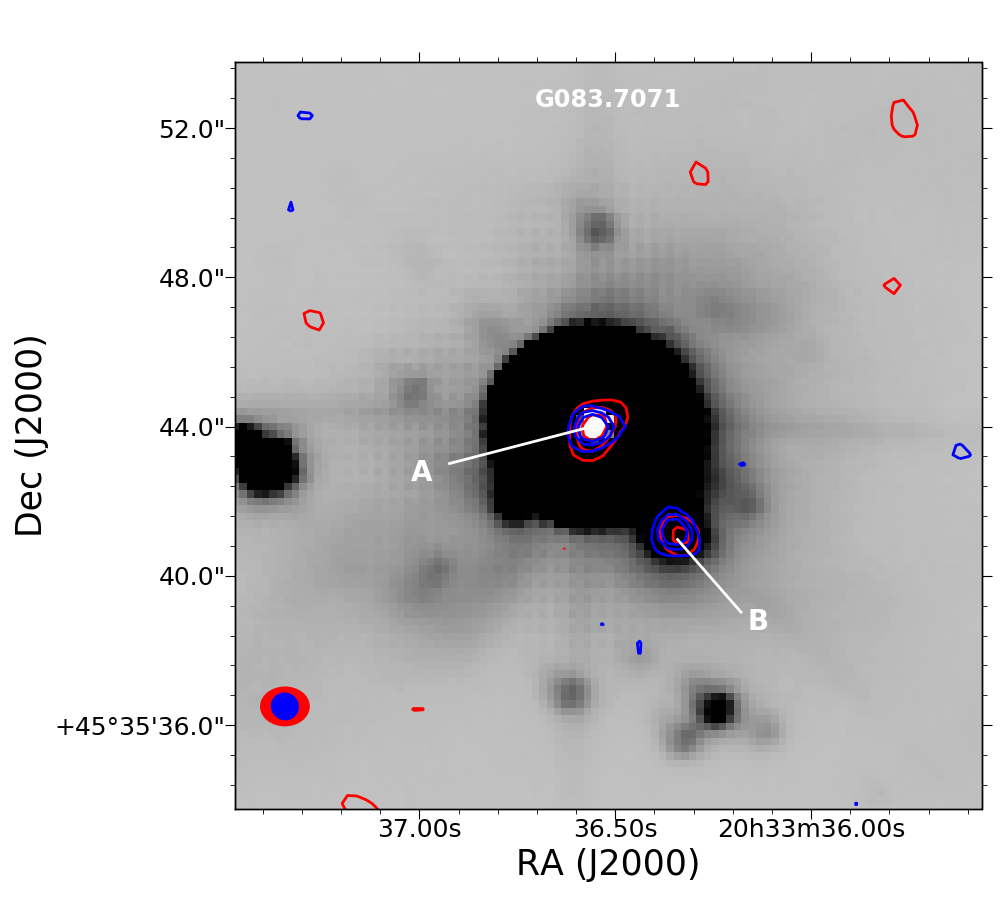}

    \caption{UKIDSS K band IR image of the field with L- and C-band contours shown in red and blue lines respectively.
    }
    \label{G083_7071plots}
\end{figure}

Two sources, A and B were detected in the field (see Figure \ref{G083_7071plots}). Source A whose spectral index $\alpha = $0.44$\pm$0.10 is the MYSO. The L-band emission of A is elongated, perhaps signifying presence of a jet. 
Apart from the MYSO, source B, located $\sim 3.5^{\prime\prime}$ to the SW of A was also detected in the field. This source was detected in UKIDSS's IR K band just like A but not at Q-band where the rms noise $\sim 29\,\mu$Jy/beam. It is a thermal radio source of L-C spectral index $\alpha_{LC} \sim 0.60\pm0.13$, perhaps another MYSO.

\subsection{G094.2615}
\begin{figure}
    \centering
    
    \includegraphics[width = 0.5\textwidth]{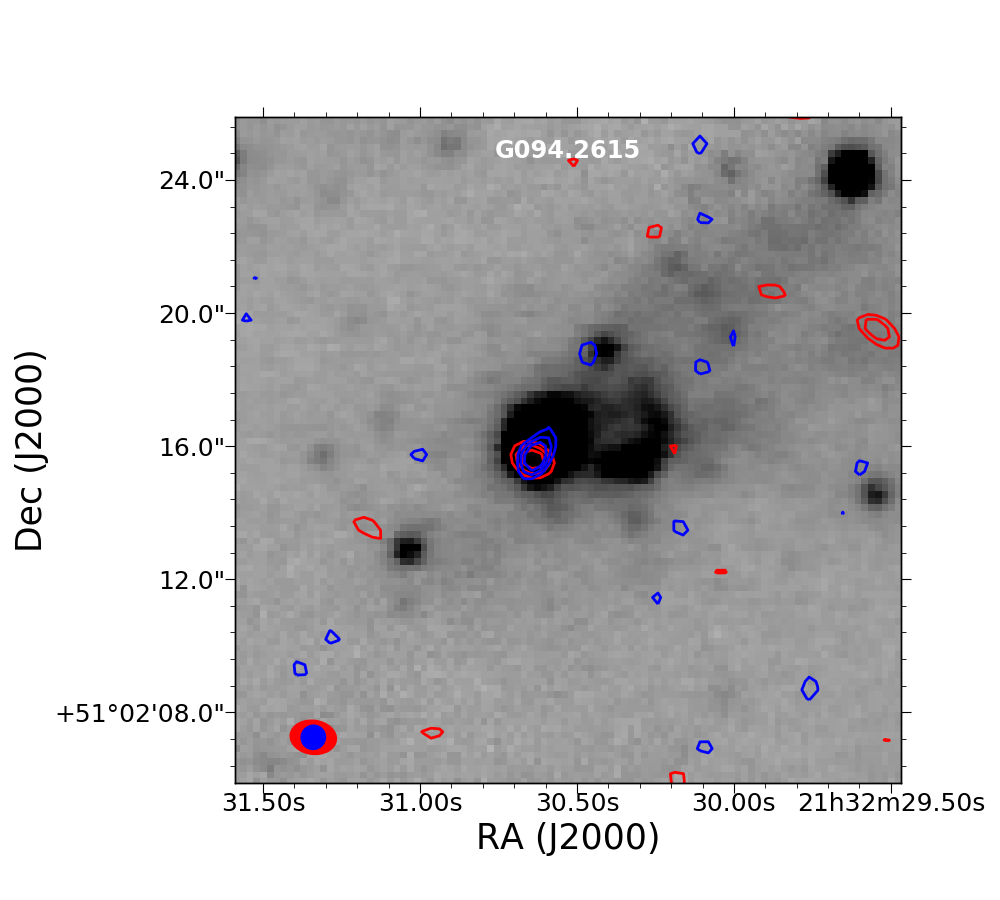}

    \caption{G094.2615 UKIDSS K band map with L- and C- band contours shown in red and blue lines respectively.}
    \label{G094_2_IRplot}
\end{figure}

G094.2615 is made up of two sources, A1 and A2 (Purser et al 2019 in prep), clearly seen in the C-band map of high resolution (see Figure \ref{L_band_images}). Its L-C band spectral index is lower than that of C-Q band which could be due to non-thermal contribution at L even-though the overall spectral index shows that it is thermal i.e $\alpha_\nu = 0.23\pm0.18$. 
A1 was detected at both C- and Q bands giving it a C-Q spectral index of 0.47$\pm$0.24, consistent with the MYSO core.
UKIDSS map of the field (Figure \ref{G094_2_IRplot}) shows emission that is largely aligned in a SE-NW direction similar to that of C-band. The UKIDSS emission, 2.12\,$\mu$m H$_2$ line emission (\citealt{2017ApJ...844...38W}, \citealt{2010MNRAS.404..661V}) and CO molecular outflow \citep{2004A&A...424..179F} all points to a source that drives a jet in a SE-NW direction. There is also a NIR emission to the west of the MYSO but its source is unclear. 

\subsection{G094.4637}
Two sources, G094.4637(A+B+C+D) and a lobe (source E), of spectral indices 0.53$\pm$0.09 and 0.21$\pm$0.45 respectively were detected at L-band. The spectral index of G094.4637(A+B+C+D) suggests that it is thermal but one of its component, D show non-thermal property (Figure \ref{L_band_spix_images}).
Lobe E is aligned with components A, B and D in such a way that they exhibit a central thermal core and associated jet lobes. Its C- and L-band emission are slightly offset in a NE-SW direction, similar to that of the outflow, consequently giving it a non-thermal property towards the SW where L-band emission is stronger and a thermal characteristic towards the NE. The displacement of its L-band emission is consistent with that of a jet originating from A and directed towards E.

\begin{figure}
    \centering
    
    \includegraphics[width = 0.48\textwidth]{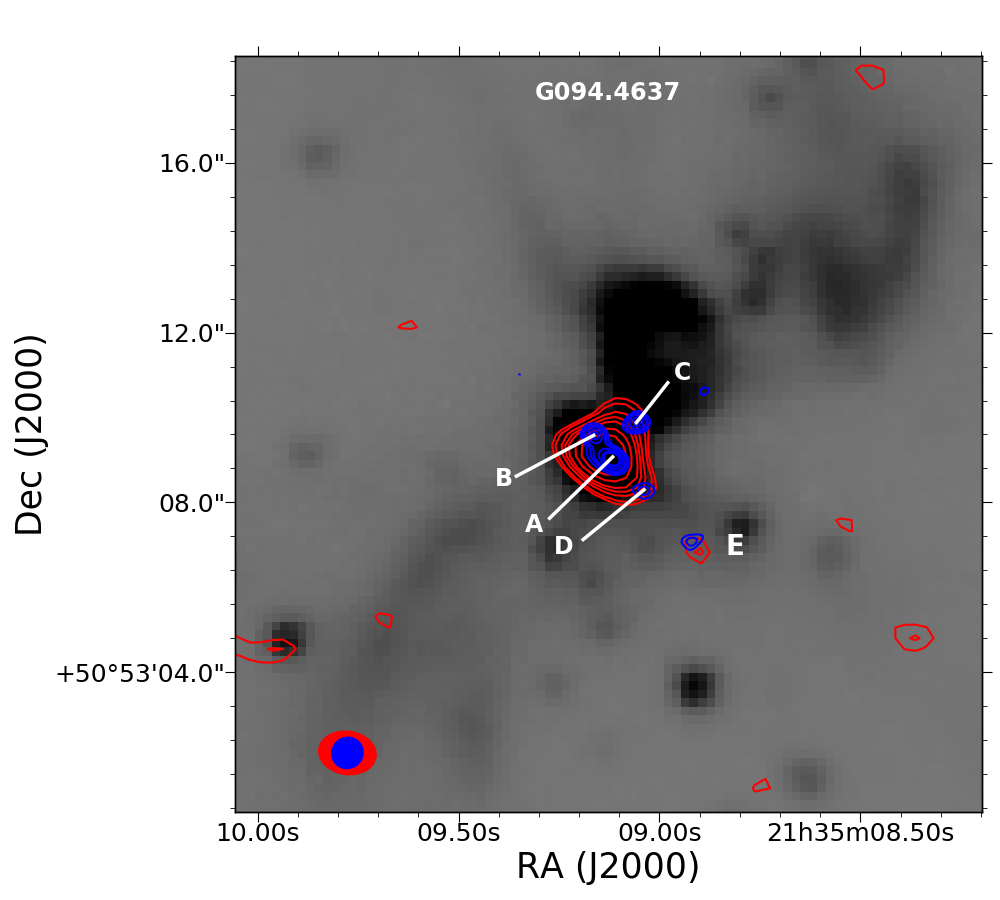}

    \caption{G094.4637 UIST map together with L- and C- band contours shown in red and blue respectively. Both L- and C- band contours start from $3\sigma$  and increase in steps of $2\sigma$ and $4\sigma$ respectively.}
    \label{G094_4plots}
\end{figure}

Of the four sources, only A appears to be the core given its proximity to the IR source. Again, it is the only source in the group that was detected at Q-band \citep{2017PhDT.......147P} where it splits into two thermal sources, A and A2 of C-Q band spectral indices 0.39$\pm$0.12 and $>$1.47$\pm$0.17, suggesting that it is a binary and A or A2 drives the jet. The morphology of L-band emission as well as the alignment of the sources in C-band also suggests presence of a jet. The results of \citet{1992ApJ...398L..99S} and \citet{2015MNRAS.450.4364N} indicate the presence of multiple jets. \citet{2015MNRAS.450.4364N}, for example, detected a monopolar and two bipolar outflows in the field. Two of their outflows, 2a and 3a, have position angles that are comparable to the position angles of D and C from the location of the IR source i.e $\sim$ 210$^o$ and 330$^o$ respectively. Despite the evidence of multiple outflows in the field, the driving sources are not clear from both L- and C-band emission.
Source C, detected at Q-band \citep{2017PhDT.......147P} may also be a thermal core. Its C-Q band spectral index is $\alpha_{CQ} \sim 0.86\pm0.05$. The source is located at a position angle of $\sim -30^o$ from A and is aligned with the main K-band NIR emission (Figure \ref{G094_4plots}) hence is a potential thermal core. A 12.5\,$\mu$m image of the field also display two sources \footnote{$\mathrm{http://rms.leeds.ac.uk/cgi\-bin/public/RMS\_DATABASE.cgi}$}, possibly A and C, aligned in a SE-NW direction, confirming that there are two cores.

\subsection{G094.6028}
G094.6028, also known as V645 Cyg, has a thermal radio core of spectral index 0.44$\pm$0.15. The source, named A is a known variable protostar of spectral type $\sim $O7 - O9 \citep{1977ApJ...215..533C}. Earlier observation of the source by \citet{1997ApJ...482..433D} at 8.3\,GHz registered a higher flux, perhaps confirming its variability \citep{2006A&A...457..183C}. Apart from the MYSO, a nearby source, B, of flux density 0.07$\pm$0.02\,mJy was also detected $\sim 4^{\prime\prime}$ away to the NW at L- but not C-band. It appears to be a non-thermal lobe of spectral index $<-0.42$. 
The MYSO is known to emit Br$\gamma $, accrete material, drive bipolar CO outflow and harbour a circumstellar disk (\citealt{2015MNRAS.447..202E}, \citealt{1989ApJ...339.1078H}, \citealt{2006A&A...457..183C},  \citealt{1989ApJ...341..288S}). 

\begin{figure}
    \centering
    
    \includegraphics[width = 0.5\textwidth]{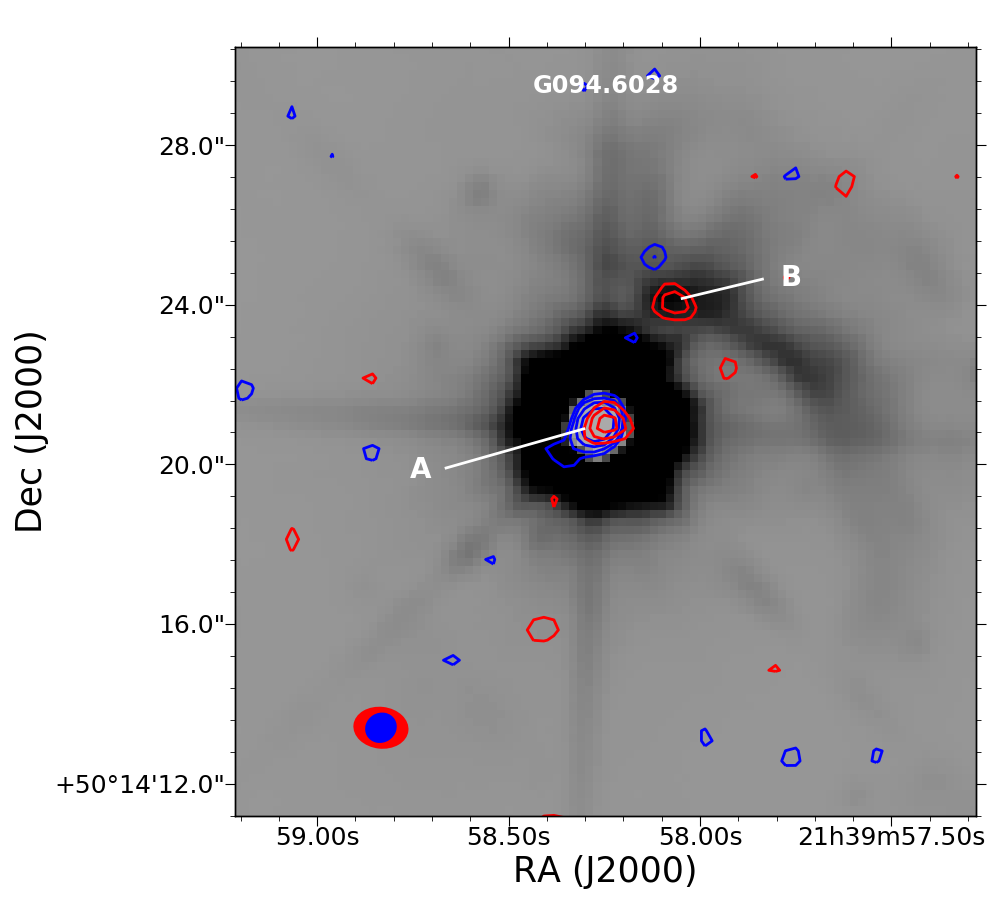}
    
    \caption{G094.6028 UKIDSS K band map with L (3,4,5\,$\sigma$) - and C(3,5, 9,15\,$\sigma$)- band contours shown in red and blue continuous lines respectively.}
    \label{G094_6plots}
\end{figure}

\subsection{G103.8744}
\begin{figure}
    \centering
    \includegraphics[width = 0.5\textwidth]{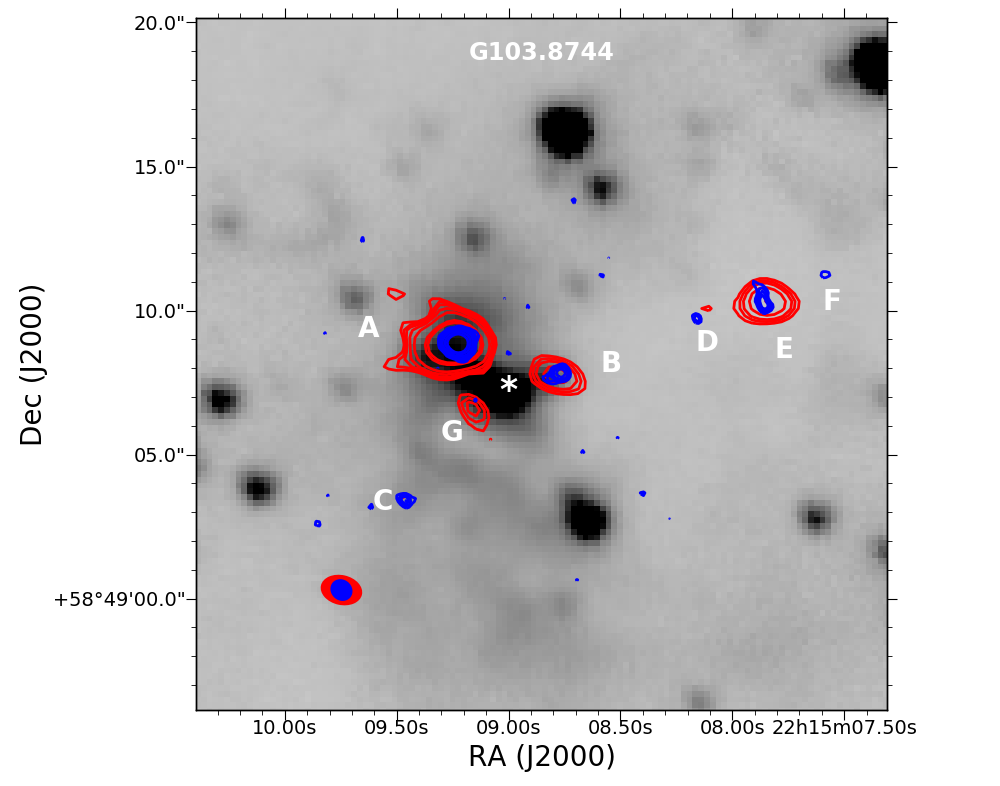}
    \caption{UKIDSS K band image of G103.8744 field with L- and C- band contours shown in red and blue lines respectively. The contour levels both start at 3$\sigma$. C-band ones increase in steps of 4$\sigma$ while L-band in steps of 2$\sigma$.}
    \label{G103_plots}
\end{figure}

\begin{figure}
    \centering
    \includegraphics[width = 0.5\textwidth]{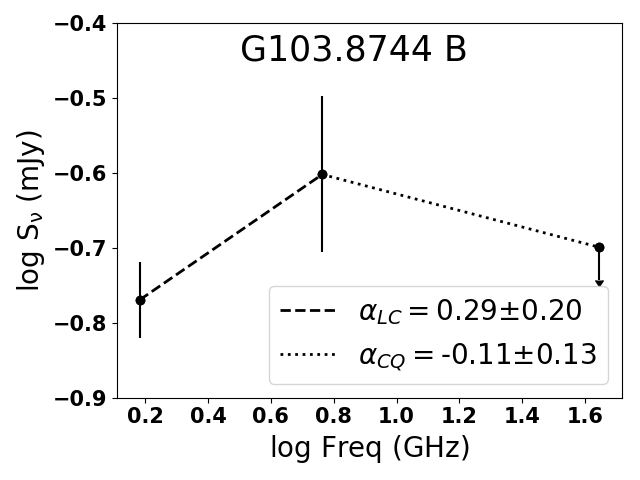}
    \caption{Spectrum of G103.8744-B.}
    \label{HIIregionG103}
\end{figure}

G103.8744 is associated with the IRAS source 22134+5834. It has five radio sources in its field, three of which are within a radius of $\sim 3^{\prime\prime}$ from the location of the IR source (white asterisk in Figure \ref{G103_plots}). Both A and  B are thermal and are close to the location of the MYSO but their spectra suggests that they are HII-regions (see Figure \ref{HIIregionG103}; spectrum of G103.8744-B). Indeed, A is a known ultra-compact HII-region \citep{2016A&A...587A..69W} whose turn-over frequency seems to lie between 5 and 8.3\,GHz (see Figure \ref{spectral_index_plots}). B has a fainter but compact IR counterpart suggesting that it is a source, perhaps a low mass YSO or an obscured massive protostar. Source G, on the other hand was not detected at C-band where rms noise is $\sim$15\,$\mu$Jy/beam implying that it is a non-thermal radio source whose spectral index is $<-$0.44. 

Seeing that B has an IR counterpart and sources B, D, E, F and G are all aligned in a SE-NW direction, reminiscent of jet lobes, B could be the jet driver. \citet{2016A&A...587A..69W} imaged the source at radio frequencies but could not separate the MYSO from the HII region. They estimated a combined flux for the sources as 2.1$\pm$0.2\,mJy, 3.5$\pm$0.11\,mJy and 2.9$\pm$0.70\,mJy at 1.3\,cm, 6.0\,cm and 20\,cm respectively. These fluxes were also used in generating the spectrum of source A as it is the dominant emitter in the field. A map of $H_2$ emission line at 2.12$\mu$m does not reveal any clear evidence of shocks or photo-dissociation locations
\citep{2002ApJ...576..313K}, however its $C^{18}O\,(J=1-0)$ map displays an outflow \citep{2001PASJ...53..799D} in a direction similar to that of the IR emission i.e a position angle PA$\sim$45$^o$.

\subsection{G108.5955C}
L- and C- band emission of G108.5955C show different morphologies. Its L-C band spectral index $\alpha_{LC} = 0.20\pm0.16$ shows that it is thermal but the spectral index map displays evidence of non-thermal emission on the eastern part. Both \citet{2009RAA.....9.1343W} and \citet{2010RAA....10..151X} detected molecular outflows in the field, however, $\mathrm{2.12\,\mu m}$ $\mathrm{H_2}$ line map \citep{2017ApJ...844...38W} does not show any evidence of knots which may be associated with the source.

\subsection{G110.0931} 

G110.0931's field contains two L-band sources, the MYSO and source D (see Figure \ref{L_band_images}), whose L-C band spectral indices are $0.36\pm0.14$ and $-0.08\pm0.15$ respectively. The MYSO encompasses three sources; A, B and C of spectral indices $\alpha_\nu$ = -0.14$\pm$1.49, 0.33$\pm$0.14 and -0.11$\pm$0.07 respectively (\citealt{2012ApJ...755..100R}, \citealt{2017PhDT.......147P}).

Unlike B and C, A was not detected at Q-band \citep{2017PhDT.......147P} consequently giving G110.0931 overall an HII region-like spectrum. The
detection of sources B and C at Q-band suggests that they could be
cores, though we note that only B is detected by NOEMA\footnote{NOrthern Extended Millimeter Array} at 1.3mm (Bosco et al 2019 in prep), whereas we would expect any core at this evolutionary stage to be detectable in the mm. Overall it seems likely therefore that B is the core and A and C are jet lobes. Source D, detected on the eastern part of G110.0931 was not detected at Q-band where the field rms is $\sim 0.044$\,mJy/beam, setting a limit on its spectral index as $\alpha \leq -0.12$, suggesting that it may be an optically thin HII region. This source does not show any clear association with G110.0931 which is approximately $3.3^{\prime\prime}$ away from it.

\citet{2015MNRAS.450.4364N} detected a bipolar outflow that is aligned in a direction similar to that of both L- and C- band emission. 2MASS IR- emission as well as the K-band continuum emission shown in \citet{2015MNRAS.450.4364N} are also oriented in the same direction. \citet{2014ApJ...790...84L} detected $\mathrm{NH_3 (1,1)}$ emission that is aligned in a direction perpendicular to the jet and passing through the location of the central source, B. The orientation of $\mathrm{NH_3 (1,1)}$ emission, which is a tracer of dense molecular gas \citep{2013A&A...553A..58L} and that of the molecular line at 2.12$\,\mu$m suggests that the source may be a disk-jet system. Finally, source C has higher radio flux compared to A, perhaps an indicator of a E-W density gradient.

\subsection{G111.2552}
G111.2552, also known as I23139 \citep{2006AJ....132.1918T} is a thermal radio source of spectral index $\alpha_\nu \sim 0.67\pm0.04$. Its L- and C-band emission are largely aligned in a SE-NW direction (see Figure \ref{L_band_spix_images}). 
Its Q-band flux appears to be lower than expected (see Figure \ref{spectral_index_plots}) when compared to estimates from \citet{2006AJ....132.1918T} i.e a flux density of 0.98$\pm$0.24\,mJy and 0.53$\pm$0.13\,mJy at  23\,GHz and 8.5\,GHz respectively. Furthermore, \citet{2006AJ....132.1918T} detected a nearby source of flux density 0.22$\pm$0.07\,mJy, approximately 0.5$^{\prime\prime}$ to the south, which they suppose is I23139's companion. Finally, molecular outflows \citep{1989A&A...215..131W} and water maser emission (\citealt{1995A&AS..112..299T}; \citealt{2006AJ....132.1918T}) detected in the field suggests presence of outflow activity \citep{2005ApJS..156..179D}.

\subsection{G111.5671}
Three sources, A, B and C were detected in the field at L-band. Source A, also known as NGC 7538 IRS9 \citep{1991ApJ...371L..33M} seems to be the core of G111.5671 while B and C, located $\sim 2.8^{\prime\prime}$ and $ 1.7^{\prime\prime}$  away to the SW and NE of the core respectively are its lobes. L-C spectral indices ($\alpha_{LC}$) of B and C are $<-0.6$ and $0.60\pm0.21$ suggesting that B is a non-thermal source while C is thermal. Also, B seems to be more diffuse, bow shaped and extended towards A as seen in the C-band image of higher resolution, suggesting that it is a bow shock.

A has a flux density of 1.00$\pm$0.44\,mJy, 0.76$\pm$0.15\,mJy \citep{2005ApJ...621..839S} and $<0.51$\,mJy \citep{1996ApJ...465..363R} at 4.86\,GHz, 8.46\,GHz and 15\,GHz respectively. \citet{2005A&A...437..947V} used both C and A configurations of the VLA to estimate its flux density at 43\,GHz, obtaining $2.9\pm0.3$\,mJy and $1.1\pm0.3$\,mJy respectively. The fluxes  gives it a spectral index $\alpha = 0.84\pm0.09$. The three sources were previously detected by \citet{2008A&A...485..497S} as a single source at 1.3\,cm. The source, aligned in a NE-SW direction has a spectral index of $>+0.1$. They also detected another source to the SW of A, B and C whose spectral index is $>-1.2$. The orientation of the sources suggests the presence of a jet in a NE-SW direction.
  
Both a highly collimated HCO$^+$ outflow observed by \citet{2005ApJ...621..839S} and 2.12$\,\mu$m $H_2$ emission by \citet{1998AJ....115.1118D} are aligned in a NE-SW direction, similar to the alignment of the radio sources. An offshoot of the IR emission that is directed towards the south is also seen in both 2MASS and \citet{2005ApJ...621..839S}'s map. This emission is in the same direction as a CO outflow (\citealt{2005ApJ...621..839S}, \citealt{1998AJ....115.1118D}), perhaps implying presence of two outflow drivers. Furthermore, the $CO\, J= 2-1$ observed by \citet{1991ApJ...371L..33M} shows a blue outflow lobe that is aligned in a NE - SW direction and a red one in the SE-NW. 

\subsection{G114.0835}
Two sources, A and B were detected at L-band (see Figure \ref{L_band_images}). B is a non-thermal source of spectral index $\alpha \sim -0.42\pm0.19$ while A is a thermal source whose spectral index is 0.40$\pm$0.20. A high resolution image of A at C-band reveals two partially resolved sources, A1 and A2, that are oriented in a NE-SW direction. The two sources (Figure \ref{L_band_images}) are likely to be thermal. It is not clear if A and B are associated but B is elongated in the direction of IR emission in the field (\citealt{2015MNRAS.450.4364N}, \citealt{2006AJ....131.1163S}) signifying the likelihood of the presence of a jet. Also, A and B are aligned in the direction of IR emission. 

\subsection{G126.7114} 
G126.7114 is a thermal source of spectral index $\alpha = 0.82\pm0.02$\,mJy. Its L-band and C-band emission are largely aligned in a SE-NW direction.  
Apart from G126.7114, a weak source (4$\sigma$ detection) of flux density 0.08$\pm$0.02 was also detected at C-band, approximately $4.5^{\prime\prime}$ away to the SE at ($\alpha = 01^h23^m33.29^s; \delta = +061^\circ48^m44.7^s$).
NIR K-band emission (\citealt{2001AJ....122..313J}; \citealt{2015MNRAS.450.4364N}) of the source is aligned in a direction similar to that of L- and C-bands suggesting the presence of a SE-NW outflow. \citet{2015MNRAS.450.4364N} also detected knots that are typical of a bipolar jet with a similar orientation i.e at a position angle PA$\sim160^o$. Moreover, \citet{2001AJ....122..313J} detected a polarization disk that is oriented in a direction perpendicular to that of the outflow, suggesting a disk-jet case.

\subsection{G136.3833}
\begin{figure}
    \centering
    
    \includegraphics[width = 0.5\textwidth]{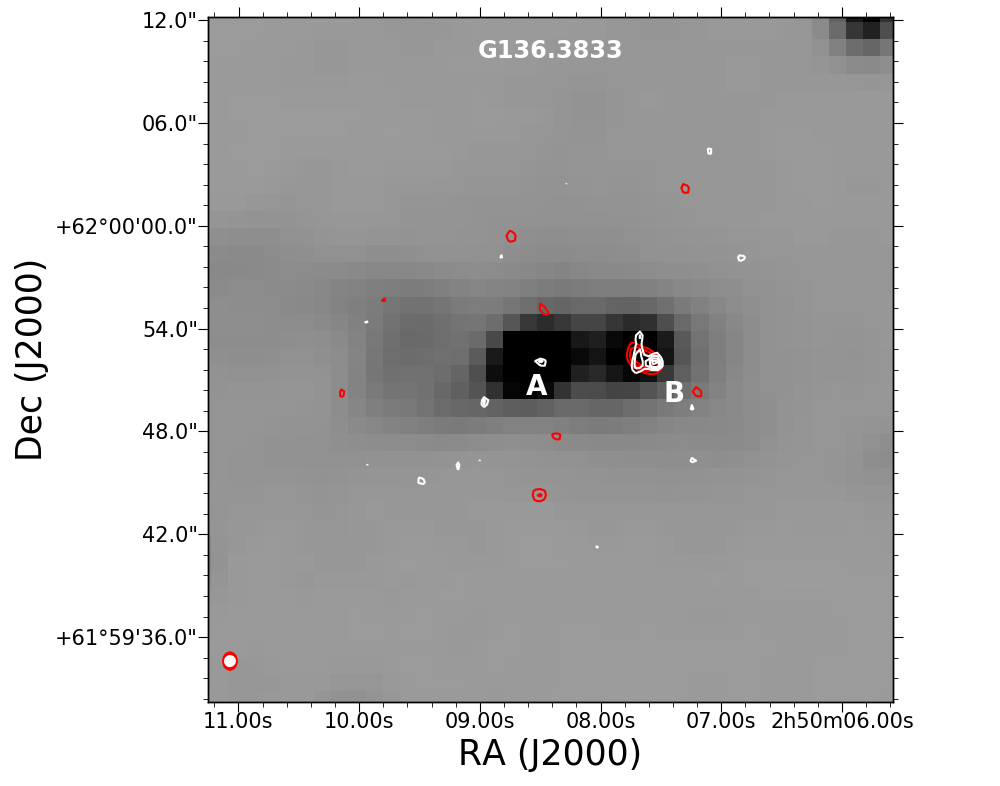}

    \caption{2MASS Ks band map of G136.3833 field and over-plots of L- and C-band contours shown in red and white lines respectively.}
    \label{G136_IR_map}
\end{figure}

G136.3833-A was detected at both C- and Q- but not L-band where the field rms is $\sim 50$\,$\mu$Jy/beam. It is a thermal radio source whose C-Q band spectral index $\alpha_{CQ}$ is $\sim1.22\pm 0.11$. A non-thermal source, B of spectral index $\alpha_{LC} = -0.68\pm0.08$, was detected approximately $6^{\prime\prime}$ to the west A at L-band. The source whose position angle is $60\pm12^o$ at L-band is elliptical at L- but irregular at C-band. The IR nebula in the field is oriented in an east-west direction \citep{2006AJ....131.1163S}, just like the alignment of sources A and B (see Figure \ref{G136_IR_map}). Both A and B have 2MASS counterparts at K-band and B may therefore be a non-thermal lobe or a star.

\subsection{G138.2957}
G138.2957, also known as AFGL 4029 \citep{1990ApJ...357L..45R} displays an extended morphology at C-band but splits into B and C at L-band. Both B and C are non-thermal lobes of flux densities 0.13$\pm$0.04\,mJy and 0.10$\pm$0.03\,mJy respectively. Polygons whose sizes are equivalent to the size of the sources at L-band were used to estimate their fluxes at C- band. The flux of B and C at C-band were determined to be approximately 0.07$\pm$0.02\,mJy and 0.06$\pm$0.02\,mJy respectively, giving them spectral indices of -0.46$\pm$0.31 and -0.38$\pm$0.33 respectively. Its spectral index map also display evidence of non-thermal emission, with some pixels having a non-thermal index of -0.6$\pm$0.2. The central core of the source was not detected at L-band. It appears to be located at the position of the peak emission at C-band (Figure \ref{L_band_spix_images}). A polygon that is equivalent to the C-band contour of level 7 (80$\mu$Jy/beam), which encloses the peak emission, was used in estimating the flux density of the core at C-band as 0.09$\pm$0.01\,mJy. A similar polygon was used at L-band, approximating its peak flux at the frequency as $<$0.06\,mJy. The fluxes suggest that the source has a thermal core of spectral index $\alpha >$0.69.

The source drives an optical jet \citep{1990ApJ...357L..45R} whose position angle is comparable to that of C-band emission ($\sim$ 75$^o$). \citet{2015MNRAS.450.4364N} also mapped a bipolar outflow of position angle PA $\sim$ 90 degrees which appears to be associated with it. \citet{1994ApJS...91..659K} and \citet{1999RMxAA..35...97C} observed the source using B and D configurations of the VLA, estimating its flux density at 3.6\,cm to be $\sim 2.1$\,mJy and $0.60\pm0.12$\,mJy, perhaps signifying variability. 





\subsection{G139.9091A}
\begin{figure}
    \centering
    
    \includegraphics[width = 0.5\textwidth]{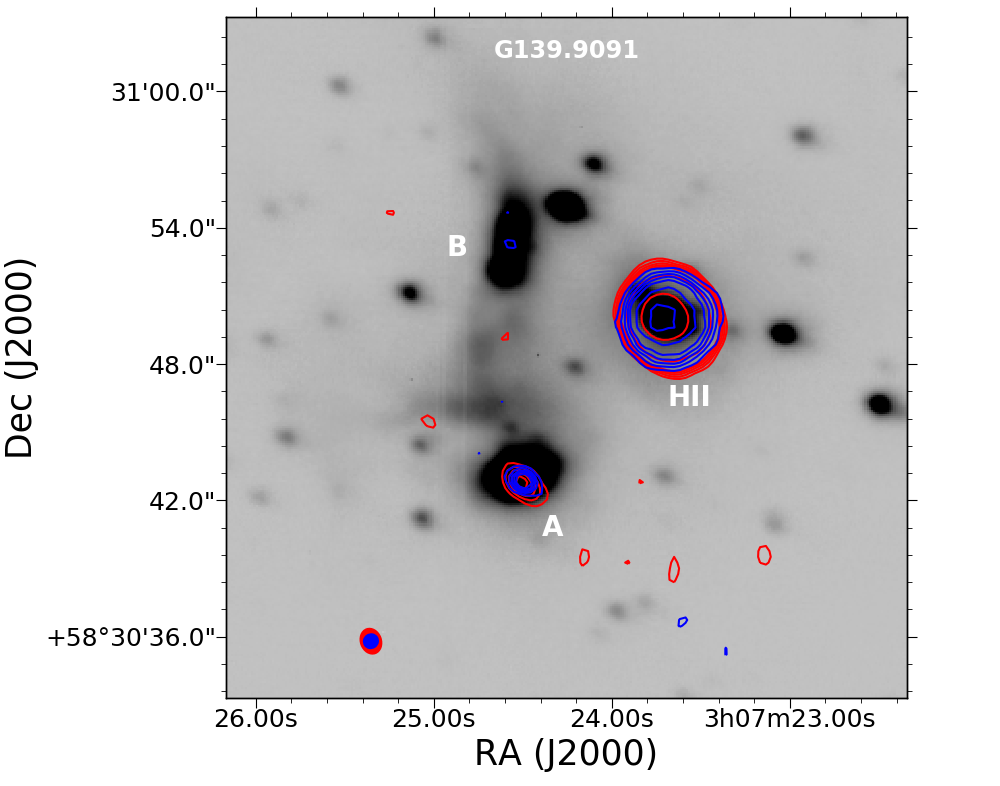}

    \caption{UKIDSS K band map of G139.9091A field and over-plots of L- and C-band contours shown in red and blue lines respectively.}
    \label{G139.9091A_spixmap_map}
\end{figure}

G139.9091A, also known as GL437S \citep{2010MNRAS.402.2583K} is a thermal radio source of spectral index $\alpha = 0.36\pm0.04$. It has a position angle of $\sim50\pm8^o$, similar to the orientation of \citet{2009A&A...494..157D}'s mid-infrared emission. Similarly, \citet{2012MNRAS.419.3338M} imaged it using archival data from JVLA's B,C and D configurations at 3.6\,cm and 2\,cm getting a comparable position angle i.e $\sim60^o$. Its flux density at 3.6\,cm and 2\,cm are 1.5$\pm$0.4\,mJy and $<2$\,mJy respectively, perhaps a sign of variability. Both infrared (Figure \ref{G139.9091A_spixmap_map}) and CO \citep{1992ApJ...397..492G} emission of the field are aligned in a N-S direction, however, its 2.12\,$\mu$m  map \citep{1998AJ....115.1118D} does not show evidence of outflow. 

Apart from G139.9091A, an optically thin HII region of spectral index -0.11$\pm$0.02, also identified as  AFGL 437W \citep{2012MNRAS.419.3338M}, was detected to the NW of A. It has flux densities of 19.0$\pm$0.5\,mJy and 16.3 $\pm$0.4\,mJy at L- and C-bands respectively. 


\subsection{G141.9996}
G141.9996, also known as AFGL 490 \citep{1998AJ....115.1118D} is an extended source that is largely elongated in a N-S direction. Two lobes, B and C, of flux densities 0.06$\pm$0.03\,mJy and 0.07$\pm$0.03 respectively were detected at L- but not at C-band. The L-C spectral indices of B and C are $\alpha_{LC} <-0.51$ and $<-0.63$ respectively, suggesting that they are non-thermal lobes.  The core, A, is thermal and is elongated in a direction that is similar to the alignment of A,B and C i.e $PA\sim 180^o$. Both CO outflow (\citealt{1995ApJ...438..794M}) and 2.12\,$\mu m$ emission (\citealt{1998AJ....115.1118D}, \citealt{2015MNRAS.450.4364N}) maps of the field are oriented in a NE-SW direction. However $CS\ J=2 \rightarrow 1$ emission of the source has a positions angle of -45$^o$ \citep{2002A&A...394..561S}.
 
Also, an optically thin source of spectral index -0.06$\pm$0.16 was detected approximately $13^{\prime\prime}$ away, at both bands. The source, located at $\mathrm{\alpha (J2000) \sim 03^h 27^{m} 37.07^{s}\ \delta \sim +058^\circ 46^{m} 59.4^{s}}$ has flux densities of 0.12$\pm$0.03\,mJy and 0.11$\pm$0.02\,mJy at L- and C-bands respectively.


\color{black}

\begin{figure}
    \centering
    
    \includegraphics[width = 0.5\textwidth]{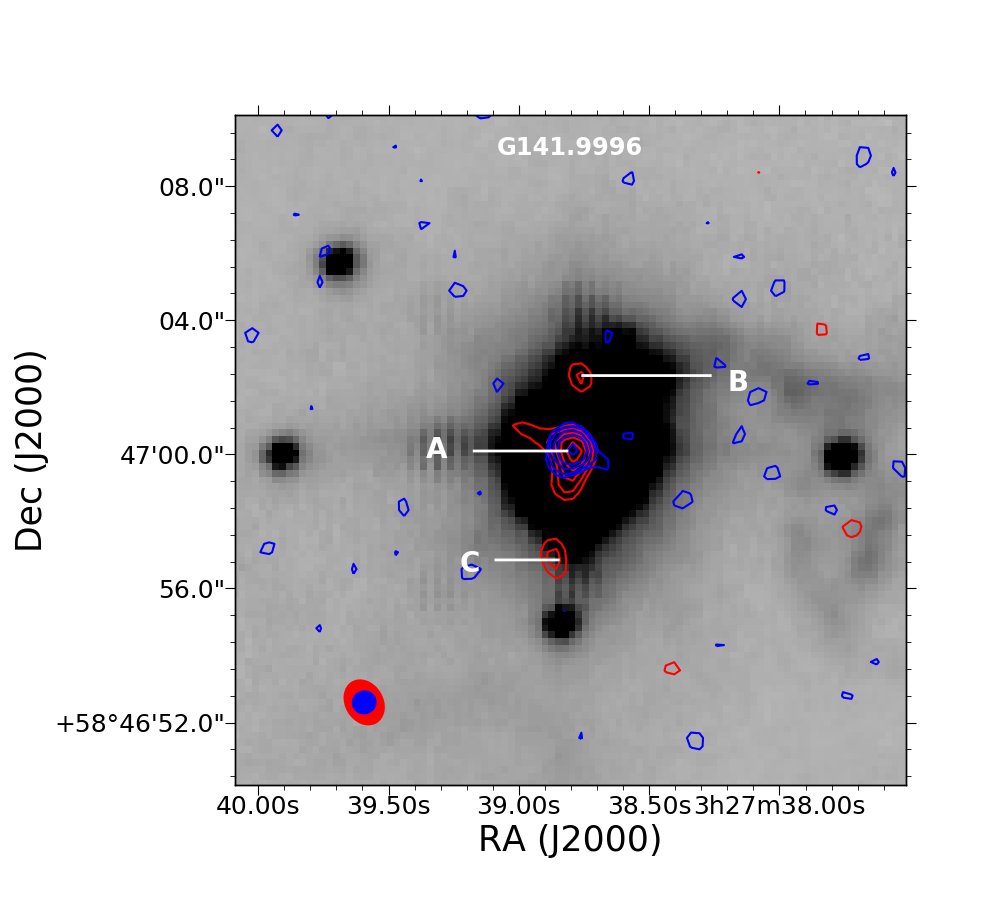}
    
    \caption{UKIDSS J band image of G141.9996 shown in grey together with L- and C-band contours in red and blue respectively.}
    \label{G141_IR_map}
\end{figure}

\label{lastpage}
\end{document}